\documentclass[12pt,jpg]{iopart}
\usepackage{graphicx}
\usepackage{dcolumn}
\usepackage{bm}
\usepackage{epstopdf}
\usepackage[dvipsnames]{xcolor}
\usepackage{soul}
\usepackage[utf8]{inputenc}
\usepackage[T1]{fontenc}
\usepackage{indentfirst}
\usepackage{har2nat}
\usepackage{enumitem}
\usepackage{xspace}
\usepackage{hyperref}
\hypersetup{breaklinks=true,colorlinks=true,linkcolor=blue,citecolor=blue,filecolor=magenta,urlcolor=cyan}
\usepackage{datetime}
\usepackage{amsfonts}

\newcommand{\newblock}{}

\renewcommand{\harvardurl}[1]{\url{#1}}

\newcommand{\tfrac}[2]{{\textstyle\frac{#1}{#2}}}
\newcommand{\text}[1]{\mbox{\scriptsize{#1}}}
\newcommand{\Fig}{Fig.}
\renewcommand{\Fig}{figure}
\newcommand{\Figs}{Figs.}
\renewcommand{\Figs}{figures}
\newcommand{\Eq}{Eq.}
\renewcommand{\Eq}{equation}
\newcommand{\Eqs}{Eqs.}
\renewcommand{\Eqs}{equations}
\newcommand{\Sec}{Sec.}
\renewcommand{\Sec}{section}
\newcommand{\Secs}{Secs.}
\renewcommand{\Secs}{sections}

\newcommand{\sumint}{%
\mathchoice%
  {\ooalign{$\displaystyle\sum$\cr\hidewidth$\displaystyle\int$\hidewidth\cr}}
  {\ooalign{\raisebox{.14\height}{\scalebox{.7}{$\textstyle\sum$}}\cr\hidewidth$\textstyle\int$\hidewidth\cr}}
  {\ooalign{\raisebox{.2\height}{\scalebox{.6}{$\scriptstyle\sum$}}\cr$\scriptstyle\int$\cr}}
  {\ooalign{\raisebox{.2\height}{\scalebox{.6}{$\scriptstyle\sum$}}\cr$\scriptstyle\int$\cr}}
}

 \def\ket{\rangle} \def\ack{|}

\begin{document}

\title[Symmetry restoration in mean-field approaches]{Symmetry restoration in mean-field approaches}

\author{J.A. Sheikh$^{1}$,
J. Dobaczewski$^{2-5}$,
P. Ring$^{6-7}$,
L. M. Robledo$^{8-9}$,
C. Yannouleas$^{10}$}

\address{$^{1}$  Department of Physics, University of Kashmir, Srinagar, 190 006, India}
\address{$^{2}$  Department of Physics, University of York, Heslington, York YO10 5DD, United Kingdom }
\address{$^{3}$  Department of Physics, P.O. Box 35 (YFL), University of Jyv\"askyl\"a, FI-40014  Jyv\"askyl\"a, Finland }
\address{$^{4}$  Institute of Theoretical Physics, Faculty of Physics, University of Warsaw, ul. Pasteura 5, PL-02-093 Warsaw, Poland }
\address{$^{5}$  Helsinki Institute of Physics, P.O. Box 64, FI-00014 University of Helsinki, Finland}
\address{$^{6}$  Physik-Department, Technische Universit\"at M\"unchen, D-85747 Garching, Germany}
\address{$^{7}$  State Key Laboratory of Nuclear Physics and Technology, School of Physics, Peking University, Beijing 100871, China}
\address{$^{8}$  Center for Computational Simulation, Universidad Polit\'ecnica de Madrid, Campus de Montegancedo, Boadilla del Monte, 28660-Madrid, Spain}
\address{$^{9}$  Departamento de F\'isica Te\'orica and CIAFF, Facultad de F\'isica, Universidad Aut\'onoma de Madrid,E-28049 Madrid, Spain }
\address{$^{10}$ School of Physics, Georgia Institute of Technology, Atlanta, Georgia 30332-0430, USA}

\date{\today{}}

\begin{abstract}
The mean-field approximation based on effective interactions or
density functionals plays a pivotal role in the description of
finite quantum many-body systems that are too large to be treated by
{\it ab initio} methods. Some examples are strongly interacting medium and
heavy mass atomic
nuclei and mesoscopic condensed matter systems. In this
approach, the linear Schr\"odinger equation for the exact many-body
wave function is mapped onto a non-linear density-dependent one-body
potential problem. This approximation, not only provides computationally very simple
solutions even for systems with many particles, but due to the
non-linearity, it also allows for obtaining solutions that break essential symmetries
of the system, often connected with phase transitions. In this way,
additional correlations are subsumed in the system. However,
the mean-field approach suffers from the drawback that the corresponding
wave functions do not have sharp quantum numbers and, therefore, many
results cannot be compared directly with experimental data. In this
article, we discuss
general group-theory techniques to restore the broken
symmetries, and provide detailed expressions on the restoration of translational, rotational,
spin, isospin, parity and gauge symmetries, where the latter
corresponds to the restoration of the particle number. In order to avoid the
numerical complexity of exact projection techniques,
various approximation methods available in the literature are
examined.  Applications of the projection methods are presented for simple
nuclear models, realistic calculations in relatively small
configuration spaces, nuclear energy density functional theory, as
well as in other mesoscopic systems. We also discuss applications of
projection techniques to quantum statistics in order to treat the
averaging over restricted ensembles with fixed quantum numbers.
Further, unresolved problems in the application of the symmetry
restoration methods to the energy density functional theories
are highlighted in the present work.

\end{abstract}


\submitto{\JPG}

\maketitle

\tableofcontents{}


\section{Introduction \label{sec:intro}}

Mean-field approaches play a central role in the description of
quantum many-body problems in areas like quantum chemistry, atomic,
molecular, condensed matter, and nuclear physics. The simplicity of the
associated wave functions, both in the fermion and the boson cases, is
the reason behind the popularity of the mean-field approaches. These product-type wave
functions allow, on the one hand, an easy implementation
of the symmetrization principle of quantum mechanics required for identical particles, and,
on the other hand, permit the application of techniques used in the field theory, like the Wick's theorem, which
enormously simplify the evaluation of the matrix elements of
the many-body operators.

The optimal mean field, generating the single-particle orbitals, is
usually determined through the application of the variational principle.
In the fermion case, the variational principle, performed in the space of Slater
determinants, leads to the familiar Hartree-Fock (HF) method. It is common to find situations
where short-range attractive interactions induce correlations leading
to the superfluidity or superconductivity phenomena that are well described
by the BCS theory. The quasiparticles introduced in the BCS theory can be combined
with the concepts of the HF theory to give the Hartree-Fock-Bogoliubov (HFB) mean-field theory,
which is widely employed in nuclear physics.

Another facet of the mean-field approach is revealed within the
Density Functional Theory (DFT). By using one-body densities as
efficient and relevant degrees of freedom, DFT aims to map the exact
wave functions of many-body systems onto the product states, which
leads to the dynamical equations becoming formally identical to those
given by the HF method. Although the foundations and approximations
leading to HF and DFT methods are different, the similar structure of
dynamical equations allows us to use for both, HF and DFT, the common name
of mean-field approach.

One of the most salient characteristics
of mean-field approaches is the fact that solutions often spontaneously break
symmetries of the Hamiltonian. This is the case, for instance, in the BCS
and HFB theories, where the associated mean-field wave functions, which are the
vacua of the corresponding quasiparticle operators, do not represent
states with good particle number. The spontaneous symmetry-breaking mechanism
provides a way to incorporate nontrivial dynamic correlations on top of the simple Slater determinants,
while preserving
the simplicity of the mean-field description.

It is the nonlinearity of the mean-field equations that
favors the spontaneous symmetry-breaking mechanism, and may constitute the simplest
description of the symmetry-breaking effects occurring at a more fundamental level.
In this way, the mean-field symmetry breaking leads to interesting
perspectives to understand the physics of a given problem.
For example, it leads to an easy and efficient description of different
collective effects, such as the appearance of rotational bands being the
result of the rotational-symmetry breaking, so common in nuclear or molecular physics.
Symmetry breaking can also be a useful concept
in the presence of stationary degenerate symmetry-conserving states, whereupon the
famous Jahn-Teller effect becomes effective.

Nevertheless, advantages of the spontaneous mean-field symmetry
breaking come at a price:
the resulting wave functions are not invariant or covariant with respect to
broken-symmetry groups and, therefore, they cannot be labeled with
the symmetry quantum numbers such as angular momentum, parity, etc. This
represents a serious drawback if quantities like electromagnetic transition
probabilities, with their selection rules, are to be computed. Another
drawback of the symmetry-breaking mechanism is connected with sharp
transitions observed between the symmetry-conserving and symmetry-breaking
solutions, which may occur as a function of some parameters of the Hamiltonian. Such
sharp transitions are typical for infinite systems, but cannot characterize finite
many-body or mesoscopic systems.

A way to overcome the disadvantages of the symmetry-breaking mean field is
to restore the broken symmetries, which is the subject matter of this
review article.
A general idea of such a restoration is to take linear
combinations, properly weighted, of wave functions obtained by applying
the elements of the symmetry group to the symmetry-breaking
(often called ``deformed") mean-field wave function. As a consequence of the symmetry
of the Hamiltonian, these ``rotated'' deformed wave functions are degenerate,
and linear combinations, thereof, are expected to have lower energies. In addition,
if the weights of the linear combination are chosen according to the
rules of group theory, the obtained wave functions become invariant or covariant with respect to
the underlying symmetry group and can be labeled with proper quantum numbers. This procedure is
denoted in the literature as ``symmetry restoration" or ``projection".
As it leads to linear combinations of product states, it can be understood as
introducing correlations beyond the mean-field approach.

The theory behind the projection method is rooted in group theory as the
weights of the linear combination are given by the irreducible representations
of the symmetry group. In most of the cases, the symmetry group is a
continuous Lie group (rotation, translation, particle-number gauge, etc.)
while in other cases it is a discrete group (parity).

Once the
structure of the projector operator is fixed,
two alternatives are available to determine the deformed intrinsic state.
The simplest one is to restore the symmetry of the deformed state obtained
after solving the HF, HFB, or Kohn-Sham equations. This procedure is called ``projection after variation'',
where variation refers to the minimization of the mean-field energy.
Another approach, fully self-consistent and variational, called ``variation after projection", determines
the deformed state through minimization of the
projected energy, separately for each quantum numbers. In this way, different
deformed states are obtained for a given quantum number. It turns out
that extra flexibility brought in by the variation after projection method
is able to smear out the sharp transitions mentioned before.

To compute basic quantities involved in the symmetry restoration,
one takes advantage of the generalized Wick's theorem, which allows for calculating matrix elements of operators
between mean-field states.
This theorem can be applied to projection, because the rotated mean-field states
are mean-field states again -- they simply correspond to rotated (quasi)particles. This property
is a direct consequence of the Thouless theorem, and of the fact that the Lie
algebras of the relevant symmetry groups can be represented in terms of one-body operators.

The program to perform the projected calculations can be directly
implemented when the problem is defined in terms of a Hamiltonian operator.
However, this is not always the case, and in many applications the Hamiltonian
is replaced by a density functional in the spirit of the Hohenberg-Kohn or Kohn-Sham
approach. In this situation, one is forced
to introduce some sort of prescriptions on how to compute the energy
kernels. However, these recipes are plagued with conceptual problems
that were not resolved yet in the most general case.

Further, it is also common to use different (effective) interactions for each of the three contributions
to the energy coming from a two-body operator, namely the direct, exchange,
and pairing contributions (the most typical case is probably the use of the
Slater approximation for the Coulomb exchange contribution and the neglect of
the Coulomb antipairing field). In this case, a naive use of the
generalized Wick's theorem can lead to spurious contributions and specific ways to
deal with this problem have to be devised. These difficulties
represent serious impediments for the practical implementation of the symmetry-restoration
methods in nuclear physics.

Since the effective interactions that are most often used to construct density functionals
usually contain density-dependent terms, the corresponding mean-field approaches are referred to as
based on the energy density functionals (EDFs),
see \citeasnoun{(Dug14),(Sch19),(Fur20)} for further reading. The main
benefit of such approaches is their applicability over the entire Segr\'e chart.
Very successful effective interactions were developed over the years with
great success in describing bulk nuclear properties. The success of the
EDF methods motivated the introduction
of the symmetry restoration (beyond mean field step) aimed
to gain access to symmetry-conserving observables and to increase the accuracy of
the bulk properties.

The authors of this review are fully aware of the fact
that nuclear physics specialists may find it difficult to follow the
sections on electronic wave-function-based methods and vice versa.
The two domains were
evolving separately indeed. However, as far as the symmetry
restoration is concerned, apart from apparent differences in the
notation and language, the similarities are abundant. Obviously we
were confronted with the question of unification of notation. We came
to the conclusion that a single review would never change what has
been deeply encrusted in the historical development of both domains.
Therefore, we think that promoting a unified notation and language
would not be a good idea, because then the text would become alien to
everybody. We thus decided to always keep the original terminology
and as clearly as possible explain what various names and symbols
mean.

The present review article is organized in the following manner. Basic ideas regarding
the symmetry breaking and restoration are presented in
{\Sec}~\ref{sec:mfmodels}. Section~\ref{sec_srgf} outlines
the general formalism of the symmetry restoration and {\Sec}~\ref{sec:approximate} discusses
the applicability of the approximate projection methods.
Symmetry restoration methods in simple nuclear models are discussed in {\Sec}~\ref{sec:simple},
nuclear DFT in {\Sec}~\ref{sec:functional}, and other mesoscopic systems
in {\Sec}~\ref{sec:mesoscopic}.
A brief description of the projection in statistical approaches is
given in {\Sec}~\ref{sec:statistics}
and finally, we provide a summary and concluding remarks in {\Sec}~\ref{sec:summary}.

\section{Symmetry breaking in simple illustrative models}
\label{sec:mfmodels}

In this section, we introduce the subject matter of symmetry breaking
and restoration by presenting three very simple examples. Firstly, in
{\Sec}~\ref{DSPW}, we discuss a solution of a one-particle problem in
one dimension, where neither mean-field nor many-body complications
appear. Secondly, in {\Sec}~\ref{exh2}, we present a two-particle problem
in two dimensions, where one can illustrate the role of the mean-field
approximation. And third, in {\Sec}~\ref{seniority}, we discuss the
case of a many-body setting. These three simple models are exactly
solvable, which allow us to analyze the problem of symmetry breaking
and restoration in the quantum mechanical context and to clearly
delineate the role of approximations that unavoidably have to be made
in realistic situations.

\subsection{Doubly symmetric potential well}
\label{DSPW}

\begin{figure}
\centering\includegraphics[width=0.7\columnwidth]{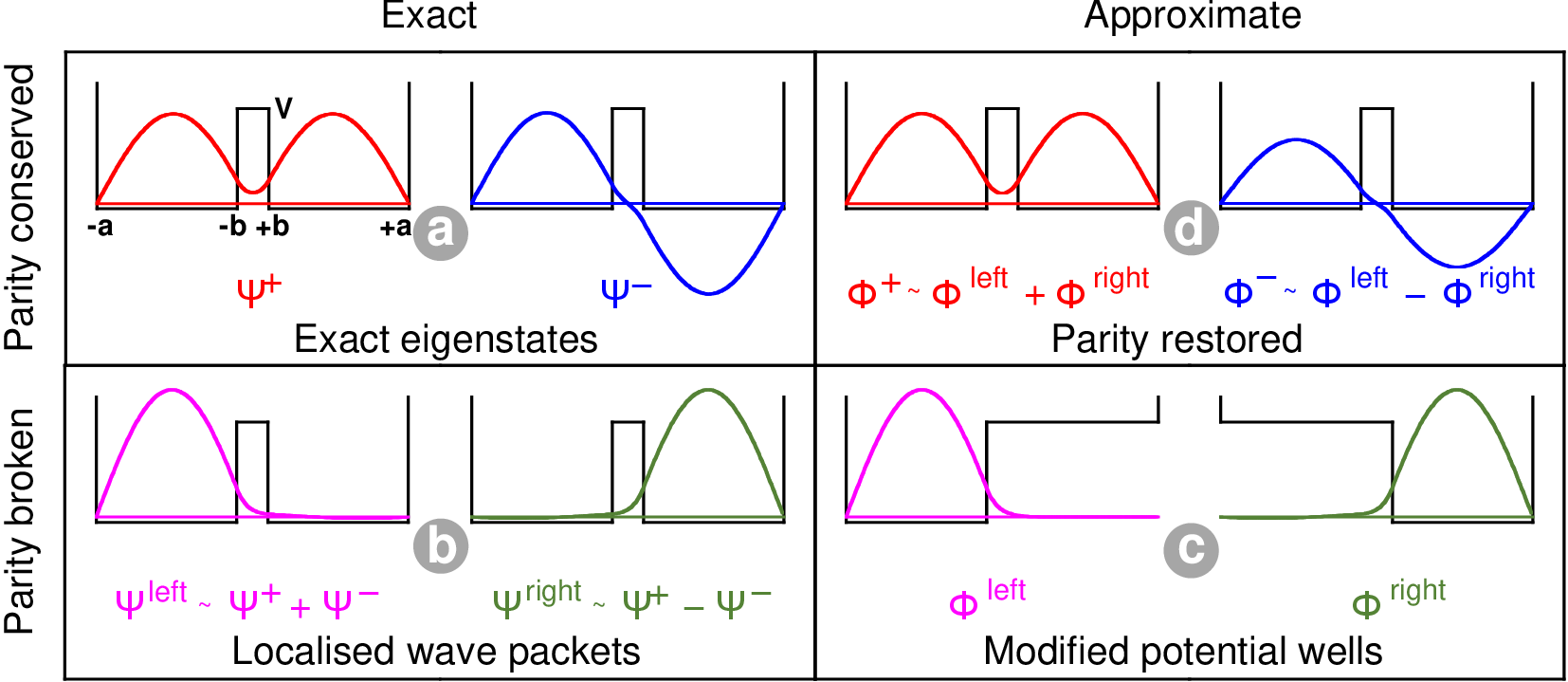}
\caption{\label{DSPW1}Wave functions of the two lowest eigenstates of a particle
moving in the double symmetric potential well. Left (right) panels show
exact (approximate) solutions of the Schr\"odinger equation. Top (bottom) panels
show parity-conserving (broken-parity) solutions. The approximate broken-parity
wave functions shown in the bottom-right panel are obtained by filling in one of the
wells.}
\end{figure}

Consider the doubly symmetric potential well~\cite{(Sak94)},
that is, a one-dimensional infinite potential well of width $2a$
with a step-like potential barrier of width $2b$ and height $V$ placed in the middle.
To link this example to nuclear-physics scales of mass, distance, and energy,
let us use the parameters of $\hbar^2/2m=20$\,MeV\,fm$^2$, $a=10$\,fm, $b=1$\,fm,
and  $V=40$\,MeV. In this model, exact wave functions can be very easily determined;
those of the two lowest eigenstates are plotted in {\Fig}~\ref{DSPW1}(a).

The model is symmetric with respect to the middle of the well, and
thus the eigenstates are either symmetric or antisymmetric,
$\hat{\Pi}|\Psi^\pm\rangle=\pm|\Psi^\pm\rangle$, where $\hat{\Pi}$ is the
inversion operator $x\rightarrow{-x}$. The parameters of the model
are chosen in such a way that the two lowest states reside
predominantly within the left and right well, and not in the barrier
region.

The two lowest eigenstates of opposite parity can be expressed as
linear combinations of two localized configurations, that is,
\begin{eqnarray}
|\Psi^\pm\rangle&=&\tfrac{1}{\sqrt{2}}\left(|\Psi^{\text{left }}\rangle\pm|\Psi^{\text{right}}\rangle\right) ,
\end{eqnarray}
for
\begin{eqnarray}
|\Psi^{\text{left }}\rangle&=&\tfrac{1}{\sqrt{2}}\left(|\Psi^+\rangle+|\Psi^-\rangle\right) ,\\
|\Psi^{\text{right}}\rangle&=&\tfrac{1}{\sqrt{2}}\left(|\Psi^+\rangle-|\Psi^-\rangle\right) ,
\end{eqnarray}
see {\Fig}~\ref{DSPW1}(b). That is, the localized configurations are wave packets
built of the two lowest eigenstates of the system. In these configurations, the particle resides either
in the left or in the right well. The four states are pairwise orthogonal,
$\langle\Psi^+|\Psi^-\rangle=0$ and $\langle\Psi^{\text{left }}|\Psi^{\text{right}}\rangle=0$,
and both, the pair of exact states, $|\Psi^+\rangle$ and $|\Psi^-\rangle$,
and that of localized wave packets, $|\Psi^{\text{left}}\rangle$ and $|\Psi^{\text{right}}\rangle$,
span the same subspace of the two lowest eigenstates.

If we denote the exact Hamiltonian of the doubly symmetric potential well by $\hat{H}$, we obviously have
\begin{equation}\label{Eloc2}
\begin{array}{rcl}
\hat{H}|\Psi^+\rangle &=& E^+|\Psi^+\rangle  \\
\hat{H}|\Psi^-\rangle &=& E^-|\Psi^+\rangle  \\
\end{array}
\end{equation}
and
\begin{equation}\label{Eloc3}
\begin{array}{rcl}
\langle\Psi^{\text{left }}|\hat{H}|\Psi^{\text{left }}\rangle &=& E^{\text{loc}}  \\
\langle\Psi^{\text{right}}|\hat{H}|\Psi^{\text{right}}\rangle &=& E^{\text{loc}}  \\
\langle\Psi^{\text{left }}|\hat{H}|\Psi^{\text{right}}\rangle &=& -\tfrac{1}{2}\delta{E}  \\
\langle\Psi^{\text{right}}|\hat{H}|\Psi^{\text{left }}\rangle &=& -\tfrac{1}{2}\delta{E}  \\
\end{array}
\end{equation}
for
\begin{equation}\label{Eloc}
\delta{E}=E^--E^+,\quad\quad
E^{\text{loc}}=\tfrac{1}{2}\left(E^++E^-\right) ,
\end{equation}
where in our model the exact eigenstates
are split in energy by $\delta{E}=75.6$\,keV and the average energies of
both wave packets $E^{\text{loc}}$ are, of course, the same and located exactly in the
middle between the two eigenenergies.
In the matrix notation, {\Eqs}~(\ref{Eloc2}) and (\ref{Eloc3}) can be represented as
\begin{equation}\label{Eloc4}
\Big(\langle\Psi^+|,\langle\Psi^-|\Big)\hat{H}
\left(\begin{array}{l}|\Psi^+\rangle\\|\Psi^-\rangle\end{array}\right)
 = \left(\begin{array}{ll} E^+ & 0  \\
                           0 & E^- \\
\end{array}\right)
\end{equation}
and
\begin{equation}\label{Eloc5}
\Big(\langle\Psi^{\text{left }}|,\langle\Psi^{\text{right}}|\Big)\hat{H}
\left(\begin{array}{l}|\Psi^{\text{left }}\rangle\\|\Psi^{\text{right}}\rangle\end{array}\right)
 = \left(\begin{array}{rr} E^{\text{loc}} & -\tfrac{1}{2}\delta{E}  \\
                      -\tfrac{1}{2}\delta{E} & E^{\text{loc}} \\
\end{array}\right) ,
\end{equation}
respectively.

It is now very important to realize that by breaking the symmetry of
the problem,  we can build a very
reasonable model of the localized wave packets,
see {\Fig}~\ref{DSPW1}(c). Indeed, by keeping only
the left or right potential well, we obtain the left and right {\em
broken-symmetry} states, $|\Phi^{\text{left}}\rangle$ and
$|\Phi^{\text{right}}\rangle$. The broken-symmetry states are the
exact eigenstates in the modified potential wells, but at the same
time they approximate exact localized states in the original doubly symmetric potential well. In
the scale of {\Fig}~\ref{DSPW1}, they cannot really be distinguished
from the exact wave packets $|\Psi^{\text{left}}\rangle$ and
$|\Psi^{\text{right}}\rangle$. Note that $|\Phi^{\text{left}}\rangle$
and $|\Phi^{\text{right}}\rangle$ are stationary in the modified
potential wells, whereas we use them to model non-stationary wave
packets $|\Psi^{\text{left}}\rangle$ and
$|\Psi^{\text{right}}\rangle$ of the original doubly symmetric potential well.

At this point, we arrive at the very heart of the subject matter of this
article. Namely, the symmetry-broken solutions, which pertain
to a different problem than the original one, can serve us as approximate solutions of the
original problem.
This is achieved by restoring their symmetry, that is, by considering the normalized symmetric
and antisymmetric combinations of the normalized states $|\Phi^{\text{left}}\rangle$ and $|\Phi^{\text{right}}\rangle$,
see {\Fig}~\ref{DSPW1}(d),
\begin{equation}\label{Psipm}
|\Phi^\pm\rangle=\tfrac{1}{\sqrt{2\pm2\epsilon}}
                 \left(|\Phi^{\text{left}}\rangle\pm|\Phi^{\text{right}}\rangle\right)
\quad\mbox{for}\quad
\epsilon= \langle\Phi^{\text{left}}|\Phi^{\text{right}}\rangle.
\end{equation}
Since
the inversion transforms the two broken-symmetry states one
into another,
\begin{equation}\label{Psipm2}
\hat{\Pi}|\Phi^{\text{left}}\rangle=|\Phi^{\text{right}}\rangle
\quad\mbox{and}\quad
\hat{\Pi}|\Phi^{\text{right}}\rangle=|\Phi^{\text{left}}\rangle,
\end{equation}
states (\ref{Psipm}) have correct symmetry properties of
$\hat{\Pi}|\Phi^\pm\rangle=\pm|\Phi^\pm\rangle$ with their normalization
factors that depend on the overlap $\epsilon$ between the approximate
broken-symmetry states. Note that since the inversion
is a hermitian operator, the overlap $\epsilon$,
\begin{equation}\label{Psipm3}
\epsilon= \langle\Phi^{\text{left}} |          \Phi^{\text{right}}\rangle
        = \langle\Phi^{\text{left}} |\hat{\Pi}|\Phi^{\text{left}} \rangle
        = \langle\Phi^{\text{right}}|\hat{\Pi}|\Phi^{\text{right}}\rangle
        = \langle\Phi^{\text{right}}|          \Phi^{\text{left}} \rangle,
\end{equation}
is real.

For one-dimensional representations of the symmetry group, as is the
case of the parity symmetry discussed in this section (see also
discussion in {\Sec}~\ref{sec_pmfvs}), the symmetry restoration of the
broken-symmetry states, {\Eq}~(\ref{Psipm}), automatically brings the
Hamiltonian matrix to its diagonal form. Then, using the symmetry
condition $[\hat{H},\hat{\Pi}]=0$, the analogues of
{\Eqs}~(\ref{Eloc4}) and (\ref{Eloc5}) read
\begin{equation}\label{Eloc6}
\Big(\langle\Phi^+|,\langle\Phi^-|\Big)\hat{H}
\left(\begin{array}{l}|\Phi^+\rangle\\|\Phi^-\rangle\end{array}\right)
 = \left(\begin{array}{ll} {\cal E}^+ & 0  \\
                                    0 & {\cal E}^- \\
\end{array}\right)
\end{equation}
and
\begin{equation}\label{Eloc7}
\Big(\langle\Phi^{\text{left }}|,\langle\Phi^{\text{right}}|\Big)\hat{H}
\left(\begin{array}{l}|\Phi^{\text{left }}\rangle\\|\Phi^{\text{right}}\rangle\end{array}\right)
 = \left(\begin{array}{rr} {\cal E}^{\text{loc}} & \Delta  \\
                                          \Delta & {\cal E}^{\text{loc}} \\
\end{array}\right) ,
\end{equation}
respectively, where the approximate eigenenergies ${\cal E}^\pm$ and their splitting $\delta{\cal E}$,
\begin{eqnarray}\label{Erest}
{\cal E}^\pm &=& \frac {{\cal E}^{\text{loc}}\pm\Delta} {1\pm\epsilon} , \\
\label{Ediff}
\delta{\cal E} &=& {\cal E}^--{\cal E}^+=\frac {2\epsilon{\cal E}^{\text{loc}}-2\Delta} {1-\epsilon^2} ,
\end{eqnarray}
depend on the matrix elements of the Hamiltonian,
\begin{eqnarray}\label{Erest1}
{\cal E}^{\text{loc}}
&=& \langle\Phi^{\text{left }}|\hat{H}|\Phi^{\text{left }}\rangle
 =  \langle\Phi^{\text{right}}|\hat{H}|\Phi^{\text{right}}\rangle , \\
\label{Erest2}
\Delta
&=& \langle\Phi^{\text{left }}|\hat{H}|\Phi^{\text{right}}\rangle
 =  \langle\Phi^{\text{right}}|\hat{H}|\Phi^{\text{left }}\rangle ,
\end{eqnarray}
where, by the argument analogous to that used when deriving {\Eq}~(\ref{Psipm3}), $\Delta$ is real.

\begin{figure}
\centering\includegraphics[width=0.7\columnwidth]{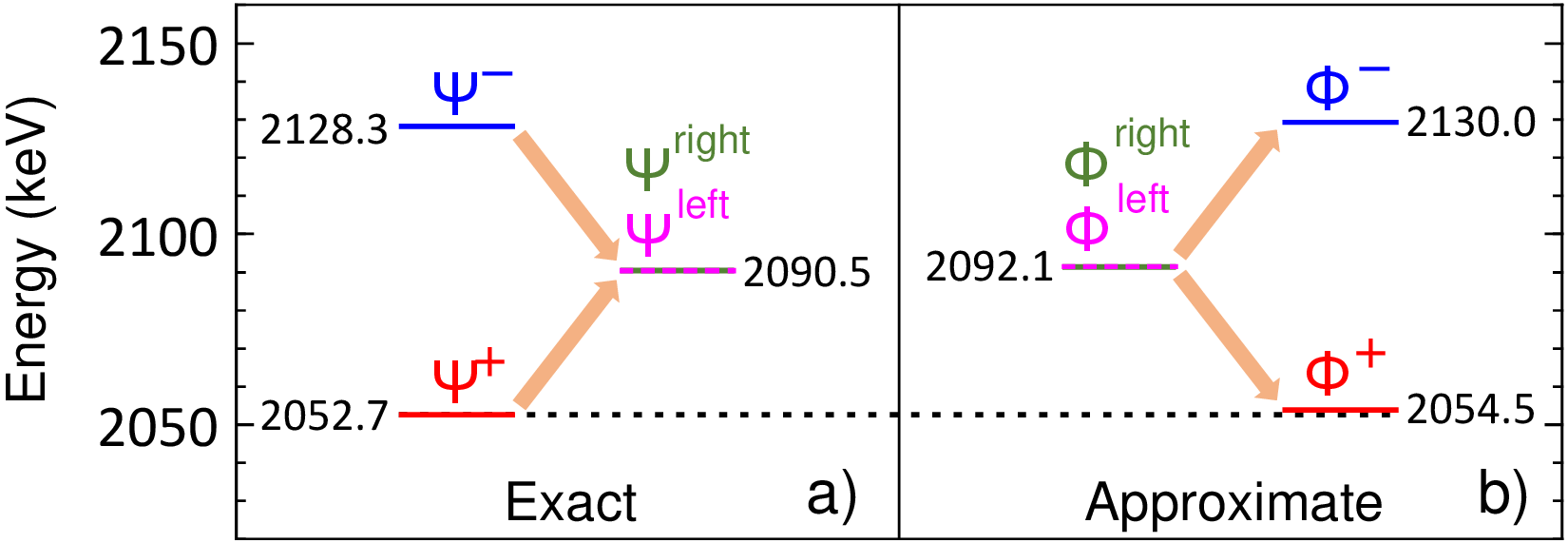}
\caption{\label{DSPW2}Energies of the exact (left) and approximate (right)
states shown in {\Fig}~\protect\ref{DSPW1}.}
\end{figure}

Figure~\ref{DSPW2} summarizes the logic of the construction presented
above and depicts energies of all discussed states. In the left
panel, we show how the pair of exact eigenstates,
$|\Psi^+\rangle$ and $|\Psi^-\rangle$, is transformed into the pair
of exact localized wave packets, $|\Psi^{\text{left}}\rangle$ and
$|\Psi^{\text{right}}\rangle$.
In the right panel of {\Fig}~\ref{DSPW2}, we show how the pair of approximate localized
states, $|\Phi^{\text{left}}\rangle$ and
$|\Phi^{\text{right}}\rangle$, is by the symmetry restoration
transformed into the pair of approximate eigenstates $|\Phi^+\rangle$
and $|\Phi^-\rangle$.

It is gratifying to see that the model energies
of approximate localized states ${\cal E}^{\text{loc}}$ are only
1.6\,keV higher than the average energies of the localized exact wave
packets $E^{\text{loc}}$ (\ref{Eloc}). It is even more gratifying to
see that energies of the symmetry-restored states,
are only 1.8 and 1.7\,keV above their exact counterparts.
The energy splitting between the
two symmetry-restored states,
which equals 75.5\,keV, almost exactly reproduces the exact result.

We should mention at this point that the exact localized wave packets,
$|\Psi^{\text{left}}\rangle$ and $|\Psi^{\text{right}}\rangle$, which
represent non-stationary solutions of the Schr\"odinger equation in the doubly symmetric potential well,
evolve in time in such a way that after the time of
$T=\pi\hbar/\delta{E}$, the left wave packet will appear on the
right-hand side and vice versa. For the selected parameters of the
model, this left-right quantum oscillation time is very short,
$T\simeq3\times10^{-21}$\,s, and thus, a localized particle created
in the left or right well will not really keep its identity. However,
if the barrier width is increased from 2 to 45\,fm, this oscillation
time becomes $T\simeq40$ days, and the particle created in one of the
wells would remain there as a classical system would do. In the
case of the wide barrier, the approximate localized states,
$|\Phi^{\text{left}}\rangle$ and $|\Phi^{\text{right}}\rangle$,
become extremely good representations of the localized exact wave
packets, and the symmetry restoration becomes an extremely efficient
method to obtain perfect approximations of the symmetry-conserving
exact eigenstates.

The simple model discussed in this section shows that the symmetry
breaking and restoration is a useful concept of describing the physical
reality of quantum mechanics, and that it is not inherent to
complicated many-body systems. Nevertheless, it is for these
complicated and difficult systems that it finds its most prominent
and successful applications.\footnote{Analyses performed in schematic models
can be found, e.g., in~\citeasnoun{(Rob92a),(Yan02a),(Yan02b)}.}
In particular, in nuclear and molecular
physics, there is overwhelming evidence that symmetry-restored
mean-field states provide for a global understanding of multiple
phenomena and experimental observations.

At this point, to relate the symmetry
restoration to the rigorous
DFT~\cite{(Hoh64a),(Koh65a),(Bar10c),(Bec14b),(Jon15a)},
a few comments are in order. The basis for
existence theorems of exact DFT is the variational principle, whereby
one reaches the exact ground-state of the system and its density.
Within our simple example above, it would mean that the DFT is bound
to yield the exact, symmetry conserving, positive-parity ground-state
wave function $\Psi^+(x)=\langle{x}|\Psi^+\rangle$ and its density
$\rho(x)=|\Psi^+(x)|^2$, and not the localized wave functions
$\Psi^{\text{left}}(x)=\langle{x}|\Psi^{\text{left}}\rangle$ or
$\Psi^{\text{right}}(x)=\langle{x}|\Psi^{\text{right}}\rangle$ and
their respective densities.

However, it is obvious that densities (be
they average or maximum) of exact and localized wave functions differ by
about a factor of two, compare {\Figs}~\ref{DSPW1}(a) and
(b). This simple observation creates an important issue for
systems, like nuclei, for which the equilibrium local density
(the so-called saturation
density of about 0.16\,fm$^{-3}$) is an
important physical parameter determined by the nature of the
underlying interaction. Indeed, for such systems we build (or derive)
functionals that describe infinite saturated systems, which leads to the local density approximation, or finite self-bound systems,
within a single potential well (typical for a drop of matter). Such
functionals then have minimum energies at saturation density and thus can
properly work only for localized wave functions
and not for the exact symmetry-conserving ones.

To bring the discussion above away from the simple example, which we
introduced only to illustrate basic concepts, and towards a realistic
case, consider a positive-parity ground state of an alpha-particle
emitting nucleus. Before the decay, the density of nucleons is almost
constant within the nucleus and equal to the saturation
density. After the decay, the exact
parity-conserving wave function would correspond to a symmetric
combination of a recoil nucleus moving right, with the alpha particle
moving left, and that of the recoil moving left and the alpha moving
right. It is obvious that such a state cannot be modeled by the same
density functional as that used to model the nucleus before the
decay, because densities are now twice smaller than the saturation
density. However, it is also obvious that a symmetry-broken state,
e.g., the one with the recoil moving left and alpha moving right, is
entirely within the remit of that functional, because both subsystems
do have similar local densities, not very different from the
saturation density.

The case of the alpha-emitting nucleus illustrates crucial points of
nuclear DFT, whereupon the symmetry breaking plays a fundamental
role. It also tells us that the symmetry restoration is an equally
fundamental piece of the description. Indeed, after modeling the DFT
state of the recoil moving left and alpha moving right, we must
symmetrize the obtained solution, because the alpha-particle
detectors will, of course, never see any left-right asymmetry of the
decay process. In this sense, the DFT description of many-body
systems gives us immediate access to physical localized states
describing specific configurations, by which we mean specific
arrangements of constituents of composite objects. However, it is now
clear that these configurations should never be confused with exact
eigenstates, as they simply represent specific wave packets thereof,
whereas a reasonable modeling of the exact eigenstates is then
accessible via the symmetry restoration.

There remains, nevertheless, one troubling element of the link
between the DFT and symmetry restoration. Indeed, to restore the
symmetry, we need to have access not only to the average energies of
the localized broken-symmetry states $|\Phi^{\text{left}}\rangle$ or
$|\Phi^{\text{right}}\rangle$, which are within the remit of DFT, but
also to the overlaps,
$\epsilon=\langle\Phi^{\text{left}}|\Phi^{\text{right}}\rangle$, and
matrix elements,
$\Delta=\langle\Phi^{\text{left}}|\hat{H}|\Phi^{\text{right}}\rangle$,
thereof, neither of which is. Within the nuclear-DFT applications, there is
overwhelming evidence that $\epsilon$ and $\Delta$ can be evaluated
using the corresponding Kohn-Sham states and generalized Wick's
theorem. This gives us a rich and reasonable
description of numerous experimental data. However, such an approach
constitutes a hybrid mix of the DFT and wave-function approaches and,
to our knowledge, it has as yet no justification in any solid
formalism. It appears that the many-body-physics community has executed a
spectacular triple Axle jump into a pool without really verifying whether
the water is there or not. Nevertheless, the obtained excellent results
indicate that we may rather worry about finding a justification
than about questioning the method itself.

The reader is begged to excuse us for the partly simplistic and partly
philosophical narrative of this introductory section. We thought
that exposing these basic notions could constitute a useful background
of the following sections, where we move right on to the forefront
description pertaining to the subject matter of this review. However,
the advanced discussion that is coming up should not obscure the vision
of the forest behind trees.

\subsection{Dissociation of the natural molecular hydrogen and other similar
two-dimensional artificial dimers}
\label{exh2}

The second illustrative example is taken from the areas of condensed-matter physics and
chemistry, where the long-range interparticle Coulombic repulsion plays an essential role in
inducing symmetry breaking at the level of mean-field treatments of interacting electrons.
In this context, the simplest nontrivial examples involve at a minimum a pair of electrons.
Here we will describe the case of an artificial two-dimensional (2D) quantum dot molecule
(QDM), specifically a manmade system denoted as H$_2$-QDM. The H$_2$-QDM system
consists of two electrons trapped inside two parabolic quantum dots (each with a
harmonic-potential confinement specified by a frequency $\hbar \omega_0$) separated by an
interdot distance $d$ and an interdot barrier $V_b$. The overall confinement potential is
effectively that of a 2D two-center oscillator.
Increasing the interdot separation $d$, or the interdot barrier $V_b$, generates a process
bearing analogies to the dissociation of the natural H$_2$ Hydrogen molecule. This process
offers an immediate illustration of the symmetry dilemma facing the mean-field
approaches. This expression [coined by L\"{o}wdin  \cite{lyko63}] succinctly conveys
the fact that imposing the Hamiltonian symmetries on the
mean-field treatment [exemplified by the restricted Hartree Fock (RHF) method for electrons;
see {\Sec}~\ref{sec:mesoscopic}], provides wave functions with the proper symmetries, but often
the corresponding total energy is higher than that obtained when the symmetry requirements are
relaxed [using the unrestricted Hartree Fock (UHF) method for electrons; see
{\Sec}~\ref{sec:mesoscopic}]. In the double-dot example here, the UHF lowers the total energy by
breaking in the ensuing wave functions the symmetries of total spin and of the parity along the
separation axis.

The stretched natural H$_2$ in both the RHF and the UHF mean-field levels and the ensuing
correct-symmetry/higher-energy versus lower-energy/wrong-symmetry dilemma are
described in detail in Chapter~3.8.7. of \citeasnoun{so89}.
In wave-function-based approaches describing electronic systems, the symmetry dilemma
can be overcome by using symmetry restoration.\footnote{In the Kohn-Sham DFT
treatment of electronic systems, the symmetry dilemma remains an open question,
with a main obstacle being the large self-interaction error; see, e.g.,~\citeasnoun{(Per95)}.}
An explicit illustration to this effect for the case of the artificial H$_2$-QDM was provided in
\citeasnoun{(Yan01),(Yan02b),(Yan02a)}.
Before proceeding further, we mention two points that will be crucial in grasping the overall
picture behind the mathematical formalism and the
numerical details below. Namely, (I) The UHF equations do not provide a symmetry-broken solution in all
circumstances. Given  a specific many-body state, whether a ground state or an excited one, symmetry
breaking appears at well-defined regions of the parameters characterizing the many-body electronic
problem. In the absence of symmetry breaking, the UHF solutions coincide with the RHF ones, and thus
subsequent symmetry restoration has no effect. Anticipating the specific numerical results below, we
mention that, in the case of the singlet state of the H$_2$-QDM, smaller values of Coulombic repulsion,
interdot separation, and interdot barrier tend to suppress symmetry breaking.
(II) Two electrons correspond to a closed electronic shell in both two and three dimensions. However,
unlike the nuclear experience where shell closures prevent symmetry breaking (associated with nuclear
shape deformations and nuclear pairing), the two-particle shell closure in electronic systems is not immune
to symmetry breaking (associated with electron localization).

\begin{figure}[t]
\centering\includegraphics[width=0.7\columnwidth]{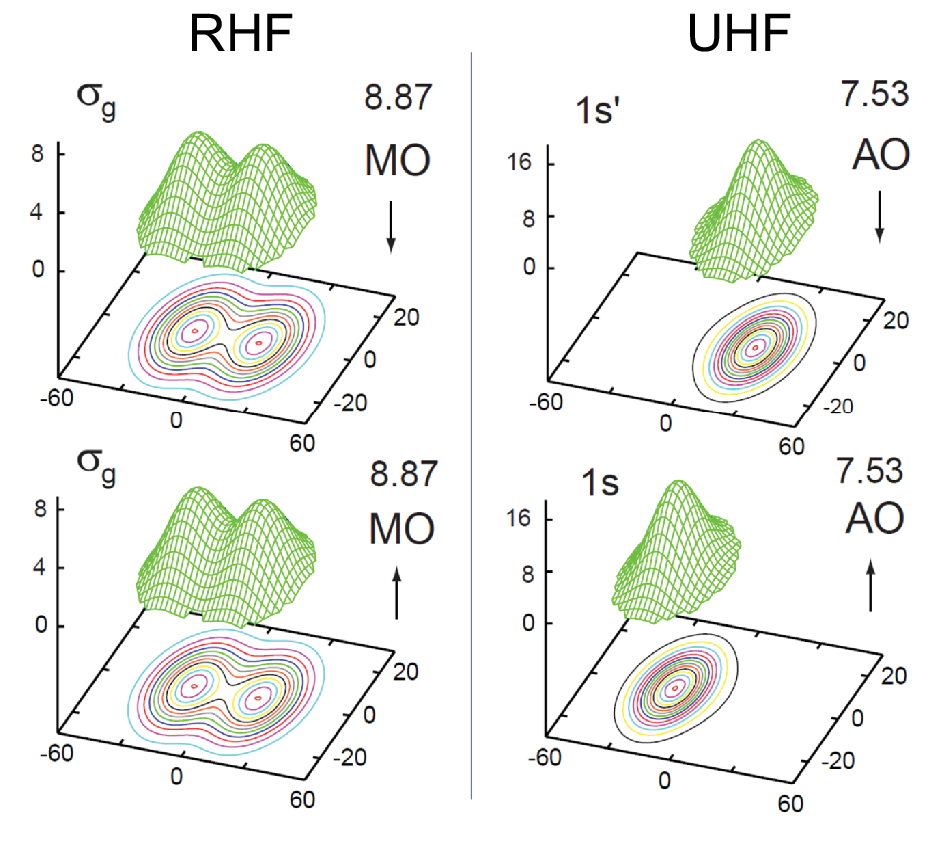}
\caption{
Lateral H$_2$-QDM: Occupied orbitals (modulus square)
for the spin unpolarized case ($S_z=0$). Left column: Restricted Hartree-Fock (RHF).
Right column: Unrestricted Hartree-Fock (UHF) results with
breaking of the space symmetry (parity). The numbers with each orbital
are their eigenenergies in meV. Up and down arrows indicate
up or down spin. The RHF orbitals extend over both
quantum dots; they are denoted as molecular orbitals (MO). The UHF orbitals are localized
on one quantum dot (either left or right), and are denoted as atomic orbitals (AO).
Distances along the $x$ and $y$ axes are in nm and the electron densities (vertical axes)
are in $10^{-4}$ nm$^{-2}$. The parameters are: effective mass of the electron
$m^*=0.067 m_e$ ($m_e$ is the free-electron mass),
trapping frequency for each quantum dot $\hbar \omega_0=5$\,meV, interdot separation
$d=30$\,nm, interdot barrier $V_b=4.95$\,meV, and material dielectric constant $\kappa=20$.
Reprinted by permission from Springer Nature and Copyright Clearance Center from \citeasnoun{(Yan01)}.
}
\label{new1}
\end{figure}

RHF and UHF numerical results for the singlet state of the H$_2$-QDM are displayed in {\Fig}~\ref{new1}
for a case with trapping frequency of $\hbar \omega_0=5$\,meV,
interdot distance of $d=30$\,nm and barrier of $V_b=4.95$\,meV. In the RHF result
(left column), a single bonding molecular orbital ($\sigma_g$)
is occupied by both the spin-up and spin-down electrons.
In the UHF case (right column), however, the total spin and parity symmetries are broken, and
the spin-up electron occupies an atomic-like ($1s$)
orbital located on the left quantum dot, while the spin-down electron occupies another different
atomic-like ($1s^\prime$) orbital centered on the right quantum dot. For the total
energies, the RHF result is $E_{\rm RHF}=13.68$\,meV, while the UHF
energy yields $E_{\rm UHF}=12.83$ meV, amounting to a lowering in energy of 0.85
meV. Furthermore, the energy of the triplet state is $E_{\rm UHF}=
13.01$\,meV, and thus the singlet state conforms to the requirement that
for two electrons at zero magnetic field the singlet is always the
ground state; in sharp contrast, the RHF molecular-orbital solution fails to
fulfill this exact requirement.

Next we show how to construct an improved wave function by using the spin projection method to
restore the broken symmetry of the UHF step. For clarity, we explicitly write down the UHF
determinant,
cf.~{\Eq}~(\ref{psiuhf}),
\begin{equation}
\Phi^{S_z=0}_{\rm UHF}(1,2)=  \frac{1}{\sqrt{2}}
\left|
\begin{array}{cc}
\varphi_l({\bm r}_1) \alpha_1 \; & \; \varphi_l({\bm r}_2) \alpha_2 \\
\varphi_r ({\bm r}_1) \beta _1 \; & \; \varphi_r({\bm r}_2) \beta _2
\end{array}
\right|,
\label{det}
\end{equation}
where $\varphi_l({\bm r})$ and $\varphi_r({\bm r})$ are the $1s$ (left) and $1s^\prime$ (right)
localized orbitals of the UHF solution displayed in the right column of
{\Fig}~\ref{new1}, and $\alpha$ and $\beta$ denote the up and down spins, respectively.
We also found useful to introduce a shorthand notation for the determinant above as
$\sqrt{2}\Phi^{S_z=0}_{\rm UHF} (1,2) \equiv | \varphi_l(1)\overline{\varphi}_r(2) \rangle$, where a bar
over a space orbital denotes a spin-down electron; absence of a bar denotes a spin-up electron.

$\Phi^{S_z=0}_{\rm UHF}(1,2)$ is an eigenstate of the projection $S_z$ of the
total spin ${\bm S} = {\bm s}_1 + {\bm s}_2$, but not of ${\bm S}^2$.
A many-body wave function which is an eigenstate of ${\bm S}^2$ with eigenvalue $s(s+1)$ (here $s=0$)
can be generated by applying the singlet-state spin projection operator, i.e.,
\begin{equation}
\hat{P}_{\rm spin}^{s}=(1-\varpi_{12})/2,
\label{prjp2}
\end{equation}
where $\varpi_{12}$ is an operator that interchanges {\it opposite\/} spins of the two electrons;
see the general formula in {\Eq}\ (\ref{prjp}) below.

Upon the spin-symmetry restoration, the singlet state of two electrons (with $s=0$) is  given by:
\begin{equation}
2\sqrt{2} \hat{P}_{\rm spin}^{s} \Phi^{S_z=0}_{\rm UHF}(1,2)
= |\varphi_l(1)\overline{\varphi}_r(2)  \rangle  - |  \overline{\varphi}_l(1)\varphi_r(2)  \rangle.
\label{prj0}
\end{equation}
We note that the beyond-mean-field projected many-body wave function (\ref{prj0}) is a linear
superposition of two Slater determinants, in contrast to the single-determinant wave functions
of the RHF and UHF methods.

A further expansion of the determinants in {\Eq}~(\ref{prj0}) produces the equivalent expression
\begin{eqnarray}
2 \hat{P}_{\rm spin}^{s} \Phi^{S_z=0}_{\rm UHF}(1,2)  &=&
(\varphi_l({\bm r}_1)\varphi_r({\bm r}_2)
+\varphi_l({\bm r}_2)\varphi_r({\bm r}_1))
\nonumber \\
&& \times \chi(s=0,S_z=0),
\label{hl1}
\end{eqnarray}
where $\chi(s=0,S_z=0)$ is the spin eigenfunction for the singlet state, and is given by
\begin{equation}
\chi(s=0,S_z=0)=(\alpha_1\beta_2-\alpha_2\beta_1)/\sqrt{2}~.
\label{spin0}
\end{equation}
The expression in the right-hand side of {\Eq}~(\ref{hl1}) is similar
to the Heitler-London \cite{hl27} or valence-bond wave function.
However, at variance with the Heitler-London approach, which employs the
left and right orbitals of the fully separated atoms, this expression
involves the UHF orbitals that are optimized self-consistently at any
separation $d$ and potential barrier height $V_b$.

Considering the
normalization of the spatial part, one arrives at the following
improved wave function of the singlet state:
\begin{equation}
\Psi^{\rm s}_{\text{PRJ}}(1,2) = N_+ \sqrt{2}\hat{P}_{\rm spin}^s \Phi^{S_z=0}_{\rm UHF}(1,2),
\label{gvb}
\end{equation}
where the normalization constant is given by
\begin{equation}
N_+ = 1/\sqrt{1+S^2_{lr}}~,
\end{equation}
$S_{lr}$ being the overlap integral of the spatial orbitals $\varphi_l({\bm r})$
and $\varphi_r({\bm r})$,
\begin{equation}
S_{lr}= \int  {\rm d}^2\bm{r} \varphi_l({\bm r})\varphi_r({\bm r}).
\end{equation}
We stress again that the improved wave function $\Psi^{\rm s}_{\text{PRJ}}(1,2)$
exhibits all the symmetries of the original
many-body Hamiltonian.

The total energy of the symmetry-restored singlet state $\Psi^{\rm s}_{\text{PRJ}}(1,2)$
[{\Eq}~(\ref{gvb})] reads
\begin{equation}
E^{\rm s}_{\text{PRJ}}=N_+^2 \left[h_{ll}+h_{rr}
+2S_{lr}h_{lr}+J_{lr}+K_{lr}\right],
\label{engvb}
\end{equation}
where $h$ is the single-particle contribution to the total Hamiltonian defined in {\Sec}~\ref{mbh},
and $J$ and $K$ are the direct and exchange matrix elements
of the $e-e$ repulsion $e^2/\kappa r_{12}$, where $\kappa$ is the material dielectric constant.
In {\Eq}\ (\ref{engvb}), the subscripts $l$ and $r$ indicate that the matrix elements are associated with the
 left and right space orbitals, $\varphi_l({\bm r})$ and $\varphi_r({\bm r})$, respectively.
For comparison, we provide also here the corresponding formula for the HF total energy, which is valid
for both the RHF [with $\varphi_l({\bm r})=\varphi_r({\bm r})=\varphi({\bm r})$] and UHF
[with either $\varphi_l({\bm r})\neq\varphi_r({\bm r})$ or $\varphi_l({\bm r})=\varphi_r({\bm r})=
\varphi({\bm r})$] cases, i.e.,
\begin{equation}
E^{\rm s}_{\text{HF}}=h_{ll}+h_{rr}+J_{lr}.
\label{enghf}
\end{equation}

For the triplet state, the UHF solution (with $S_z=1$) does not break any symmetries for all
set of values of the parameters $d$, $V_b$, and strength of the $e$-$e$ repulsion.
For the triplet HF determinant, one has  $\Phi^{\rm t}_{\rm RHF}(1,2)=\Phi^{\rm t}_{\rm UHF}(1,2)
=|\varphi_{\rm 0n}({\bm r})\varphi_{\rm 1n}({\bm r})\rangle$,
with  the absence of bars over the $\varphi$ orbitals reflecting that both electrons have $\alpha$ (up)
spins. The indices  "0n" and "1n" denote zero-node and one-node space orbitals, respectively, and,
for the set of parameters used here, they have the general form of molecular orbitals,
similar to the $\Psi^+$ and $\Psi^-$  in {\Fig}\ \ref{DSPW1}(a); for the nodeless $\varphi_{\rm 0n}$,
see also the MO orbital (modulus square) in the left column of {\Fig}\ \ref{new1}.
In this case, the spin-projection operator, $\hat{P}_{\rm spin}^{{\rm t},S_z=1} \equiv 1$, has no effect
[no opposite spins to be interchanged, see the general formula in {\Eq}\ (\ref{prjp}) below],
 and the projected wave function coincides with the starting
HF determinant. As a result the energies for the triplet state in all three
approximations are equal, i.e., $E^{\rm t}_{\text{PRJ}}=E^{\rm t}_{\text{UHF}}=
E^{\rm t}_{\rm RHF}=E^{\rm t}_{\rm HF}$, with
\begin{equation}
E^{\rm t}_{\rm HF}=h_{\rm 0n,0n}+h_{\rm 1n,1n}+J_{\rm 0n,1n}-K_{\rm 0n,1n}.
\label{entri}
\end{equation}

\begin{figure}[t]
\centering\includegraphics[width=0.80\columnwidth]{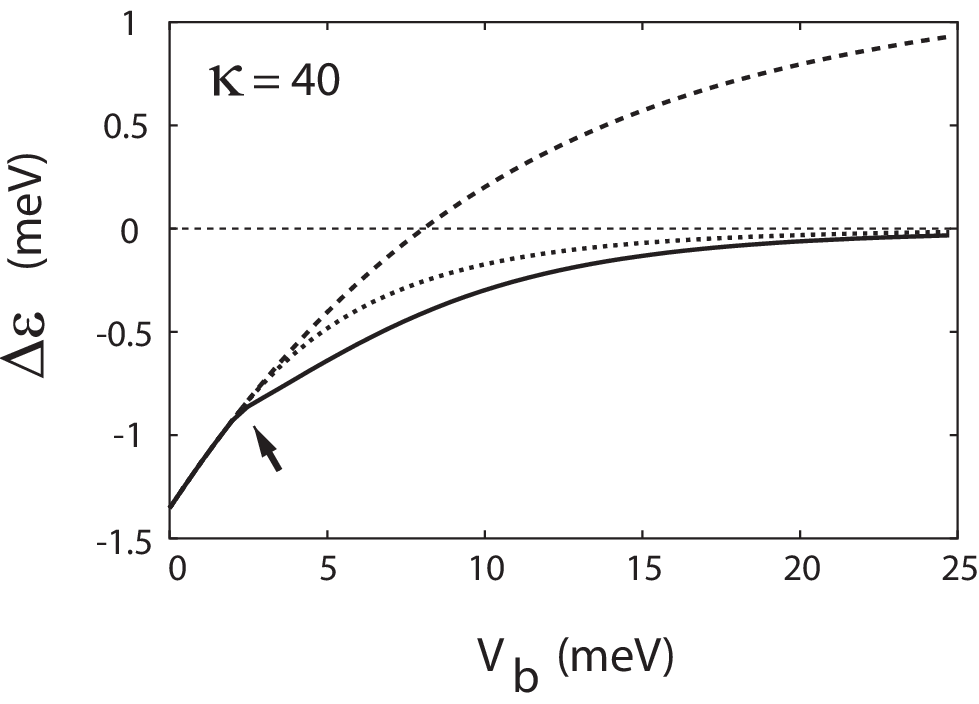}
\caption{
Lateral H$_2$-QDM: The energy difference $\Delta \varepsilon=E^{\rm s}-E^{\rm t}$ between
the singlet and triplet states according to the RHF (molecular-orbital theory, top line), the UHF
(broken symmetry, middle line), and the Projection Method (symmetry restoration,
bottom line) as a function of the interdot barrier $V_b$. For $V_b=25$\,meV, complete separation of
the two dots is reached, signaled by a value of $\Delta \varepsilon \rightarrow 0$ in both
the UHF and Projection-technique approaches.
The remaining parameters are: electron effective mass $m^*=0.067 m_e$, trapping frequency
$\hbar \omega_0=5$\,meV, interdot distance $d=30$ nm, and dielectric constant $\kappa=40$.
The arrow marks the onset of symmetry breaking for the singlet state.
Reprinted by permission from Springer Nature and Copyright Clearance Center from \protect\citeasnoun{(Yan01)}.
}
\label{new2}
\end{figure}

Like in the case of the natural H$_2$, the energy difference, $\Delta \varepsilon=E^{\rm s}-E^{\rm  t}$,
between the singlet and the triplet states of the artificial H$_2$-QDM, must vanish from below as the
separation increases. Equivalently one can keep the interdot distance $d$ constant and vary the height
of the interdot barrier $V_b$.

Figure~\ref{new2} illustrates the evolution of $\Delta \varepsilon$ as a function of $V_b$ in zero magnetic
field, and for all three successive approximation steps, i.e., the RHF (molecular-orbital theory, top line),
the UHF (broken symmetry, middle line), and the projection approach (symmetry restoration, bottom line).

An inspection of {\Fig}~\ref{new2} shows that both the UHF and projection-technique wave functions
describe the energetics of the separation limit ($\Delta \varepsilon \rightarrow 0$ for
$V_b \rightarrow \infty$) rather well, while the RHF approach fails. Furthermore, in this limit,
the projected (symmetry restored) singlet wave function [see {\Eqs}\ (\ref{prj0}) and (\ref{hl1})]  has the
advantage of reproducing the fully spin-entangled two-fermion wave function introduced by David Bohm
\cite{bohm57,bohmbook} as a simpler example for demonstrating the Einstein-Podolsky-Rosen
quantum-mechanical entanglement paradox \cite{epr35}. This exemplifies the potential for utilizing the
symmetry restoration methodologies in the context of the emerging field of quantum information
science \cite{cloe19}.

\subsection{The seniority model}
\label{seniority}

So far we discussed symmetry breaking and restoration for one or two particles.
However, an essential point of the symmetry
breaking discussed in this article is the approximate treatment of
correlations in a many-body system by introducing the mean-field
approximation, that is, by describing the many-body system in terms of a product
state $|\Phi\rangle$ of uncorrelated particles (or quasi-particles) moving in
a single-particle potential with a broken symmetry. To elaborate this aspect
in more detail, we now briefly discuss the seniority model introduced by
\citeasnoun{Kerman1961_APNY12-300} as an example. We consider $N$ fermions in
a degenerate single $j$-shell (with $\Omega=j+\frac{1}{2}$) interacting
through a monopole pairing force with the corresponding many-body Hamiltonian $\hat{H}$,
\begin{equation}
\hat{H}=-G\hat{S}_{+}\hat{S}_{-}\label{eq:seniority-model},
\end{equation}
where $G$ is the strength of the interaction and the operator,
\begin{equation}
\hat{S}_{+}=(\hat{S}_{-})^{\dag}=\sqrt{\frac{\Omega}{2}}[a_j^{\dag}a_j^{\dag
}]_{J=0},
\end{equation}
where $a_j^{\dag}$ creates a single particle in the j-th shell and $\hat{S}_{+}$
creates a Cooper-pair of particles coupled to angular momentum $J=0$. Together
with the operator $\hat{S}_{0}$ connected with the particle-number operator $\hat{N}$ by
the relation%
\begin{equation}
\hat{S}_{0}=\frac{1}{2}(\hat{N}-\Omega),
\end{equation}
operators $\hat{S}_{+},\hat{S}_{0},$ and $\hat{S}_{-}$ form the algebra of
the group $SU(2)$ of the quasi-spin. It has the Casimir operator $\mathbf{\hat{S}}^{2}$,
\begin{equation}
\mathbf{\hat{S}}^{2}=\hat{S}_{+}\hat{S}_{-}+\hat{S}_{0}^{2}-\hat{S}_{0},
\end{equation}
with the eigenvalue $S(S+1)$. The Hamiltonian (\ref{eq:seniority-model})
can be expressed as
\begin{equation}
\hat{H}=-G(\mathbf{\hat{S}}^{2}-\hat{S}_{0}^{2}+\hat{S}_{0}),
\end{equation}
and thus it is diagonal in the basis characterized by the quantum numbers $S,S_{0}$ or $S,N$.
The particle vacuum $|-\rangle$ with $N=0$ is given by $S=\frac{\Omega}{2}$
and $S_{0}=-\frac{\Omega}{2}$. Starting from this vacuum $|-\rangle$, and
applying the raising operator $\hat{S}_{+}$, one finds the exact ground
states of the system with even particle numbers $N$ \cite{Hara1967_NPA95-385},
\begin{equation}
|\frac{\Omega}{2},N\rangle\propto\hat{S}_{+}^{N/2}|-\rangle .
\label{eq:exact-solution}%
\end{equation}
This is a condensate of $n=\frac{N}{2}$ Cooper pairs. In terms of the
original fermions this is a highly correlated state.

We now use the mean-field approximation, i.e., we start with the BCS-state
\begin{equation}
\label{BCSstate}
|\Phi\rangle=%
{\displaystyle\prod\limits_{m>0}}
(u_{m}+v_{m} a_{jm}^{\dag}a_{j\bar{m}}^{\dag})|-\rangle\propto%
{\displaystyle\prod\limits_{m}}
\alpha_{jm}|-\rangle ,
\end{equation}
where $\alpha_{jm}^{\dag}=u_{m}a_{jm}^{\dag}-v_{m}a_{j\bar{m}}^{{}}$.
$\bar{m}$ denotes the time-reversed state. $|\Phi\rangle$ has the
form of a generalized product state and it is the vacuum for the
quasiparticle operators $\alpha_{jm}^{\dag}$. We, therefore, can
apply the Wick theorem (see {\Eq}~(7.47) of Ref.~\cite{ring2000}) for the
two-body operator (\ref{eq:seniority-model}). Neglecting, as usual in
the BCS approximation, the terms $a^\dag a$ we obtain the BCS
Hamiltonian
\begin{equation}
\hat{H}_{BCS}= const. - \Delta(\hat{S}_{+}+\hat{S}_{-}),
\end{equation}
with the gap parameter $\Delta=G\langle\Phi|S_{+}|\Phi\rangle$. The
BCS state $|\Phi\rangle$ is an eigenstate of $\hat{H}_{BCS}$ and the
BCS coefficients $u_{m}$ and $v_{m}$ are determined by the
diagonalization of the mean-field Hamiltonian $\hat{H}_{BCS}$.

It is evident that $\hat{H}_{BCS}$ breaks the particle-number symmetry; therefore one looks for
eigenfunctions of $\hat{H}_{BCS}^{\prime}=\hat{H}_{BCS}-\lambda\hat{N}$, where the
Lagrange parameter $\lambda$ is determined by the subsidiary condition
$\langle\Phi|\hat{N}|\Phi\rangle=N$. Since the BCS Hamiltonian depends on the gap
parameter $\Delta$, one ends up with a non-linear problem, which
has to be solved iteratively. In the seniority model discussed here, the single-particle part (${\sim}a^{\dag
}a$) of $\hat{H}_{BCS}$ vanishes and thus does not depend on the quantum number $m$, and hence the BCS
amplitudes $u_{m}$ and $v_{m}$ do not depend on $m$ either.

The BCS Hamiltonian $\hat{H}_{BCS}$ breaks the particle-number symmetry, and so it also breaks the
gauge symmetry, see {\Sec}~\ref{sec-PNPN} below.
Through this symmetry breaking, it is possible to represent the wave function
as a product state in terms of quasiparticles, cf.~{\Eq}~(\ref{BCSstate}). In the exact solution, these
quasiparticles are not independent and there are additional correlations,
which are not taken into account in the product state. However, we can
bring these correlations back by the restoration of the symmetry. This is achieved
by the particle-number projection, that is, by neglecting in the wave function $|\Phi\rangle$
all the contributions with particle numbers different from $N$. For this
purpose we express the BCS state $|\Phi\rangle$ (\ref{BCSstate}) in terms of the operator
$\hat{S}_{+}$ as
\begin{equation}
|\Phi\rangle\propto{\displaystyle\prod\limits_{m>0}}\exp(\frac{v}{u}%
a_{jm}^{\dag}a_{j\bar{m}}^{\dag})|-\rangle=\exp(\frac{v}{u}%
{\displaystyle\sum\limits_{m>0}}
a_{jm}^{\dag}a_{j\bar{m}}^{\dag})|-\rangle\propto\exp(\frac{v}{u}\hat{S}%
_{+})|-\rangle,
\end{equation}
where we have used the fact that the squares of fermion creation operators
are equal to zero and thus only the first two terms of the exponential
remain, as in {\Eq}~(\ref{BCSstate}).
The particle-number projection then leads to%
\begin{equation}
\hat{P}^{N}|\Phi\rangle\propto\hat{S}_{+}^{N/2}|-\rangle .
\end{equation}
We find that the restoration of the symmetry leads to the exact solution
(\ref{eq:exact-solution}) of the system. Of course, this is a very specific
model and the fact, that symmetry restoration brings us back to the exact
solution, depends definitely on the fact, that the operator $\hat{S}_{+}$ does
not depend on the particle number nor on other properties of the model such as
the strength parameter $G$. Nonetheless, also in more general
cases, where the symmetry restoration does not lead to the exact solution, we will find that
by restoring symmetries, one can improve the mean-field approximation considerably.

\section{Symmetry Restoration - General Formalism}
\label{sec_srgf}

As discussed in the previous section, mean-field solutions may break
symmetries that the original many-body Hamiltonian obeys. For strong
symmetry breaking, approximate methods can be used to evaluate the
observable quantities to a good accuracy, and these are discussed in
{\Sec}~\ref{sec:approximate}. For weak symmetry breaking, wave
functions defined in the intrinsic frame of reference should have
their broken symmetries restored. In the sixties and seventies of the
last century, considerable efforts were made to decouple the total
Hamiltonian in terms of intrinsic and collective degrees of freedom \cite{(Lipkin3),(Lipkin4),Villars1957_ARNS7-185,(Lip58),(Vil70a)}.
Owing to the Galilean invariance, this method is well defined only
for the simplest case of the linear-momentum, where the center of mass
coordinate separates exactly from the intrinsic degrees of freedom,
described, for instance, by Jacobi coordinates.
This is used for few-body systems, i.e., in very light nuclei~\cite{Navratil2000_PRC61-044001}. For heavier
nuclei this technique becomes quite cumbersome because the wave functions
are far from being product states in these coordinates. Only recently a method
based on Quantum Monte Carlo techniques has been introduced, which allows
the exact treatment of Galilean invariance even for density functionals with
finite range~\cite{(Mas20)}. Because of its numerical complexity,
however, applications are also limited to light nuclei and it cannot be generalized to
other symmetry violations, where Galilean invariance does not apply.

Moreover, the popular terminology referring to the intrinsic and
laboratory reference frames can be either confusing, or useless, or
both. Indeed, in the language of the symmetry restoration, no
reference frame is ever changed, namely, both the broken-symmetry and
restored-symmetry wave functions reside in the same unique Hilbert
space with one unique reference frame conveniently predefined and
used. Then, the restored-symmetry wave function is obtained from the
broken-symmetry one by acting on it with a specific
symmetry-restoration operator, which does not change any reference
frame either, but rather rotates the broken-symmetry wave function in
the predefined reference frame.

In this article, we use the notion of an {\em active} rotation,
whereby not the reference frame but the states are rotated.
However, even if we used a {\em passive} rotation scheme, whereby the
wave function stays the same and the reference frame is rotated, this
would not have been equivalent to any change of the reference frame
from intrinsic to laboratory. Although we may occasionally slip into
the traditional terminology of the intrinsic and laboratory frames,
the reader should always follow the correct description by translating the
term "wave function in the intrinsic (laboratory) frame" as
"broken-symmetry (restored-symmetry) wave function".

For a continuous symmetry group, an infinite number of solutions, which are
degenerate in energy, are obtained by applying the
elements of the Lie group on the intrinsic state. For instance,
all Nilsson intrinsic states that differ only by rotation in space have
the same energy. \citeasnoun{(Pei57a)} employed a linear
superposition of these degenerate states with the weight
functions determined through a second variational procedure.
In this way, the variational procedure is performed in two steps. In
the first one, the intrinsic state is determined, and in the
second one, the energy is minimized within the subspace of states,
invariant under the symmetry group, which can be projected from that
intrinsic state. This double variational approach is referred to as
projection after variation.

In a single-step variational procedure, the variation is performed by
simultaneously considering symmetry-restored trial
states that can be obtained from any intrinsic state.
In the case of rotational symmetry, this corresponds to
first projecting the intrinsic wave function onto a state with
well-defined angular-momentum and then performing the variational procedure over
the intrinsic states. This variation after projection method, proposed originally by Zeh \cite{ZEH65},
for a simple degenerate model leads to practically exact results \cite{(She00c)}.
The difference between two-step and single-step variational approaches becomes quite obvious for the case of
particle permutational symmetry, with the method of Peierls and Yoccoz giving
rise to an approximate anti-symmetrized Hartree solution, while the method of
Zeh leading to the correct HF solution.

Projection methods developed to restore the symmetries can be divided
into those pertaining to abelian and non-abelian symmetry groups \cite{Lowdin1967,MacDonald2006,ring2000}.
Restoration of the linear-momentum symmetry,
gauge symmetry associated with the particle number, and parity symmetry
pertain to the abelian groups. For this class, projection
operators have expected mathematical properties of idempotency and
hermiticity.  For the class of non-abelian symmetry groups, which includes
three-dimensional rotations, and corresponds to restoration of angular-momentum or isospin, the projection
operators do not have properties of idempotency and hermiticity. However, as we discuss below, such operators
project out the relevant quantum numbers from the intrinsic wave function, and that is what is important for physical applications.

In the following, we first construct the
projection operators using group-theory and generator-coordinate
methods. As is evident, all the projection operators
can be expressed as basis transformations in some representation.
Projection methods for various symmetries, such as
linear momentum, three-dimensional angular momentum, particle number,
and parity are discussed in {\Sec}~\ref{sec_pmfvs}.

The projection formalism using the generalized HFB basis is
discussed in {\Sec}~\ref{sec:PHFB}. In this section, the expressions for
the norm and the Hamiltonian kernels between the HFB transformed and the initial
basis are obtained using the generalized Wick's theorem. Methods for performing variation after
projection are then discussed in this section. In particular,
it is demonstrated that variation after projection of an arbitrary
symmetry-projected energy leads to  HFB-like equations, as is the case
for the unprojected energy, with the only difference that the pairing and HF
fields get modified and depend on the projected quantum numbers
as well.

\subsection{Projection Operator - Mathematical Basis} \label{sub_sec_pomb}
The mathematical structure of the projection operators
is constructed in this subsection using the group-theory approach and the
generator coordinate method.

\subsubsection{Group-Theory Approach}
\label{sec:grouptheory}
Given a symmetry group, a projection operator, $\hat{P}^\lambda$  can be constructed such that
for an arbitrary wave function $|\Phi\rangle$, $\hat{P}^\lambda |\Phi\rangle$ is
the component of $|\Phi\rangle$ belonging to the irreducible representation characterized by the
quantum number(s) $\lambda$. In the following, we construct the mathematical structure of such an
operator using the group-theory approach and closely follow
the textbooks of Hamermesh \cite{hame62} and Gilmore \cite{(Gil08)}.

Let us suppose that the states $|\lambda\mu\rangle$, enumerated by quantum number(s) $\mu$,
span an irreducible representation of the group
defined by the quantum number(s) $\lambda$.  Within
the irreducible representation, the group transformations $\{ {\hat{R}}(g)\}$ act in the following way
\begin{equation}
{{\hat{R}}(g)}~|\lambda\mu\rangle  = \sum_{\nu} |\lambda\nu\rangle~D^\lambda_{\nu\mu}(g) ,
\label{eq:29}
\end{equation}
where $g$ denotes the group element.
The matrix functions $D^\lambda_{\mu\nu}(g)$ are the continuous single-valued representations of the group,
which for the group of three-dimensional rotations are called Wigner $D$-functions \cite{Edmonds1957,(Var88)}.
They obey the orthogonality theorem of representation theory
\begin{equation}
\sumint dg~D^{\lambda\ast}_{\mu\nu^{}}(g)D^{\lambda'}_{\mu'\nu'}(g) = \frac {V}{n_I}
\delta_{\lambda\lambda'}~\delta_{\mu\mu'}~\delta_{\nu\nu'},
\label{eq:30}
\end{equation}
where $n_I$ denotes the dimension of the irreducible
representation.

For finite groups, $V$ is the order of the group and the sum in
{\Eq}~(\ref{eq:30}) runs over all group elements, whereas for continuous
compact groups, $V$ is the volume of the parameter space  of the
group, and the sum over the group elements should be replaced by an
integral over the group parameters. For example, for the rotational
group in three dimensions, where the group elements depend on three
Euler angles $\Omega=(\alpha,\beta,\gamma)$, the volume is $V=\int
d\Omega=8\pi^2$ and the dimension of the representation characterized
by the angular momentum $I$ is $n_I=2I+1$. Moreover, for continuous
non-compact groups (like the translational group), representations
are labeled by continuous parameters and the Kronecker deltas must
be replaced by Dirac deltas.

Multiplying {\Eq}~(\ref{eq:29}) by $D^{\lambda\ast}_{\mu\nu}(g)$ and summing over
the group elements, we have
\begin{equation}
\sumint dg\,D^{\lambda\ast}_{\mu\nu}(g){{\hat{R}}(g)}~|\lambda'\nu'\rangle =
\delta_{\lambda\lambda'}~\delta_{\nu\nu'}\frac{V}{n_I}|\lambda\mu\rangle~.
\label{eq:31}
\end{equation}
This allows us to define operators $\hat{P}^\lambda_{\mu\nu}$,
\begin{equation}
\hat{P}^\lambda_{\mu\nu} = \frac {n_I} {V} \sumint dg\,D^{\lambda\ast}_{\mu\nu}(g){\hat{R}}(g)
\label{eq:32}
\end{equation}
such that
\begin{equation}
\hat{P}^\lambda_{\mu\nu}~|\lambda'\nu'\rangle =\delta_{\lambda\lambda'}~\delta_{\nu\nu'}~|\lambda\mu\rangle
\label{eq:33}
\end{equation}
and
\begin{equation}
\hat{P}^\lambda_{\mu\nu}\hat{P}^{\lambda'}_{\mu'\nu'}
=\delta_{\lambda\lambda'}~\delta_{\nu\mu'}~\hat{P}^\lambda_{\mu\nu'},
~~~~\left(\hat{P}^\lambda_{\mu\nu}\right)^\dag=\hat{P}^\lambda_{\nu\mu}.
\label{eq:34}
\end{equation}
It is evident from the above equation that the diagonal operators,
$\hat{P}^\lambda_{\mu\mu}$,
project out the $\mu$-th columns of the $\lambda$-th irreducible representation, and obey
\begin{equation}
  \hat{P}^\lambda_{\mu\mu}~\hat{P}^{\lambda'}_{\mu'\mu'}
  = \delta_{\lambda\lambda'}~\delta_{\mu\mu'}~\hat{P}^\lambda_{\mu\mu}.
\label{eq:35}
\end{equation}
It means that they are hermitian and idempotent, i.e., projection
operators in the strict mathematical sense.

For non-abelian groups, the wave functions
$|\Phi^\lambda_{\mu\mu}\rangle\equiv\hat{P}^\lambda_{\mu\mu}|\Phi\rangle$ are no longer tensors with respect
to this group, i.e., under the operation $\hat{R}(g)$ they do not
behave as wave functions $|\lambda\mu\rangle$ shown in
{\Eq}~(\ref{eq:29}). However, owing to the completeness of the matrix
functions $D^\lambda_{\mu\nu}(g)$ in the space of regular functions of $g$, operators
$\hat{P}^\lambda_{\mu\mu}$ obey the following important relation,
\begin{equation}
  \sum_{\lambda\mu}\hat{P}^\lambda_{\mu\mu}=\hat{1},
\label{eq:35a}
\end{equation}
that is, they provide for the so-called resolution of unity. This
means, that every wave function $|\Phi\rangle$ can be uniquely split
into a sum of components $|\Phi^\lambda_{\mu\mu}\rangle$,
$|\Phi\rangle=\sum_{\lambda\mu}|\Phi^\lambda_{\mu\mu}\rangle$, even if
these components are not necessarily orthogonal or normalized, or
even if some of them can vanish.

\subsubsection{Generator Coordinate Method}
\label{GCM}

The expression for the projection operator in {\Eq}~(\ref{eq:32}) can also be obtained as a
special case of the generator coordinate method~\cite{hill53,grif57,ring2000}.
In this method, \citeasnoun{(Pei57a)} started from the symmetry
breaking wave function $|\Phi\rangle$ and diagonalized the many-body Hamiltonian
in the collective subspace spanned by a set of generating functions
\begin{equation}
|\Phi (g)\rangle = \hat{R}(g)|\Phi\rangle.
\label{eq:36}
\end{equation}
Using the ansatz
\begin{equation}
|\Psi\rangle = \int{dg} f(g)|\Phi(g)\rangle,
\label{eq:37}
\end{equation}
the variation of the energy expectation value with respect to the
weight function $f(g)$ leads to a generalized eigenvalue problem
with the generating functions as non-orthogonal basis states.
To restore the symmetry, the weight function is chosen in such a way
that the resulting many-body wave function transforms as in
{\Eq}~(\ref{eq:29}). It can be shown that the collective subspace is
invariant under the symmetry transformations, i.e.,
\begin{equation}
\hat{R}(g)|\Psi\rangle = \int{dg'} f(-g+g') |\Phi(g') \rangle.
\label{eq:38}
\end{equation}
where $(-g+g')$ is a short hand notation for the group element
$\hat{R}^{-1}(g)\hat{R}(g')$.
This implies that the projector $\hat{P}$ onto this subspace
commutes with the symmetry operator $\hat{R}(g)$ and it is
possible to find simultaneous eigenstates of $\hat{P} \hat{H}
\hat{P}$ and the symmetry operator. Thus, a function $f(g)$
exists that not only minimizes the energy but also has the proper
symmetry. This function $f(g)$ can be found by expanding it in terms of the
representations of the group, characterized by eigenvalues of the
Casimir operators.

\subsubsection{Generic properties of projection}

Over the years projection techniques were introduced for various symmetry groups.
In the following, we use a generic form of the projection operator (\ref{eq:32}),
\begin{equation}
\hat{P}^{I}=\int dg\,D^{I*}(g)\hat{R}(g).
\label{eq:51a}
\end{equation}
Assuming that the Hamiltonian $\hat H$ commutes with the symmetry operator $\hat R(g)$,
the projected energy is given by
\begin{equation}
E^{I}=\frac{\langle \Phi |\hat{H}\hat{P}^{I}|\Phi \rangle }{\langle \Phi
|\hat{P}^{I}|\Phi \rangle }=
\frac{\int dg\,D^{I*}(g)\langle\Phi |\hat{H}\hat{R}(g)|\Phi \rangle }{\int dg\,D^{I*}(g)
\langle \Phi |\hat{R}(g)|\Phi \rangle },
\label{eq:60}
\end{equation}
or equivalently as~\cite{(She00c)}
\begin{equation}
E^{I}=\int dg\,y(g)\langle \Phi|\hat{H}|g\rangle  ,
\label{eq:60a}
\end{equation}
where the rotated wave function $|g\rangle$ is defined as a suitably normalized
generating function (\ref{eq:36}),
\begin{equation}
|g\rangle =\frac{\hat{R}(g)|\Phi \rangle }{\langle \Phi |\hat{R}%
  (g)|\Phi \rangle }\,\,\,  {\rm for} \,\,\,\langle \Phi|g\rangle =1,
\label{eq:61}
\end{equation}
and two auxiliary functions $x(g)$ and $y(g)$ are defined as
\begin{equation}
x(g) =D^{I*}(g)\langle\Phi|\hat{R}(g)|\Phi\rangle~~~~{\rm and}~~~~y(g) =\frac{x(g)}{\int dg\,x(g)}.
\label{eq:62}
\end{equation}
Similar expressions can be found for other operators, such as the multipole operators
and for electromagnetic transition matrix elements~\cite{(Dob09g)}.

For the one-dimensional case, $I$ represents directly the quantum number on which one projects.
In the general case, $I$ represents several quantum numbers, as for instance
$K_x, K_y, K_z$ for projection onto the linear momentum or $I, M, K$ in the case of
three-dimensional rotations.

\subsection{Projection Methods for Various Symmetries}\label{sec_pmfvs}

In Tables \ref{tab:table1}--\ref{tab:table2}, we show examples of symmetry restoration corresponding
to several symmetry groups, with five specific cases discussed in
{\Secs}~\ref{sec-PNPN}--\ref{sec_rot3D} in more detail. In addition, the isospin symmetry restoration
being formally identical to the angular-momentum restoration, {\Secs}~\ref{sec_rot1D} and~\ref{sec_rot3D},
is not explicitly covered in this section, apart from listing the relevant expressions in
Tables~\ref{tab:table1} and~\ref{tab:table2}.

\begin{table}[htb]
\renewcommand{\arraystretch}{1.2}
\caption{Various symmetries and the corresponding parts of the projection operators.}
\label{tab:table1}
\begin{center}
\begin{tabular}{l|c|c|c} %
\hline
symmetry & shift & generator            & group operator \\
         & g     & $\hat{S}$            & $R(g)$         \\
\hline
rotation (1D) & $\alpha$ &~~~ang. mom. $\hat{J}_z$&$e^{-i\alpha\hat{J}_z}$ \\
rotation (3D) & ~$\alpha,\beta,\gamma$~&~~ang. mom. $\hat{\bm J}$&
~$e^{-i\alpha\hat{J}_z}e^{-i\beta\hat{J}_y}e^{-i\gamma\hat{J}_z}$          \\
translation (3D) & ${\bm a}$ & momentum $\hat{\bm P}$  & $e^{-i{\bm a\cdot\hat{P}}}$ \\
rotation in gauge space~& $\phi$ &~~part. number $\hat{N}$~&$e^{-i\phi\hat{N}}$      \\
isorotation (1D) & $\varphi$ &~~~isospin $\hat{T}_3$&$e^{-i\varphi\hat{T}_3}$        \\
isorotation (3D) & $\varphi,\theta,\psi$ &~~~isospin $\hat{\bm T} $&
$e^{-i\varphi\hat{T}_3}e^{-i\theta\hat{T}_2}e^{-i\psi\hat{T}_3}$                     \\
parity & $\phi=0,\pi$ &~~ $\hat{N}_-$& $e^{-i\phi\hat{N}_-}$                         \\
\hline
\end{tabular}
\end{center}
\end{table}

\begin{table}[htb]
\renewcommand{\arraystretch}{1.2}
\caption{Table~\protect\ref{tab:table1} continued.}
\label{tab:table2}
\begin{center}
\begin{tabular}{l|c|c|c} %
\hline
symmetry & weight function & eigenvalue&proj. operator\\
         & $D^{I*}(g)$        & I  &~in equation  \\
\hline
rotation (1D) &
$\frac{1}{2\pi}e^{i\alpha M}$& $M$&(\ref{eq:43}) \\
rotation (3D) &
$\frac{2I+1}{8\pi^2}D^{I*}_{MK}(\alpha,\beta,\gamma)$&$I,M,K$&(\ref{eq:45})\\
translation (3D) &
$\frac{1}{(2\pi)^3}e^{i{\bm a\cdot P}}$ & $\bm P$& \\
rotation in gauge space~&
$\frac{1}{2\pi}e^{i\phi N}$&~$N$&(\ref{eq:53})\\
isorotation (1D) &
$\frac{1}{2\pi}e^{i\varphi T_3}$&~$T_3=\frac{1}{2}(N-Z)$~&\\
isorotation (3D) &
$\frac{2T+1}{8\pi^2}D^{T*}_{T_3,T'_3}(\varphi,\theta,\psi)$ & $T,T_3,T'_3$& \\
parity &
$\frac{1}{2}$,$\frac{p}{2}$& $p$&(\ref{eq:54})\\
\hline
\end{tabular}
\end{center}
\end{table}

\subsubsection{Particle-number}\label{sec-PNPN}

Rotations in gauge space are defined by
\begin{equation}
\hat{R}(\phi )=e^{-i\phi \hat{N}},
\label{eq:52}
\end{equation}
where $\hat N$ corresponds to the number operator for neutrons or protons. Wave functions with
good particle number are only multiplied with a phase. This is no
longer true in the case of HFB- or BCS-wave functions, being linear
combinations of states with different particle numbers. In this case
one has to restore the symmetry by projection onto a good particle
number. This is an abelian
symmetry group and from Table~\ref{tab:table2} we find the corresponding operator
\begin{equation}
\hat{P}^{N}= \frac{1}{2\pi } \int d\phi\ e^{-i\phi (\hat{N}-N)}  .
\label{eq:53}
\end{equation}

Particle-number projection, in particular in the BCS-case is relatively simple. It was applied using different
methods~\cite{(Bay60),Die64,Fomenko1970_JPA3-8,(Jan81a),(Egi82a)},
in particular, also in the framework of approximate projections, see
{\Sec}~\ref{sec:approximate}.

\subsubsection{Linear Momentum}

In the nuclear theory, localized single-particle states are employed that
are not eigenstates of the momentum operator. Conversely, the plane waves that are
eigenstates of the momentum operator cannot describe a localized system of particles.

Apart from very light nuclei, in most of the practical applications the nuclear wave functions
are based in one way or another, on wave functions in localized potentials.
They violate the translational invariance of the underlying Hamiltonian. As shown in
Tables~\ref{tab:table1} and~\ref{tab:table2}, this is a three-dimensional group with the
three operators $\hat{{\bm P}} = \sum_i {\bm p}_i$ of the total momentum as generators.

In principle, one should carry out a projection onto good momentum~\cite{(Pei57a)}. This
is technically rather difficult and therefore, there are very few examples in the literature,
where such an exact projection was carried
out~\cite{Schmid2001_EPJA12-29,Schmid2002_EPJA14-413,Schmid2002_EPJA13-319,Schmid2003_EPJA16-475,Rod04a}. In most of the applications approximate methods are
used. As discussed in {\Sec}~\ref{sec:approximate}, it can be shown that the corrections
introduced in this way decrease with numbers of particles, $A$ as $1/A$ and therefore in most of the applications conservation of the linear momentum is only taken into account approximately (see {\Sec} \ref{sec:approximate}).

\subsubsection{Parity}\label{sec_PP}

The parity projection operator is connected to a discrete symmetry. It
is similar to the projection of spin singlet and triplet states (\ref{prjp2}),
and it can be written as \cite{Egi91},
\begin{equation}
\hat{P}^p  = \frac{1}{2}( 1 + p \hat{\Pi})~~~,
\label{eq:54}
\end{equation}
where $p=\pm 1$ and $\hat{\Pi}=\exp{(-i\pi\hat{N}_-)}$
is the standard parity (inversion) operator. Here $\hat{N}_{-}=\sum_{k}^{'}a^\dag_k a_k$ is a restricted summation over all states $k$ with negative parity. We find $\hat{\Pi}^2=1$ and $\hat{P}^p$ is a
true projection operator.

\subsubsection{One-dimensional rotation}\label{sec_rot1D}

The one-dimensional rotation by the angle $\alpha$ around the $z$-axis is given by
\begin{equation}
\hat{R}(\alpha)|\Phi\rangle = e^{-i\alpha\hat{J}_z}|\Phi\rangle,
\label{eq:39}
\end{equation}
where $\hat{J}_z$ denotes the $z$ component of the angular-momentum
operator.\footnote{In this review, we use dimensionless linear-momentum, angular-momentum,
and spin operators, which represent generators of the corresponding symmetry
groups. Whenever the corresponding quantum-mechanical operators and/or their eigenvalues are considered,
we tacitly assume that the Planck-constant prefactors $\hbar$ are also included.}
The irreducible representations of this group are given by
\begin{equation}
D^M(\alpha)=e^{-i M \alpha},
\label{eq:40}
\end{equation}
where $M$ denotes the eigenvalues of $\hat{J}_z$.
Expressing function $f(\alpha)$ of {\Eq}~(\ref{eq:37}) in terms of these irreducible
representations is equivalent to a Fourier transformation,
\begin{equation}
f(\alpha)= \sum_M g_M D^{M\ast}(\alpha)= \sum_M g_M  e^{iM \alpha}.
\label{eq:41}
\end{equation}
Using the $M$-th component $D^{M\ast}(\alpha)$ as the weight function in {\Eq}~(\ref{eq:37}), we obtain
\begin{equation}
| \Psi^M\rangle =\frac{1}{2\pi}\int\limits_0^{2\pi}d\alpha D^{M\ast}(\alpha)
e^{-i\alpha\hat{J}_z}|\Phi\rangle
=\hat{P}^M|\Phi\rangle.
\label{eq:42}
\end{equation}
with the projector
\begin{equation}
\hat{P}^M=\delta(\hat{J}_z-M)=\frac{1}{2\pi}\int\limits_0^{2\pi} d\alpha\ e^{-i\alpha(\hat{J}_z-M)}.
\label{eq:43}
\end{equation}
This is in full agreement with the general expression (\ref{eq:32}) for
the projection operator derived in {\Sec}~\ref{sec:grouptheory} from group-theory considerations.
We note here, that the one-dimensional symmetry restoration projects deformed
states on good projections $M$ of angular momentum $\hat{J}_z$, but each projected
states is still a mixture of components with good total angular momenta $I\geq{M}$.

\subsubsection{Three-dimensional rotation}\label{sec_rot3D}

As an example of a non-abelian group, we consider rotations in three
dimensions, cf.\ recent comprehensive review in Ref.~\cite{(Bal21)}.
They are characterized by the Euler angles
$\Omega=(\alpha,\beta,\gamma)$. The corresponding group element is
given by the operator
\begin{equation}
\hat{R}(\Omega)=e^{-i\alpha \hat{J}_{z}}~e^{-i\beta \hat{J}_{y}}~e^{-i\gamma \hat{J}_{z}}~~~,
\label{eq:44}
\end{equation}
and we have to introduce the generalized projection operators (\ref{eq:32}):
\begin{equation}
\hat{P}^I_{MK}=\frac{2I+1}{8\pi^2}\int d\Omega\ D^{I\ast}_{MK}(\Omega)\ \hat{R}(\Omega),
\label{eq:45}
\end{equation}
where
\begin{equation}
D^{I}_{MK}(\Omega)=e^{-i\alpha M}~d^{I}_{MK}(\beta)~e^{-i\gamma K}
\label{eq:45a}
\end{equation}
are Wigner $D$-functions \cite{Edmonds1957,(Var88)}.
Following \citeasnoun{LB68}, we define a complete and orthogonal set
of many-body wave functions $|IMi\rangle$ which are eigenstates of
the angular-momentum operators $\mathbf{\hat{J}}^{2}$ and
$\mathbf{\hat{J}}_z$, and $i$ combines all the remaining quantum
numbers in the many-body Hilbert space. Using {\Eq}~(\ref{eq:29}), we have
\begin{equation}
  \hat{R}(\Omega)|IMi\rangle = \sum_{K}D^I_{KM}(\Omega)|IKi\rangle,
\label{eq:46}
\end{equation}
and owing to the completeness relations for the states $|IMi\rangle$, we can express the generalized projectors
(\ref{eq:45}) as
\begin{equation}
\hat{P}^I_{MK} = \sum_i |IMi\rangle\langle IKi|.
\label{eq:47}
\end{equation}
This again shows that only the diagonal term $\hat{P}^I_{MM}$ is a
true projector onto the sub-space of the Hilbert space with the quantum
numbers $I$ and $M$. However, individual states $\hat{P}^I_{MM}|\Phi\rangle$
cannot be identified with the basis states $|IMi\rangle$, i.e., they do not obey
{\Eq}~(\ref{eq:46}).

For a better understanding of the additional\footnote{It is better to call
$K$ additional quantum number than to call it projection of angular momentum
on the intrinsic axis, which is the term frequently used. Indeed, states
$\langle IKi|$ are ``bra'' representations of ``ket'' states  $|IMi\rangle$,
and both correspond to projections of angular momentum on the same predefined
quantization axis.}
quantum number $K$ in the projector $\hat{P}^I_{MK}$, we start from a deformed
intrinsic (symmetry-breaking) wave function $|\Phi\rangle$, and, in analogy to {\Eq}(\ref{eq:49})
we expand the weight function $f(\Omega)$ in {\Eq} (\ref{eq:37}) in terms of the complete set of Wigner functions
\begin{equation}
f(\Omega) = \frac{2I+1}{8\pi^2}\sum_{IMK} g^I_{MK} D^{I\ast}_{MK}(\Omega).
\label{eq:48}
\end{equation}
If we restrict us in this sum to the terms with fixed $I$ and $M$, we obtain, by using {\Eqs} (\ref{eq:37}) and (\ref{eq:45}), the projected many-body state
\begin{equation}
|\Psi^I_M\rangle = \sum_K g^I_{MK} \hat{P}^I_{MK}|\Phi\rangle.
\label{eq:49}
\end{equation}
From (\ref{eq:47}) it is evident that $|\Psi^I_M\rangle$ has good quantum numbers $I$ and $M$
for arbitrary expansion coefficients $g^I_{MK}$ (for further details see~\cite{(Zeh67)}).

In contrast to abelian groups, here the weight function
$f^I_M(\Omega)$ for the generator-coordinate-method ansatz
(\ref{eq:37}) is not completely defined by the symmetry group. The
coefficients $g^{I}_{MK}$ have to be determined by the dynamics of the
system, i.e., by diagonalizing the many-body Hamiltonian or by
minimizing the projected energy.

Only in special cases, this additional diagonalization is not
necessary, e.g., if the intrinsic (symmetry-breaking) wave function $|\Phi\rangle$ is symmetric
with respect to rotations around the intrinsic $z$-axis, i.e., if
$\hat{J}_z|\Phi\rangle=K_0|\Phi\rangle$, then
$\hat{P}^I_{MK}|\Phi\rangle=0$ for $K\neq K_0$ and there is only one
coefficient $g^{I}_{MK_0}$, which is determined by the normalization. A
simple case is the intrinsic state $|\Phi\rangle_{K=0}$ of the ground
state of an axially deformed even-even nucleus. Here we find the
projected states
\begin{equation}
|\Psi^I_{M}\rangle = \hat{P}^I_{M0}|\Phi_{K=0}\rangle,
\label{eq:50}
\end{equation}
which do obey {\Eq}~(\ref{eq:46}),

Further, by integrating in {\Eq}~(\ref{eq:45}) over the Euler angles $\alpha$
and $\gamma$, and by using {\Eq}~(\ref{eq:43}), the generalized projector $\hat{P}^I_{MK}$ can be
decomposed into three steps:
\begin{equation}\nonumber
\hat{P}^I_{MK}\propto \delta(\hat{J}_{z}-M) \int\limits_{-1}^{1} d\cos(\beta)
d^{I\ast}_{MK}(\beta) e^{i\beta \hat{J}_{y}}~\delta(\hat{J}_{z}-K).
\label{eq:51}
\end{equation}
We can now describe this result in two different ways:
\begin{enumerate}
\item In the traditional language of passive transformations between the intrinsic and the laboratory
reference frames, we start with the projector,
$P^K=\delta(\hat{J}_{z}-K)$, onto the quantum number $K$, corresponding to the
component of the angular momentum $\hat{\bm J}$ along the intrinsic
$z$-axis, then we have a rotation by angle $\beta$ around the
$y$-axis from the $z$ axis in the intrinsic system to the $z$-axis in
the laboratory frame, and finally $P^M=\delta(\hat{J}_{z}-M)$
projects on an eigenstate with quantum number $M$ in the laboratory
frame.
\item In the language of symmetry-broken and
symmetry-restored states, the first operation projects out
the symmetry-broken state on the good quantum number $K$
corresponding to the $z$ quantization axis of a predefined reference
frame. The second step projects on the good total-angular-momentum
quantum number $I$, but it does that by a rotation along the $y$
axis, and thus mixes again the previously restored projections of the
angular momentum. Then, the third-step projector is required
to restore the projection $M$ on the $z$ quantization axis of a
predefined reference frame, and to give the fully symmetry-restored
wave function.
\end{enumerate}

The final wave function
$|\Psi^I_M\rangle$ in {\Eq}~(\ref{eq:48}) is a quantum-mechanical
superposition of all these different orientations and, as usual in the
generator coordinate method, the weight functions are related to the corresponding
probability amplitudes~\cite{ring2000}.

It is important to emphasize, that the concept of generator
coordinates, which corresponds here to the projection onto
the subspaces determined by the symmetry group, deals only with the
coordinates of the $A$ particles in the corresponding wave functions
in the intrinsic or in the laboratory frame, i.e., before or after
the symmetry restoration. The collective
coordinates, in this case the Euler angles, enter only in a
parametric way. In none of these considerations one has to introduce
"redundant" coordinates and no spurious states are involved. One
stays, from the beginning to end, completely in the quantum-mechanical
framework and no "requantization" is necessary.  However, the results are not
characterized by orbits in the collective subspace, but by probabilities
corresponding to different orientations.

It needs to be added that there is a prize to pay as these calculations become relatively complicated.
Therefore, although these concepts were around since more than half a century, many of these
calculations, in particular, those with realistic applications are possible only
nowadays using modern computing resources~\cite{(Bal14d)}. Some
of the applications still have to wait for more advanced implementations~\cite{(Rom19c)}.

\subsection{Symmetry restoration of the HFB wave function} \label{sec:PHFB}
In most of the projection studies, one starts with a broken-symmetry
mean-field wave function $|\Phi\rangle$ of the HFB
type~\cite{ring2000,(Sch19)}. These states can be expressed as
vacua of quasiparticles $\alpha_k|\Phi\rangle =0$ which are
connected by a unitary transformation to the operators
$a^\dag_n,a^{}_n$ of an arbitrary reference basis
\begin{equation}
\alpha^\dag_k=\sum_n \left(U_{nk}a^\dag_n+V_{nk}a^{}_{n}\right).
\label{eq:56}
\end{equation}
There is a one-to-one correspondence between the HFB wavefunction and the corresponding one-body
densities,
\begin{equation}
\rho_{nn'}=\langle\Phi|a^\dag_{n'}a^{}_{n^{}}|\Phi\rangle~~~{\rm and }~~~
\kappa_{nn'}=\langle\Phi|a^{}_{n'}a^{}_{n^{}}|\Phi\rangle,
\label{eq:57}
\end{equation}
and Wick's theorem allows us to evaluate the matrix elements of arbitrary many-body operators
$\langle\Phi|\hat{O}|\Phi\rangle$  in terms of these densities, see Appendix \ref{sec:AppA}.
In the conventional HFB theory, wave function $|\Phi\rangle$, i.e., the quasiparticle amplitudes
$U_{nk}$ and $V_{nk}$, are determined by a variation of the energy
$E=\langle\Phi|\hat{H}|\Phi\rangle$ with respect to the densities.
This leads to two single-particle fields
\begin{equation}
h_{nn^{\prime}}=\frac{\partial E}{\partial\rho_{n^{\prime}n}}\text{
\ \ and \ \ \ }
\Delta_{nn^{\prime}}=-\frac{\partial E}{\partial
\kappa_{n^{\prime}n}^{\ast}},
\label{eq:58}%
\end{equation}
and the corresponding HFB-equations
\begin{equation}
{\cal H}^\prime
\left(
\begin{array}
[c]{c}%
U\\
V
\end{array}
\right)%
=
\left(
\begin{array}
[c]{cc}%
h-\lambda & \Delta\\
-\Delta^{\ast} & -h^{\ast}+\lambda%
\end{array}
\right)  \left(
\begin{array}
[c]{c}%
U\\
V
\end{array}
\right)=
\left(
\begin{array}
[c]{c}%
U\\
V
\end{array}
\right)E,%
\label{eq:59}%
\end{equation}
where ${\cal H}^\prime={\cal H}-\lambda {\cal N}$, $E$ is the diagonal matrix of quasiparticle energies and the
chemical potential $\lambda$ is determined by adjusting the average
particle number.

\subsubsection{Rotated Norm and Energy Kernels}\label{rnek}
Standard expressions for the norm overlap are given in Appendix \ref{sec:AppA},
whereas those that show the explicit dependence on the densities $\rho$ and $\kappa$ are given
in~\citeasnoun{(She00c)}:
\begin{equation}
\langle\Phi|\hat{R}(g)|\Phi\rangle^{2}=\,\,\det\left(R_{g}\rho^{-1}A_{g}\right).
\label{eq:63}
\end{equation}
Here $R_{g}^{{}}$ is the matrix representing the group element $g$ in the original basis
\begin{equation}
(R_g)^{}_{nn'}=\langle n|\hat{R}(g)|n'\rangle~~~{\rm and}~~~
A_g=\rho R^{}_g\rho - \kappa R^\ast_g\kappa^\ast.
\label{eq:64}
\end{equation}

For $|g\rangle$ given by {\Eq}~(\ref{eq:61}), the generalized Wick's theorem allows us to express matrix elements of the form
$\langle\Phi |\hat{O}|g\rangle$ in terms of the corresponding transition densities~\cite{ONISHI1966367,(Bal69a),Hara1979_NPA332-61,ring2000}:
\begin{eqnarray}
(\rho_g)_{nn'}&=&\langle\Phi|a^\dag_{n'}a^{}_n|g\rangle=
\left(R_{g}^{} \rho A_{g}^{-1} \rho\right)_{nn'},
\label{eq:65}\\
(\kappa_g)_{nn'}&=&\langle\Phi |a^{}_{n'}a^{}_n|g\rangle=
\left(R_{g}^{{}}\rho A_{g}^{-1}\kappa\right)_{nn'},
\label{eq:66}\\
(\overline{\kappa}^\ast_g)_{n^{}n'}&=&\langle\Phi|a^\dag_{n}a^\dag_{n'}|g\rangle=
\left(R_{g}^{\ast}\kappa^{\ast}A_{g}^{-1}\rho\right)_{nn'}.
\label{eq:67}%
\end{eqnarray}
It is important to note that here all the matrix elements are expressed in terms of the
intrinsic densities $\rho$ and $\kappa$ and the matrix representation $R_g$ of the group element $g$.

As an example, let us consider a Hamiltonian with a two-body interaction of the form
\begin{equation}
\hat{H}=\sum_{nn'}e_{nn'}a^\dag_n a^{}_{n'} +
\frac{1}{4}\sum_{nn'mm'}\overline{v}_{nn'mm'}a^\dag_n a^\dag_{n'}a^{}_{m'} a^{}_{m},
\label{eq:68}
\end{equation}
for which we obtain the transition matrix element:
\begin{equation}
\langle\Phi|H|g\rangle = {\rm Tr}\left(e\rho_g\right)
+\frac{1}{2}{\rm Tr}\left(\Gamma_g\rho_g\right)
-\frac{1}{2}{\rm Tr}\left(\Delta_g\overline\kappa^*_g\right)
\label{eq:69}
\end{equation}
with the rotated fields
\begin{eqnarray}
(\Gamma_g)_{nm}^{} &=& \sum_{n'm'}\overline{v}_{nn'mm'}(\rho_g)_{m'n'}
\label{eq:70}\\
(\Delta_g)_{nm}^{}&=& \sum_{m'<n'}\overline{v}_{nmn'm'}(\kappa_g)_{n'm'}
\label{eq:71}
\end{eqnarray}

In principle, the evaluation of projected matrix elements is
relatively straightforward. One only has to replace the normal
density matrices $\rho$, $\kappa$, and $\kappa^*$ by the transition
densities $\rho(g)$, $\kappa(g)$, and $\overline\kappa^*(g)$ and
integrate over the parameter space of the group. In practice,
however, depending on the dimension of the single-particle space and
the number of mesh-points in parameter space this can require a large
computational effort, in particular, for triaxial nuclei. At each
point $g$ in parameter space one has to invert a large (often
complex) matrix $A_g$ (\ref{eq:64}) with the dimension of the
single-particle space [see for instance \citeasnoun{Yao2014_PRC89-054306}].

As the Madrid group~\cite{Ang01b} showed, it may happen in regions
of level-crossings that $\langle\Phi|\hat{R}(g)|\Phi\rangle$ vanishes
at certain values of $g$. This leads to poles in
certain parts of the Hamiltonian matrix element (\ref{eq:69}), e.g.,
in ${\rm Tr}\left(\Gamma_g\rho_g\right)$, see discussion in {\Sec}~\ref{EDF:Self-energy}.

\subsubsection{Variation after projection}

With the techniques discussed in the previous section, it is relatively simple to
carry out a projection after variation. However, such a procedure is not
variationally optimal. Therefore, the method of variation after projection was proposed
\cite{ZEH65,Yoccoz1966_Varenna,Rouhaninejad1966_NP78-353},
where the mean-field wave function $|\Phi\rangle$ is determined by minimizing
the projected energy, i.e., by solving the equation:
\begin{equation}
\langle\delta\Phi|\hat{P}^{I}(\hat{H}-E)\hat{P}^{I}|\Phi\rangle=0.
\label{eq:72}%
\end{equation}
For the exact solution of this problem, the following two methods were
proposed:

\vspace{1ex}\noindent\underline{Gradient Method}\newline
A particularly powerful method to minimize the projected energy with
respect to the product state is the gradient method, which was
introduced in~\citeasnoun{(Man76a)}, and which was applied
for variation-after-projection calculations
in~\citeasnoun{(Egi82a),(Egi82b)}. In this method,
in the neighborhood of an arbitrary point $|\Phi_{0}\rangle$, the
manifold of the HFB wave functions $|\Phi\rangle$ is parameterized by
the Thouless theorem,
\begin{equation}
|\Phi\rangle\propto\exp(\sum_{k<k^{\prime}}Z_{kk^{\prime}}\alpha_{k^{{}}%
}^{\dag}\alpha_{k^{\prime}}^{\dag})|\Phi_{0}\rangle,
\label{eq:73}%
\end{equation}
where operators $\alpha_{k}^{\dag}$ are the quasiparticle operators with
respect to the quasiparticle vacuum $|\Phi_{0}\rangle$, i.e., $\alpha_{k}%
|\Phi_{0}\rangle=0$. The
gradient of the projected energy with respect to parameters $Z_{kk^{\prime}}$ is given by
\begin{equation}
\gamma_{kk^{\prime}}=\left.  \frac{\partial E^{I}}{\partial Z_{kk^{\prime}%
}^{\ast}}\right\vert _{Z=0}=\frac{\langle\Phi_{0}|\alpha_{k^{\prime}}%
\alpha_{k}(\hat{H}-E^{I})\hat{P}^{I}|\Phi_{0}\rangle}{\langle\Phi_{0}|\hat
{P}^{I}|\Phi_{0}\rangle}.
\label{eq:74}%
\end{equation}
These matrix elements can be evaluated using the generalized Wick's theorem, in a
similar way as it is done for the average energy. Following the direction of
steepest descent on this manifold by a step size of $\eta$, we obtain
in the next step of the iteration, the following wave function
\begin{equation}
|\Phi_{1}\rangle\propto\exp(-\eta\sum_{k<k^{\prime}}\gamma_{kk^{\prime}}%
\alpha_{k^{{}}}^{\dag}\alpha_{k^{\prime}}^{\dag})|\Phi_{0}\rangle.
\label{eq:75}%
\end{equation}
The resulting HFB coefficients of $|\Phi_{1}\rangle$ have to be orthogonalized
[for details see \citeasnoun{(Egi82a)}]. Calculating the new projected
energy $E_{1}^{I}$ and changing the step size $\eta$ in the next step
accordingly, the minimum of the projected energy surface can be found without
diagonalizing any matrix. This method is particularly useful, if one wants to
minimize the energy surface with additional constraints. In this case the
method of Lagrange multipliers is used, where the total gradient is projected onto the
gradient along the hyper-surface determined by the constraining operator. The
speed of convergence of this method can be considerably improved by using the
conjugate gradient method~\cite{(Egi95a)}.

It is evident that the gradient method can only be applied for cases, where the final
solution corresponds to a minimum in the energy surface. In all the
applications of the Covariant Density Functional Theory, see {\Sec}~\ref{Relativistic_EDFs},
because of the "no-sea approximation",
the solution of the corresponding mean-field equations do not correspond to a minimum,
but rather to a saddle point on the energy surface. In this case, the projected HFB equations are more useful.

\vspace{1ex}\noindent\underline{Projected HFB Equations }\newline
In a similar way as the normal HFB-equations in {\Eq}~(\ref{eq:59}) were derived by a variation
of the unprojected energy $E=\langle\Phi|\hat{H}|\Phi\rangle$ with respect to the intrinsic densities
$\rho$ and $\kappa$, projected HFB-equations were derived in \citeasnoun{(She00c)} by a variation
of the projected energy $E^{I}$ in {\Eq}~(\ref{eq:60a}). These equations have the same form as the unprojected HFB equations (\ref{eq:59}). However, in this case the HF and pairing fields depend on the quantum number $I$:     \begin{equation}
h^I_{nn^{\prime}}=\frac{\partial E^{I}}{\partial\rho_{n^{\prime}n}}\text{
\ \ and \ \ \ }\Delta^I_{nn^{\prime}}=-\frac{\partial E^{I}}{\partial
\kappa_{n^{\prime}n}^{\ast}}.
\label{eq:76}%
\end{equation}
In order to write down these quantities explicitly, we need the
analytic form of the projected $E^{I}$ in {\Eqs} (\ref{eq:60a}) and (\ref{eq:69})
in terms of the intrinsic densities $\rho$ and $\kappa$ as given in {\Eqs}~(\ref{eq:64})--(\ref{eq:67}).
The detailed expressions for $\hat{h}^I$ and $\hat{\Delta}^I$ were derived
in~\citeasnoun{SH02} and are not repeated here.

It needs to be mentioned that several other methods were developed to perform the
particle-number projection \cite{ring2000} in the BCS case. In particular, the method of
residuum integrals was introduced to perform the exact particle-number
projection \cite{Die64} before the variation. In this approach,
the particle-number projected method is cast into a set of non-linear equations,
which are similar in structure to those of BCS equations.

\section{Approximate projection methods}
\label{sec:approximate}

The general formalism of symmetry restoration, presented in the previous
section, leads to expressions that involve multi-dimensional integrals of norm overlaps and
Hamiltonian transition matrix elements over the symmetry group parameters, see
{\Eq}~(\ref{eq:60}).
In the standard numerical approach these integrals are replaced
with finite sums,\footnote{An alternative method based on solving
linear equations was recently proposed in~\citeasnoun{(Joh17),(Joh19)}.}
which entails evaluations of the norm overlaps and Hamiltonian
transition matrix elements for each mesh point in the multi-dimensional space of group
parameters.  For each mesh point, the norm overlap is evaluated using
the Onishi or Pfaffian method and the Hamiltonian transition matrix elements are
expressed in terms of the one-body rotated densities  by employing the
generalized Wick's theorem. In general,
the numerical cost of calculating one point of the projected integrand
is somewhat larger than the cost of performing one iteration of the
self-consistent method, required to determine
the broken-symmetry state being projected.

Estimates of the number of mesh points vary a lot depending
on the mass or deformation of the nucleus, and on the precision
desired for the final results. On the one hand, for a strong symmetry
breaking case, a larger number of eigenstates of the symmetry generators
are contained within the broken-symmetry state, and thus more mesh
points are required to resolve them. On the other hand, conserved
symmetries allow us to limit the integration domain and thus to
decrease the number of mesh points.

Typically, for a medium
heavy nucleus with  moderate deformation, about 10 integration points
are needed to project on good proton and neutron numbers, and
about 50 integration points, to project on good angular momentum of
an axial nucleus. Restoration of symmetry then requires a
numerical expense somewhat larger than that required
to perform 5000 self-consistent iterations. It is evident that even
with such a modest mesh size, the restoration of symmetries lead to numerical
cost largely exceeding the typical 100 iterations required
to converge the broken-symmetry state itself. For a triaxial state,
where a three-dimensional integration over the Euler angles
is needed, the number of integration points would increase to
12,500,000 and thus would become unmanageable. Therefore, up to now,
calculations of this type were restricted to lighter systems only,
see~\citeasnoun{(Bal14d)}.

In the early days of projection theory, well
before the above {\it tour-de-force} achievements were
envisioned and when the adequate computing power was not yet
available, several approximate methods for symmetry restoration were
proposed and implemented. Here we discuss in detail the most popular
one, based on the so-called Lipkin method \cite{(Lip60b)} or Kamlah
expansion \cite{Kamlah1968}, along with the variant of the former one
proposed by Nogami \cite{(Nog64)}. In fact, the Lipkin and Kamlah
ideas were basically identical, although Kamlah did not apparently
know about, and he did not cite the much earlier work of Lipkin.\footnote{However, Kamlah
did cite \citeasnoun{(Goo66)} who had cited \citeasnoun{(Lip60b)}.}

The main objective of both approaches is to obtain approximate expressions
for symmetry-projected energies and to employ them in the
implementation of the variation-after-projection approach. However, the two approaches differ on the
physical quantities to be described: Kamlah is primarily concerned
with determining the projected energy, which is then varied; whereas
Lipkin aims to model the entire spectrum of collective states related
to a given broken symmetry. Another difference between them is
that Kamlah does and Lipkin does not consider the effects of
collective motion brought about by the so-called pushing or cranking terms.

The baseline of the Lipkin and Kamlah approaches is the observation
that the average values and matrix elements of operators calculated
between symmetry-projected states always involve transition matrix elements of operators
between symmetry-transformed states, {\Eq}~(\ref{eq:69}). It is thus
obvious that a meaningful approximation of the latter may lead to a
useful approximation of the former.

Finally, we should stress the fact that although the Lipkin and Kamlah approaches
are primarily concerned with identifying the variation-after-projection symmetry-breaking state, they do not
actually determine the projected state. For that, an explicit projection
of the variation-after-projection symmetry-breaking state is always necessary. Only then
one can calculate correct transition probabilities respecting all
symmetry properties of the transition operators.

\subsection{The Lipkin method}
\label{sec:Lipkin}

The main idea of the Lipkin approach \cite{(Lip60b)} can be formulated as a proposal to
flatten the spectrum of projected energies $E^{I}$ (\ref{eq:60}) \cite{(Dob09a),(Wan14c),(Gao15b)},
\begin{equation}
E^{I} - K^{I} = const.~~~, \label{eq:Eflat}
\end{equation}
where
\begin{equation}
K^{I}=\frac{\langle \Phi |{\hat K}{\hat P}^{I}|\Phi \rangle }{\langle \Phi
|{\hat P}^{I}|\Phi \rangle } \quad\mbox{for}\quad [{\hat K},{\hat P}^{I}]=0~~~, \label{eq:KLipkin}
\end{equation}
are average values of the so-called Lipkin operator ${\hat K}$, evaluated between the projected states,
and the constant on the right-hand side of {\Eq}~(\ref{eq:Eflat})
does not depend on labels $I$ of the projected states.
Equivalently, as it is evident from {\Eq}~(\ref{eq:60}), one can flatten the
reduced energy kernels, i.e.,
\begin{equation}
\frac{\langle\Phi |\left({\hat H}-{\hat K}\right)\hat{R}(g)|\Phi \rangle }
     {\langle\Phi |\hat{R}(g)|\Phi \rangle }  = const.~~~, \label{eq:redflat}
\end{equation}
where the constant on the right-hand side, which is the same constant
as in {\Eq}~(\ref{eq:Eflat}), does not depend on the group element
$g$. Indeed, since the analogue of {\Eq}~(\ref{eq:60}) relates
$E^{I} - K^{I}$ to ${\hat H}-{\hat K}$, we can plug there the
numerator of {\Eq}~(\ref{eq:redflat}) expressed though its denominator,
which immediately gives {\Eq}~(\ref{eq:Eflat}). Otherwise, by plugging
{\Eq}~(\ref{eq:Eflat}) into the same analogue of {\Eq}~(\ref{eq:60}), we
see that the function on the group defined by$ \langle\Phi|({\hat
H}-{\hat K})\hat{R}(g)|\Phi \rangle- \langle\Phi |\hat{R}(g)|\Phi
\rangle \times const.$ is orthogonal to all functions $D^{I}(g)$,
and thus must be equal to zero, proving {\Eq}~(\ref{eq:redflat}).

How to find the Lipkin operator ${\hat K}$ that does the job as desired? The strategy is obvious,
namely, since the quantum numbers $I$ are related to the symmetry generators,
one can build ${\hat K}$ as a function of the symmetry generators. For example:
\begin{itemize}
\item
For the particle-number symmetry, projected energies depend on the number of particles, so
${\hat K}$ may depend on the particle-number operators, ${\hat N}$ and ${\hat Z}$, for neutrons and protons:
\begin{equation}
{\hat K}  = \sum_{n+m>0}k_{nm}({\hat N}-N_0)^n ({\hat Z}-Z_0)^m ,
\label{eq:KLipkin-numb}
\end{equation}
where $N_0$ and $Z_0$ are numbers of protons and neutrons of the state we want to describe.
\item
For the translational symmetry, projected energies depend on total momenta, so
${\hat K}$ may depend on the components of the total momentum, ${\hat P}_x$, ${\hat P}_y$, and ${\hat P}_z$:
\begin{equation}
{\hat K}  = \sum_{n+m+l>0}k_{nml} {\hat P}_x^n {\hat P}_y^m {\hat P}_z^l .
\label{eq:KLipkin-trans}
\end{equation}
\item
For the rotational symmetry, projected energies depend on total angular momentum, so
${\hat K}$ may depend on the components of the total angular momentum,
${\hat J}_x$, ${\hat J}_y$, and ${\hat J}_z$:
\begin{equation}
{\hat K}  = \sum_{n+m+l>0}k_{nml} {\hat J}_x^n {\hat J}_y^m {\hat J}_z^l .
\label{eq:KLipkin-rot}
\end{equation}
\item
For the rotational symmetry and axial nucleus oriented along the $z$
axis, projected energies depend on total angular momenta, so ${\hat
K}$ may depend on the total angular momentum $\bm{{\hat J}}^2$:
\begin{equation}
{\hat K}  = \sum_{m>0}k_{m} \left({\bm{\hat J}}^2-I(I+1)\right)^m ,
\label{eq:KLipkin-axial}
\end{equation}
where $I$ is the total angular momentum of the state we want to describe.
\end{itemize}
In {\Eqs}~(\ref{eq:KLipkin-numb})--(\ref{eq:KLipkin-axial}) and below, constants $k$
represent adjustable parameters (the Lipkin parameters), which should best approximate
$E^{I}$ by $K^{I}$, {\Eq}~(\ref{eq:Eflat}).

The main idea behind building the Lipkin operators is to have the best possible
description of spectra $E^I$ of projected energies in terms of averages
of group generators $K^I$. We stress that we do not deal here
with real spectra of the system, but with energies $E^{I}$ of symmetry-conserving components
$\hat{P}^I|\Phi\rangle$ derived from the symmetry-breaking state $|\Phi\rangle$.
In fact, among all the projected states, we are interested in only one of them, that with energy $E^{I_0}$
and wave function $\hat{P}^{I_0}|\Phi\rangle$. Of course, if the
flattening, {\Eqs}~(\ref{eq:Eflat}) or (\ref{eq:redflat}), is perfect -- this does not matter;
however, if it is not perfect,
we better do the best possible job for the one state $I_0$ that we want to describe. Then,
the Lipkin operator constructed so that
\begin{equation}
K^{I_0}\equiv0 \label{eq:Eflat2}
\end{equation}
gives us obviously
\begin{equation}
E^{I} - K^{I} = E^{I_0} \label{eq:Eflat1} .
\end{equation}
In fact, for the examples of the Lipkin operators presented in
{\Eqs}~(\ref{eq:KLipkin-numb}) and~(\ref{eq:KLipkin-axial}), condition
(\ref{eq:Eflat2}) is fulfilled, whereas those in
{\Eqs}~(\ref{eq:KLipkin-trans}) or~(\ref{eq:KLipkin-rot}) apply to
states at rest ($\bm{P}=0$) or non-rotating ($\bm{J}=0$),
respectively.

At this point, by evaluating {\Eq}~(\ref{eq:redflat}) at $g=0$, we obtain the ``magic''
Lipkin formula:
\begin{equation}
E^{I_0} = \langle\Phi| {\hat H}-{\hat K}|\Phi \rangle ,  \label{eq:magic}
\end{equation}
namely, the projected energy $E^{I_0}$ can be obtained as an average value
of ${\hat H}-{\hat K}$ calculated for the symmetry-breaking state $|\Phi \rangle$
{\em without performing any projection at all}. Of course, we can benefit from the
magic formula only if we can find appropriate Lipkin operators that correctly
flatten the spectrum, and the precision of it is dictated by the precision of
the flattening.

Therefore, the main thrust of the method now lies in finding the numerical coefficients
in {\Eqs}~(\ref{eq:KLipkin-numb})--(\ref{eq:KLipkin-axial}) that define the
Lipkin operators in terms of the symmetry generators. Before going into
details of specific applications, let us introduce a generic form of the Lipkin
operator as a linear combination of different terms:
\begin{equation}
{\hat K}  = \sum_{m=1}^M k_{m} {\hat K}_{m} .
\label{eq:KLipkin-generic}
\end{equation}
Following the original idea of \citeasnoun{(Pei57a)}, we now evaluate {\Eq}~(\ref{eq:redflat})
at $M+1$ group elements, $g_i$, for $i=0,\ldots,{M}$, $g_0=0$, which
leads to a set of linear equations that determine the Lipkin parameters $k_{m}$,
\begin{equation}
\sum_{m=0}^M A_{im} k_{m} = h_i ,
\label{eq:KLipkin-linear}
\end{equation}
where we extended the list of symmetry generators by defining,
\begin{equation}
{\hat K}_{0}\equiv{\hat 1} \quad\mbox{and}\quad k_{0}\equiv{}E^{I_0}.
\label{eq:KLipkin-zero}
\end{equation}
Then, coefficients in (\ref{eq:KLipkin-linear}) are
defined by the following reduced kernels:
\begin{eqnarray}
A_{im} &=& \frac{\langle\Phi |{\hat K}_m\hat{R}(g_i)|\Phi \rangle }
                {\langle\Phi |\hat{R}(g_i)          |\Phi \rangle },
\label{eq:KLipkin-reduced1} \\
h_{i}  &=& \frac{\langle\Phi |{\hat H}  \hat{R}(g_i)|\Phi \rangle }
                {\langle\Phi |\hat{R}(g_i)          |\Phi \rangle },
\label{eq:KLipkin-reduced2}
\end{eqnarray}
and the Lipkin parameters can be obtained by inverting matrix $A$:
\begin{equation}
k_{m} = \sum_{i=0}^M A^{-1}_{mi} h_i .
\label{eq:KLipkin-inverse}
\end{equation}
In doing so, we can always adjust the choice of group elements $g_i$
so as to obtain a non-singular matrix $A$. It is noted here that one can
simply ignore the value of $k_0$ given by
{\Eq}~(\ref{eq:KLipkin-inverse}). Indeed, since it is by definition
equal to $E^{I_0}$, {\Eq}~(\ref{eq:KLipkin-zero}), one can always evaluate it
from the magic Lipkin formula (\ref{eq:magic}).

Based on the Peierls-Yoccoz prescription to determine the Lipkin parameters, one has
to calculate a few norm overlaps and energy transition matrix elements -- the same
ones that are required for the execution of the full projection,
{\Eq}~(\ref{eq:60}). Their number is, however, significantly smaller than
that required for a full projection, and thus the Lipkin method is
computationally much less intensive.

The biggest advantage of the Lipkin method manifests itself when we attempt to
obtain the variation-after-projection solution. Indeed, an exact implementation leads then to
difficult programming and lengthy calculations,
cf.~{\Sec}~\ref{sec:PHFB}. On the other hand,
variation of the projected energy obtained from the magic Lipkin
formula (\ref{eq:magic}) is as easy as a direct variation of the energy of
the symmetry-breaking state. Clearly, the Lipkin method gives only
the projected energy, whereas, if average values of other observables are to be
calculated, the full projection has to be anyhow
performed. Then, the Lipkin method allows for obtaining
variation-after-projection results at the expense of a single projection-after-variation calculation, which still
constitutes a substantial gain in computing time and efficiency.

One should stress one important aspect of the Lipkin method, namely,
when varying state $|\Phi\rangle$ that appears in the magic formula
(\ref{eq:magic}) to obtain the variation-after-projection result, one should treat the
Lipkin parameters $k_{m}$ as constants that do not undergo variation.
Indeed, even if their values parametrically depend on $|\Phi\rangle$
through {\Eqs}~(\ref{eq:KLipkin-reduced1})--(\ref{eq:KLipkin-inverse}), their role is to provide the best
flattening of the final spectrum, so during the variation
they should be kept fixed and equal to the final variation-after-projection values.
For iterative solutions of self-consistent equations, the algorithm
to achieve such a goal can easily be implemented. Indeed, it is enough
to treat Lipkin parameters at every iteration as constants, and
at the end of every iteration recalculate them using
{\Eqs}~(\ref{eq:KLipkin-reduced1})--(\ref{eq:KLipkin-inverse}). Once
the convergence is reached, such an algorithm yields the desired result.
In the following sections, for specific cases we discuss solutions of
 {\Eqs}~(\ref{eq:KLipkin-reduced1})--(\ref{eq:KLipkin-inverse}).

\subsection{The Lipkin-Nogami method}
\label{sec:LN}

In a series of papers, Nogami and collaborators
\cite{(Nog64),(Nog64a),(Nog65),(Goo66),(Pra73a)} developed a
variant of the Lipkin method that replaces the calculation of norm overlaps
and Hamiltonian transition matrix elements by a calculation of several average values.
This replacement can be derived by first rewriting
{\Eqs}~(\ref{eq:redflat}), (\ref{eq:KLipkin-generic}), and
(\ref{eq:KLipkin-zero}) as
\begin{equation}
\langle\Phi |\left({\hat H}-\sum_{m=0}^M k_{m} {\hat K}_{m}\right)\hat{R}(g)|\Phi \rangle = 0 .
\label{eq:LN}
\end{equation}
Since for any $g$, and in particular for $g$ near 0, the group
operators $\hat{R}(g)$ are equal to the exponents of linear
combinations of symmetry generators ${\hat K}_{i}$, it follows from
{\Eq}~(\ref{eq:LN}) that for any $i$ we have,
\begin{equation}
\langle\Phi |\left({\hat H}-\sum_{m=0}^M k_{m} {\hat K}_{m}\right){\hat K}_{i}|\Phi \rangle = 0 .
\label{eq:LN2}
\end{equation}
Now the Lipkin parameters can be evaluated in an analogous way to how it was done
in the  Lipkin approach (\ref{eq:KLipkin-inverse}):
\begin{equation}
k_{m} = \sum_{i=0}^M B^{-1}_{mi} l_i ,
\label{eq:LN-inverse}
\end{equation}
but for
\begin{eqnarray}
B_{im} &=& \langle\Phi |{\hat K}_m{\hat K}_i|\Phi \rangle , \\
l_{i}  &=& \langle\Phi |{\hat H}  {\hat K}_i|\Phi \rangle .
\label{eq:LN-matrix}
\end{eqnarray}

Had the Lipkin method been exact, the Lipkin-Nogami expressions
would also be same, and would lead to exact results.
Otherwise, the Lipkin and Lipkin-Nogami methods may give
different results, and it is {\it a priori} difficult to
say which one is superior. Nevertheless, if the calculation
of transition matrix elements, and not only of the average values, is available, the Peierls-Yoccoz
method is certainly easier to implement. Indeed, in case the Hamiltonian
is a 2-body operator and the Lipkin operator is an $n$-body operator,
the Peierls-Yoccoz method requires calculating transition matrix elements of these 2-body and $n$-body
operators only, whereas the Lipkin-Nogami calls for calculating
averages of $n$+2-body and $2n$-body operators. In spite of that,
at second order ($n=2$) and for the particle-number
projection, the Lipkin-Nogami method was applied quite widely,
see {\Sec}~\ref{sec:Second-order}.

\subsection{The Kamlah method}
\label{sec:Kamlah}

The principal idea of the Kamlah expansion \cite{Kamlah1968} is
that the Hamiltonian transition matrix elements can by efficiently expanded into
a series of derivative operators ${\cal K}_{m}$ acting on the norm overlaps:
\begin{equation}
\langle\Phi |{\hat H}\hat{R}(g)|\Phi \rangle =
\sum_{m=0}^M k_{m} {\cal K}_{m}\langle\Phi |\hat{R}(g)|\Phi \rangle.
\label{eq:Kamlah}
\end{equation}
This expansion is supposed to work best in the limit of strong symmetry
breaking, for example, at large deformations. In this limit, the norm overlaps and
Hamiltonian transition matrix elements are both strongly peaked near
$g=0$, and therefore, the expansion of the former in a series of
derivatives of the latter may have a chance to converge rapidly.

Since for every continuous group, polynomials of symmetry generators
${\hat K}_m$ can always be represented by derivatives ${\cal K}_{m}$
with respect to the group parameters,
\begin{equation}
{\hat K}_m\hat{R}(g) \equiv {\cal K}_{m}\hat{R}(g),
\label{eq:differ}
\end{equation}
Kamlah expansion (\ref{eq:Kamlah}) is strictly equivalent to the
Lipkin flattening condition (\ref{eq:redflat}) applied for the Lipkin
operator of {\Eqs}~(\ref{eq:KLipkin-generic}) and
(\ref{eq:KLipkin-zero}). Since, in addition, Kamlah proposes to
determine coefficients $k_{m}$ by evaluating derivatives at $g=0$,
his method gives equations for $k_m$ that are strictly equivalent to
the Lipkin-Nogami method (\ref{eq:LN2}).

There are, nevertheless, two important differences. First, the Kamlah
proposal involves variation of the projected energy (\ref{eq:magic})
``as it is'', i.e., a variation over symmetry-breaking states
$|\Phi\rangle$ should also involve variation of $k_m$, see discussion
in {\Sec}~\ref{sec:Second-order}. Second, the
Kamlah expansion may contain terms that are not invariants of the
symmetry group, and therefore, they are not really within the realm of
the Lipkin method. This latter property mostly relates to the
so-called pushing and cranking terms discussed in detail in
{\Sec}~\ref{sec:First-order}.

\subsection{Applicability and Applications}
\label{sec:Applications}
The Lipkin, Kamlah, and Lipkin-Nogami methods, discussed in the previous
sections, all rely on polynomial expansions of collective
spectra or reduced kernels. This principal assumption creates two main
limitations of these approaches. First, obviously, the expansions
have to be carried out up to a sufficiently high order, see
{\Sec}~\ref{sec:Higher-order}. And second, and most importantly,
these methods cannot really be applied to spectra that have a
non-analytical dependence on the quantum numbers. Unfortunately, the latter
situation occurs in two physically meaningful cases, namely, when
particle-numbers are restored in (semi)magic nuclei
\cite{(Dob93a),(Wan14c)} and when the angular-momentum is restored in
weakly-deformed systems \cite{(Gao15b)}.

\subsubsection{First-order terms}
\label{sec:First-order}

\noindent{\it The Fermi energy:}\newline
The simplest application of the  Lipkin, Kamlah, or Lipkin-Nogami methods
concerns the approximated restoration of the particle-number symmetry
up to the first order in the particle number (\ref{eq:KLipkin-numb}), i.e.,
\begin{equation}
{\hat K}  = k_{1}({\hat N}-N_0) ,
\label{eq:KLipkin-numb1}
\end{equation}
where we can treat numbers of protons and neutrons separately.
Then, the Lipkin-Nogami equations (\ref{eq:LN-inverse}) and (\ref{eq:LN-matrix}) give
\begin{eqnarray}
\hspace*{-5mm}
k_0 &=& \frac{\langle\hat{H}              \rangle\langle({\hat N}-N_0)^2\rangle
             -\langle\hat{H}({\hat N}-N_0)\rangle\langle {\hat N}-N_0   \rangle}
             {\langle({\hat N}-N_0)^2\rangle
             -\langle {\hat N}-N_0   \rangle^2} ,
\label{eq:LN-first0} \\
\hspace*{-5mm}
k_1 &=& \frac{\langle\hat{H}({\hat N}-N_0)\rangle
             -\langle\hat{H}              \rangle\langle({\hat N}-N_0)\rangle}
             {\langle({\hat N}-N_0)^2\rangle
             -\langle {\hat N}-N_0   \rangle^2} ,
\label{eq:LN-first1}
\end{eqnarray}
where brackets $\langle\rangle$ denote average values calculated
for the particle-number-symmetry-breaking state $|\Phi\rangle$.
The variation-after-projection equation, which is derived from (\ref{eq:magic}), now reads
\begin{equation}
\delta_\Phi E^{N_0} = 0  \label{eq:magic-numb1}
\end{equation}
for
\begin{equation}
E^{N_0} = \langle\Phi| {\hat H}-k_{1}({\hat N}-N_0)|\Phi \rangle.  \label{eq:magic-numb2}
\end{equation}
According to Lipkin's methodology, variation over $|\Phi\rangle$ has
to be carried out at constant Lipkin coefficient $k_1$, and according
to the Kamlah's methodology, expression for $k_1$
(\ref{eq:LN-first1}) should be inserted into (\ref{eq:magic-numb1})
and then varied.

In this sense, at first order, the Lipkin-Nogami and Kamlah
prescriptions lead to the same result. Moreover, the Lipkin
coefficient $k_{1}$ can now be reinterpreted as a Lagrange multiplier
$\lambda_1$, that is, as a Fermi energy, which has to be adjusted so
as to obtain the correct average particle number. Then, the
Lipkin-Nogami expressions (\ref{eq:LN-first0}) and
(\ref{eq:LN-first1}) simplify tremendously, and give
\begin{eqnarray}
k_0 &=&  \langle\hat{H}\rangle,
\label{eq:LN-first0a} \\
k_1 &=& \frac{\langle\hat{H}({\hat N}-N_0)\rangle}
             {\langle({\hat{N}}-N_0)^2\rangle} .
\label{eq:LN-first1a}
\end{eqnarray}
The expression for $k_0$ is thus compatible with
(\ref{eq:magic-numb2}) and that for $k_1$ stems from
(\ref{eq:magic-numb1}), provided $\delta_\Phi|\Phi\rangle=({\hat
N}-N_0)|\Phi\rangle$ is an allowed variation.

In a similar way, we can evaluate the Lipkin expressions
(\ref{eq:KLipkin-inverse}), which gives
\begin{eqnarray}
\hspace*{-5mm}
k_0 &=& \frac{n_1(\phi_1)h(0)-n_1(0)h(\phi_1)}
             {n_1(\phi_1)-n_1(0)} ,
\label{eq:KLipkin-first0} \\
\hspace*{-5mm}
k_1 &=& \frac{h(\phi_1)-h(0)}
             {n_1(\phi_1)-n_1(0)},
\label{eq:KLipkin-first1}
\end{eqnarray}
where
\begin{equation}
n_m(\phi_i) = \frac{\langle\Phi |({\hat N}-N_0)^m\exp(i\phi_i\hat{N})|\Phi \rangle }
                     {\langle\Phi |                \exp(i\phi_i\hat{N})|\Phi \rangle }  \label{eq:kernel-numb}
\end{equation}
are reduced kernels of the shifted particle-number operator, evaluated at gauge angle $\phi_i$,
and $h(\phi_i)=h_i$ are the analogous reduced kernels of the Hamiltonian (\ref{eq:KLipkin-reduced2}).
Again it is beneficial to carry out variation (\ref{eq:magic-numb1}) with the average
particle number kept correct, $\langle{\hat N}\rangle=N_0$, which gives $n_1(0)=0$. In this case,
expression (\ref{eq:KLipkin-first0}) reduces again to $k_0=\langle\hat{H}\rangle$
and expression (\ref{eq:KLipkin-first1}) stems from (\ref{eq:magic-numb1}), provided
the finite-difference derivatives are allowed as variations $\delta_\Phi|\Phi\rangle$.

We conclude, that the Lipkin, Lipkin-Nogami, and Kamlah
symmetry restoration at first-order are completely equivalent to using
Lagrange multipliers for adjusting average values of symmetry
generators.

\bigskip\noindent{\it The pushing model and Thouless-Valatin mass:}\newline
For the restoration of translational symmetry, at first order the Lipkin operator reads,
\begin{equation}
{\hat K}  = \sum_{n=x,y,z} k_{1n}   ({\hat P}_n-P_{n0})
                     = \bm{k}_1\cdot({\hat{\bm{P}}}-\bm{P}_{0}) ,
\label{eq:KLipkin-trans1}
\end{equation}
where (\ref{eq:KLipkin-trans}) is generalized to the case
of a nucleus moving with the average total momentum $\bm{P}_0$.
We note here that the lowest-order invariant of the translational
group is equal to ${\hat{\bm{P}}}^2$, and therefore, the flattening
of the spectrum requires using the second-order Lipkin operator,
see {\Sec}~\ref{sec:Second-order}. Therefore, the first-order model
(\ref{eq:KLipkin-trans1}) rather pertains to the Kamlah approach.

Since components of the momentum operator ${\hat{\bm{P}}}$ commute,
we can treat them independently. Then, following the derivations
presented for the particle number, we conclude that the restoration
of translational symmetry is, at first order, equivalent to performing
minimization of the total energy, constrained to the given momentum $\bm{P}_{0}$,
\begin{equation}
E^{\bm{P}_{0}} = \langle\Phi| {\hat H}-\bm{v}\cdot({\hat{\bm{P}}}-\bm{P}_{0})|\Phi \rangle,  \label{eq:magic-trans1}
\end{equation}
where the vector Lipkin coefficient $\bm{k}_1$ acquires interpretation of the Lagrange multiplier
$\bm{v}$, i.e., of the velocity of the system.

For the translational symmetry, variation of the total energy $E^{\bm{P}_{0}}$ constrained
to the momentum $\bm{P}_{0}$ is particularly simple. Indeed, suppose we found
the state $|\Phi_{\bm{P}_{0}=\bm{0}}\rangle$, which is at rest,
$\langle\Phi_{\bm{P}_{0}=\bm{0}}|\hat{\bm{P}}|\Phi_{\bm{P}_{0}=\bm{0}}\rangle=\bm{0}$,
and fulfills the variation-after-projection equation $\delta{}E^{\bm{P}_{0}=\bm{0}}=0$ for $\bm{v}=\bm{0}$.
Then, the Galilean invariance,
\begin{equation}
\left[\hat{H},\hat{\bm{R}}\right]=-\frac{i\hbar\hat{\bm{P}}}{Am},  \label{eq:Galilean}
\end{equation}
where $\hat{\bm{R}}$ is the center-of-mass coordinate and $Am$ is the total mass of the system,
allows us to boost state $|\Phi_{\bm{P}_{0}=\bm{0}}\rangle$ to momentum $\bm{P}_{0}$,
\begin{equation}
|\Phi_{\bm{P}_{0}}\rangle = \exp\{\tfrac{i}{\hbar}\bm{P}_{0}\cdot\hat{\bm{R}}\}|\Phi_{\bm{P}_{0}=\bm{0}}\rangle,  \label{eq:boost}
\end{equation}
so that
\begin{eqnarray}
\bm{P}_{0} &=& \langle\Phi_{\bm{P}_{0}}|\hat{\bm{P}}|\Phi_{\bm{P}_{0}}\rangle ,
\label{eq:boost1} \\
\bm{v} &=& \frac{\bm{P}_{0}}{Am} ,
\label{eq:boost2} \\
E^{\bm{P}_{0}} &=& E^{\bm{P}_{0}=\bm{0}} +  \frac{\bm{P}_{0}^2}{2Am} .
\label{eq:boost3}
\end{eqnarray}
We see that the restoration of translational symmetry at first order, that is,
the pushing model, correctly reproduces all classical-motion relations. In particular,
from the analog of the Lipkin-Nogami expression (\ref{eq:LN-first1a})
we obtain the velocity vector as,
\begin{equation}
\bm{v} = \frac{\langle\hat{H}(\hat{\bm{P}}-\bm{P}_0)\rangle}
              {\langle       (\hat{\bm{P}}-\bm{P}_0)^2\rangle} ,  \label{eq:LN-velocity}
\end{equation}
which gives the mass
\begin{equation}
{\cal{M}}^{-1} = \frac{|\bm{v}|}{|\bm{P}_{0}|} =
 \frac{|\langle\hat{H}(\hat{\bm{P}}-\bm{P}_0)\rangle|}
       {\langle(\hat{\bm{P}}-\bm{P}_0)^2\rangle|\langle\hat{\bm{P}}\rangle|} ,  \label{eq:Thouless-Valatin-mass}
\end{equation}
that, in the translational case, correctly reproduces the true mass of the system, ${\cal{M}}=Am$ \cite{(Tho62a)}.

\bigskip\noindent{\it The cranking model and Thouless-Valatin moment of inertia:}\newline
Restoration of the rotational symmetry at first order leads to the so-called
cranking model, which was introduced originally in a semiclassical time-dependent picture of
a system rotating with constant angular velocity $\bm\omega$ around a fixed axis parallel to the
angular momentum $\bm J$ \cite{Inglis1954_PR96_1059,Inglis1956_PR103-1786}.
This model was very successfully used in nuclear physics to describe
a multitude of physical phenomena related to collective rotation \cite{bomo98,ring2000}.
In this case, the Lipkin operator reads,
\begin{equation}
{\hat K}  = \sum_{n=x,y,z} k_{1n}   ({\hat J}_n-J_{n0})
                     = \bm{k}_1\cdot({\hat{\bm{J}}}-\bm{J}_{0})~~~.
\label{eq:KLipkin-rot1}
\end{equation}
Since components of the angular-momentum operator ${\hat{\bm{J}}}$ do
not commute, we cannot treat them independently.
{\Eq}~(\ref{eq:KLipkin-rot1}) has thus to be understood as
corresponding to a nucleus having a fixed projection
$J_{0}=|\bm{J}_{0}|$ of the angular momentum on a quantization axis
oriented along the Lipkin coefficient vector, $\bm{k}_1=\bm{\omega}$.
In systems with approximate axial symmetry along the $z$-axis,
for instance, for the ground-state bands in well deformed even-even
nuclei, the rotational axis is perpendicular to the symmetry axis and usually
chosen along the $x$-axis. In this case, we have $J_{0}=\langle J_x\rangle =\sqrt{I(I+1)}$ with
integer values of $I$ (for odd systems, see \citeasnoun{Ring1974_NPA225-141}).
Different directions of $\bm{k}_1$ then mean
a freedom of choosing an arbitrary direction of the quantization axis.
This defines the so-called tilted-axis cranking model
\cite{(Ker81b),(Fra93b),(Fra01b),(Shi13)}, where the vector of the average
angular momentum $\langle{\hat{\bm{J}}}\rangle$ is arbitrarily
oriented with respect to the principal axes of the mass distribution
of the rotational-symmetry-breaking state.

Following the derivations
presented for the momentum operator, we conclude that the restoration
of rotational symmetry is, at first order, equivalent to performing
minimization of the total energy,
constrained to the given projection of the angular momentum $J_{0}$,
\begin{equation}
E^{\bm{J}_{0}} = \langle\Phi| {\hat H}-\bm{\omega}\cdot({\hat{\bm{J}}}-\bm{J}_{0})|\Phi \rangle,  \label{eq:magic-rot1}
\end{equation}
where the vector Lipkin coefficient $\bm{k}_1$ acquires interpretation of the Lagrange multiplier
$\bm{\omega}$, that is, of the angular velocity of the system.

The principal difference between translational and
rotational symmetry is the fact that for rotations there is no
analogue of the Galilean invariance (\ref{eq:Galilean}), and one cannot
simply boost a non-rotating state to higher rotational frequencies
without changing its structure. Indeed, with increasing rotational
frequency, the quantum analogues of the classical Coriolis and
centrifugal forces set in, and modify the state. Therefore, a constrained
minimization of the total energy has now to be explicitly performed.

Recall that the average momentum is exactly proportional to
the translational velocity, {\Eq}~(\ref{eq:boost2}), with a constant proportionality factor (mass).
Although the average angular momentum $\langle\hat{\bm{J}}\rangle$ has to
be parallel to the angular frequency $\bm{\omega}$ \cite{(Ker81b)}, the proportionality constant
(moment of inertia) can vary along the rotational
band. Therefore, we define two important {\em local}, that
is, frequency-dependent characteristics of the band, which are called
the first ${\cal{J}}^{(1)}$ (kinematic) and the second ${\cal{J}}^{(2)}$
(dynamic) moments of inertia,
\begin{equation}
{\cal{J}}^{(1)}(\omega) = \frac{        |\langle\hat{\bm{J}}\rangle(\omega)|}{        \omega} \quad\mbox{and}\quad
{\cal{J}}^{(2)}(\omega) = \frac{{\rm{d}}|\langle\hat{\bm{J}}\rangle(\omega)|}{{\rm{d}}\omega} ,
  \label{eq:MoI}
\end{equation}
respectively.

In parallel with {\Eq}~(\ref{eq:Thouless-Valatin-mass}),
the Lipkin expression allows us to determine the kinematic moment of inertia,
\begin{equation}
{\cal{J}}^{(1)} = \frac{\langle       (\hat{\bm{J}}-\bm{J}_0)^2\rangle|\langle\hat{\bm{J}}\rangle|}
              {|\langle\hat{H}(\hat{\bm{J}}-\bm{J}_0)\rangle|} .  \label{eq:Thouless-Valatin-MoI}
\end{equation}
The dynamic moment of inertia is identical to this value only at $\omega=0$. For all other values of $\omega$,
it corresponds to the Thouless-Valatin moment of inertia ${\cal{J}}^{(2)}={\cal{J}_{\text{TV}}}$ that can be derived in linear response theory \cite{(Tho62a)}.

\bigskip\noindent{\it Isocranking:}\newline
To perform the approximate restoration of the isospin symmetry, Frauendorf and
Sheikh \cite{Fra99} introduced the
concept of cranking in isospin with a focus on the isovector
proton-neutron pair field without Coulomb interaction. In this
isospin-symmetry conserving framework, the direction of the
isocranking axis in the isospin space is not relevant. For a specific
choice of this direction, the isocranking about the isospin $z$-axis
corresponds to the standard use of the neutron and proton Fermi
energies that are not equal to one another. Satu{\l}a and Wyss
\cite{(Sat01a),(Sat01b),(Glo04a)} used the same terminology of
{\em isocranking} in their isocranking model with Coulomb interaction
included, where dynamical consequences of the isocranking about the axis tilted
with respect to the isospin $z$-axis became relevant, and where they could investigate
both isovector and isoscalar pair fields.

In the
isospin context, the Lipkin operator reads as
\begin{equation}
{\hat K}  = \sum_{n=x,y,z} k_{1n}   ({\hat T}_n-T_{n0})
                     = \vec{k}_1\circ({\hat{\vec{T}}}-\vec{T}_{0}) ,
\label{eq:KLipkin-iso1}
\end{equation}
where arrows denote vectors in the isospace (isovectors) and symbol
"$\circ$" denotes their scalar product.
Since components of the isospin operator ${\hat{\vec{T}}}$ do not
commute, we cannot treat them independently. Thus
{\Eq}~(\ref{eq:KLipkin-iso1}) has to be understood as
corresponding to a nucleus having a fixed projection
$T_{0}=|\vec{T}_{0}|$ of the isospin on the isoquantization axis oriented
along the isovector Lipkin coefficient $\vec{k}_1$. Moreover, in even
(odd) systems, projections $T_{0}$ can only equal to integer
(half-integer) numbers. Different directions of $\vec{k}_1$ then mean
a freedom of choosing an arbitrary direction of the isoquantization axis.
The Lipkin coefficient $\vec{k}_1$ is interpreted as the isovector
Fermi energy $\vec{\lambda}$~\cite{(Sat13e),(She14b)}, which fixes the average values of
components of the isospin $\langle{\hat{\vec{T}}}\rangle$.

\begin{figure}
\centering\includegraphics[width=0.40\columnwidth]{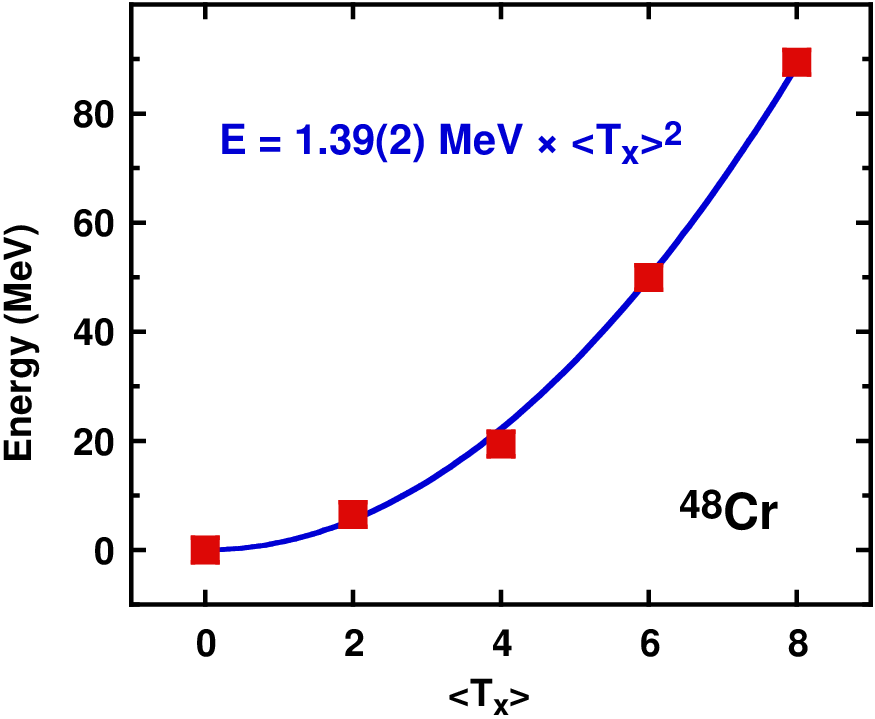}
\caption{\label{Sat13d}Excitation energies of $^{48}$Cr isocranked
to $\langle{}T_x\rangle=0$, 2, 4, 6, and 8 (squares) \protect\cite{(Sat13e)}.
The solid line represents the parabolic fit.}
\end{figure}

We note that the standard definition of the isospin implies that its
$z$ component is equal to half of the neutron excess,
$T_z=\tfrac{1}{2}(N-Z)$. Therefore, the $z$ component of the
isovector Fermi energy $\lambda_z$ along with the standard isoscalar
Fermi energy $\lambda$ simply fix the neutron $N$ and proton $Z$
numbers. When the isocranking axis is tilted away from the $z$
direction, one must use the formalism where proton and neutron
components of single-particle states are
mixed~\cite{(Sat13e),(She14b)}. Such a situation occurs when the
isospin-symmetry-breaking terms are added to the nuclear
Hamiltonian~\cite{(Bac18a)}. Figure~\ref{Sat13d} shows energies of
states in $^{48}$Cr isocranked to $\langle{}T_x\rangle=0$, 2, 4, 6,
and 8, while keeping $\langle{}T_z\rangle=0$ \cite{(Sat13e)}. We see
that one obtains a perfectly rigid isorotational band (a sequence of
states in $^{48}$Cr with increasing isospin $T$). The obtained
Thouless-Valatin moment of inertia then corresponds to the symmetry
energy coefficient $a_I$ in the symmetry energy,
$E_I(N,Z)=\tfrac{1}{2}a_I(N-Z)/A$, i.e., to
$a_I=2A \times 1.39(2)/4=33.4(5)$\,MeV.

\subsubsection{Second-order terms}
\label{sec:Second-order}

Although at first order the Lipkin or Kamlah approach provides correct
understanding of the collective effects, including the proper
determination of the collective mass, it does not really fulfill
Lipkin's requirement of flattening the spectrum of projected
states. Indeed, already from the example of the translational motion,
we see that the main component of this dependence may rather be
quadratic then linear. In this section, we thus examine the Lipkin
operators expanded up to second-order terms in the symmetry generators.
This allows us to model the spectra in terms of the quadratic Casimir
operators of the corresponding symmetry groups.

\bigskip\noindent{\it Particle-number restoration:}\newline
The main focus of the Lipkin-Nogami method \cite{(Nog64),(Nog64a),(Nog65),(Goo66),(Pra73a)} was up to now
on the approximate restoration of the particle-number symmetry, whereby the Lipkin operator
is postulated as,
\begin{equation}
{\hat K}  = k_{1}({\hat N}-N_0) + k_{2}({\hat N}-N_0)^2.
\label{eq:KLipkin-numb2}
\end{equation}
Assuming again that $k_1$ is always adjusted so as to obtain the correct
particle number, $\langle{\hat N}-N_0\rangle=0$, the Lipkin (\ref{eq:KLipkin-inverse}) \cite{(Wan14c)}
and Lipkin-Nogami (\ref{eq:LN-inverse}) \cite{(Val96)} methods give, respectively,
\begin{equation}
\label{eq:lambda2-lipkin}
k_{2}=\frac{h(\phi_2)-k_{1}n_1(\phi_2)-h(0)}{n_2(\phi_2)-n_2(0)} \, ,
\end{equation}
and
\begin{equation}
\label{eq:lambda2-LN}
\lambda_{2}=\frac{\langle\hat{H}\Delta\hat{N}^2\rangle\langle\Delta\hat{N}^2\rangle
                 -\langle\hat{H}\Delta\hat{N}  \rangle\langle\Delta\hat{N}^3\rangle
                 -\langle\hat{H}               \rangle\langle\Delta\hat{N}^2\rangle^2}
                 {\langle       \Delta\hat{N}^4\rangle\langle\Delta\hat{N}^2\rangle
                 -\langle       \Delta\hat{N}^3\rangle^2
                 -\langle       \Delta\hat{N}^2\rangle^3} \, .
\end{equation}
Here $\Delta\hat{N}\equiv\hat{N}-N_0$ is the
shifted particle-number operator, $h(\phi_2)$ and $n_i(\phi_2)$ are the Hamiltonian
(\ref{eq:KLipkin-reduced2}) and shifted particle-number
(\ref{eq:kernel-numb}) kernels, and we used the traditional
notation of $\lambda_2\equiv{}k_2$ for the second-order Lipkin-Nogami
coefficient. For an HFB vacuum $|0\rangle$, an alternative and equivalent expression for $\lambda_2$
was derived in \citeasnoun{(San78)} as
\begin{equation}
\label{eq:lambda2-LN2}
\lambda_{2}=\frac{\sum_4\langle0|\hat{H}  |4\rangle\langle4|\hat{N}^2|0\rangle}
                 {\sum_4\langle0|\hat{N}^2|4\rangle\langle4|\hat{N}^2|0\rangle} \, ,
\end{equation}
where $|4\rangle\equiv\alpha^+_\mu\alpha^+_\nu\alpha^+_{\mu^{\prime}}\alpha^+_{\nu^{\prime}}|0\rangle$
stands for all four-quasiparticle states.
After evaluating
all required matrix elements, one obtains \cite{(Flo97a),(Sto03a)}
\begin{equation}\label{ll2}
\lambda_{2}=\frac {4{\rm Tr} \Gamma^{\prime} \rho(1-\rho) + 4{\rm
Tr}\Delta^{\prime} (1-\rho)\kappa} {8\left[{\rm Tr}\rho (1-\rho
)\right]^{2}-16{\rm Tr}\rho^{2}(1-\rho)^{2}} ~,
\end{equation}
where the potentials
\begin{eqnarray}\label{eq42a}
\Gamma^{\prime}_{\mu \mu^{\prime}} &=& \sum_{\nu \nu^{\prime}}V_{\mu \nu
\mu^{\prime} \nu^{\prime}}(\rho(1-\rho))_{\nu^{\prime} \nu}, \\ \label{eq42b}
\Delta^{\prime}_{\mu \nu} &=&\tfrac{1}{2}\sum_{\mu^{\prime}
\nu^{\prime}}V_{\mu \nu \mu^{\prime} \nu^{\prime}}(\rho
\kappa)_{\mu^{\prime} \nu^{\prime}},
\end{eqnarray}
can be calculated in
full analogy to $\Gamma$ and $\Delta$ by replacing the $\rho$ and
$\kappa$ in terms of which they are defined by $\rho(1-\rho)$ and
$\rho\kappa$, respectively. In the case of the seniority pairing interaction
with strength $G$,
{\Eq}~(\ref{ll2})  simplifies to \cite{(Pra73a)}:
\begin{equation}\label{ll3}
\lambda_{2}=\frac{G}{4}\frac{\textstyle{\sum_{k>0}}
(u_{k}v_{k}^{3})\textstyle{\sum_{k>0}}(u_{k}^{3}v_{k})-
\textstyle{\sum_{k>0}}(u_{k}v_{k})^{4}}
{(\textstyle{\sum_{k>0}}u_{k}^{2}v_{k}^{2})^2-
\textstyle{\sum_{k>0}}(u_{k}v_{k})^{4}},
\end{equation}
where $k>0$ denotes the summation over one state of each canonical (or Kramers-degenerate) pair
of single-particle states.

Evaluation of the Lipkin coefficient $k_2$, as in {\Eq}~(\ref{eq:lambda2-lipkin}),
is fairly simple, see {\Sec}~\ref{sec:Higher-order}, but it was
implemented only in \citeasnoun{(Wan14c)}. A rigorous evaluation of the
Lipkin-Nogami coefficient $\lambda_2$, {\Eq}~(\ref{eq:lambda2-LN}) or
(\ref{ll2}), is for realistic Hamiltonians rather cumbersome, so it
was rarely implemented in full, see, e.g., \citeasnoun{(Val96)}. A
practical workaround, which was used quite often, see, e.g.,
\citeasnoun{(Sto03a),(Sto07e),(Kor10c)} was to use the seniority-pairing
expression {\Eq}~(\ref{ll3}) with the effective strength
$G\equiv{}G_{\text{eff}} = -\frac{\bar{\Delta}^2}{E_{\text{pair}}}$,
determined from the pairing energy $E_{\text{pair}}$ and the average
pairing gap $\bar{\Delta}$.

\begin{figure}
\centering\includegraphics[width=0.40\columnwidth]{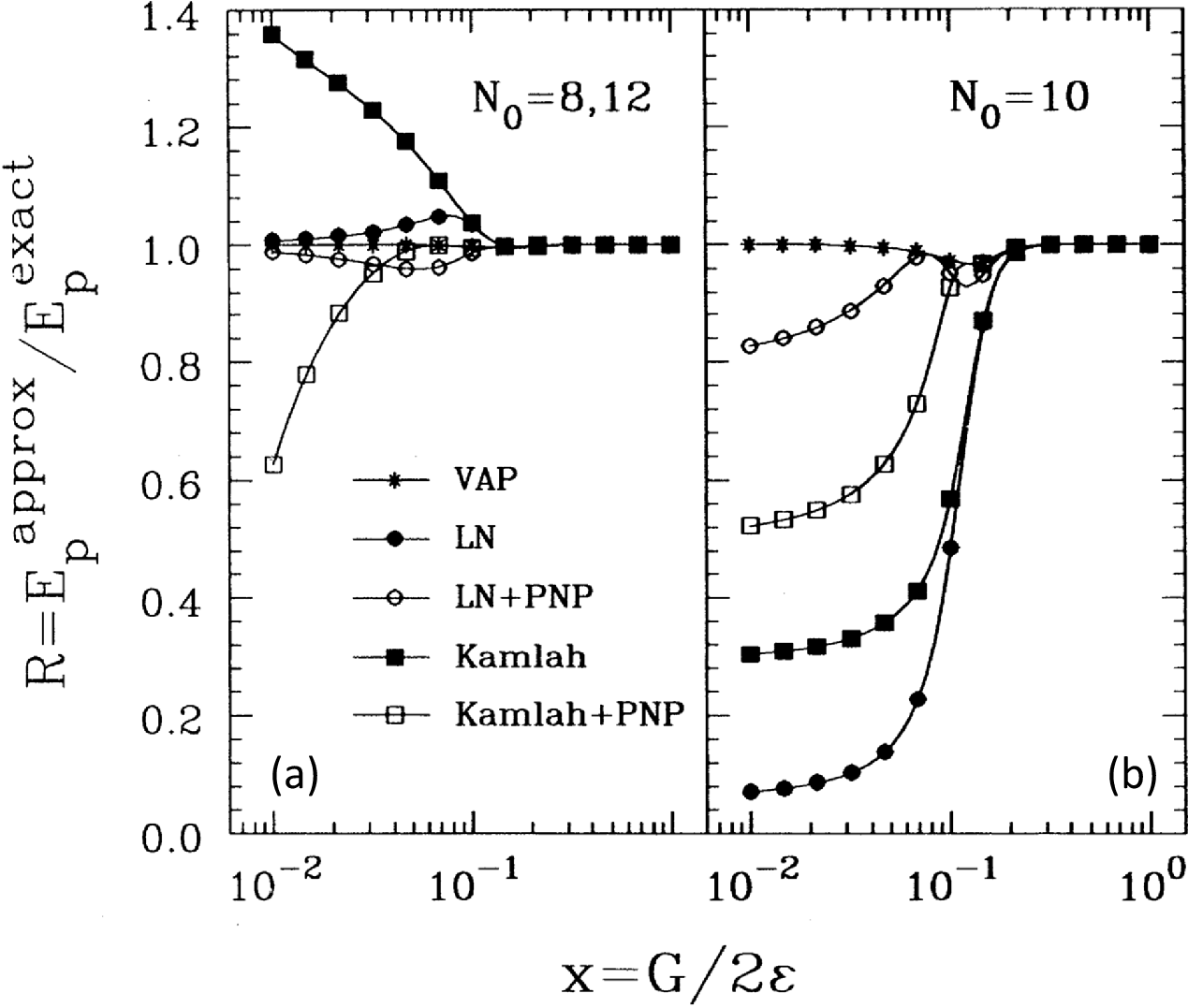}
\caption{\label{Dob93a}Pairing energies in the two-level model obtained in
\protect\citeasnoun{(Zhe92)} within the variation after projection (VAP)
(asterisks), Kamlah (full squares), and Lipkin-Nogami (full circles)
approximations, relative to the exact values and plotted for
$N_0$=$\Omega$=10 (right), and $N_0$=$\Omega$$\pm$2 (left).  The
energies resulting from the exact particle-number projection of Kamlah (open squares)
and Lipkin-Nogami (open circles) states are shown for
comparison.  The critical value of $x$ is $x_c$=1/9.
Figure reprinted with permission from \protect\citeasnoun{(Dob93a)}.
Copyrighted by the American Physical Society.}
\end{figure}

Figure~\ref{Dob93a} shows comparison of the Lipkin-Nogami and Kamlah
approaches within a simple solvable two-level model
\cite{(Zhe92),(Dob93a)}. In the strong-pairing regime, both
approaches give excellent reproduction of the exact results, however,
for the half-filled shell [{\Fig}~\ref{Dob93a}(b)], at or below the
critical pairing strength both fail. This is so because the kink in
the dependence of the exact energies on the particle number, which is
a characteristic feature of a shell gap, cannot be reproduced by the
quadratic \cite{(Dob93a)} or higher-order \cite{(Wan14c)} form of the
Lipkin operator. Away from the shell gap [{\Fig}~\ref{Dob93a}(a)], the
Lipkin-Nogami approach works well for all pairing strengths, but the
Kamlah approach fails below the critical pairing strength. This
latter feature can be attributed to the fact that by exploiting
errors of the approximation, the exact minimization over the Lipkin
coefficient $\lambda_2$ brings the approximate projected energy below
the exact result.

\bigskip\noindent{\it Peierls-Yoccoz mass:}\newline
For the translational symmetry, at second order, the general form of the
Lipkin operator reads
\begin{equation}
{\hat K}  = \bm{k}_1\cdot({\hat{\bm{P}}}-\bm{P}_{0})
          + \sum_{n=x,y,z} k_{2n} ({\hat P}_n-P_{n0})^2 .
\label{eq:KLipkin-trans2}
\end{equation}
As already discussed in {\Sec}~\ref{sec:First-order}, the
first-order terms define the pushing model, wherein the nucleus moves
in space with average momentum $\bm{P}_{0}$ and velocity
$\bm{v}\equiv\bm{k}_1$. This motion leads to an increase of energy
that is quadratic in the momentum, and the corresponding
proportionality coefficient is called the Thouless-Valatin mass
(\ref{eq:Thouless-Valatin-mass}), which is correctly equal to the
translational mass $mA$.

The role of the Lipkin coefficients $k_{2n}$ is different -- they are
meant to flatten the spectrum of energies projected on momentum
eigenstates in the direction of $n=x$, $y$, or $z$. Their values thus
characterize the momentum distributions within the
translational-symmetry-breaking state, and have nothing to do with
the physical motion of the system. This is particularly evident when
we consider the state at rest, $|\Phi_{\bm{P}_{0}=\bm{0}}\rangle$, which can be
obtained by simply conserving the time-reversal symmetry. This
state does not move, so the Lipkin coefficients $k_{2n}$ cannot
describe inertia, which is the reaction of the system under boost.

Coefficients $k_{2n}$ calculated for the $n=x$, $y$,
or $z$ directions can differ from one another \cite{(Gao15b)}. Indeed,
along the longer or shorter principal axis of the mass distribution,
the momentum distribution is narrower or wider, respectively, and the
corresponding projected energy components can thus differ from one
another.

Nevertheless, historically, quantities ${\cal M}_{\text{PYn}}=\tfrac{1}{2}k_{2n}^{-1}$,
corresponding to translational Lipkin coefficients $k_{2n}$,
are called Peierls-Yoccoz \cite{(Pei57a)} or Yoccoz \cite{ring2000} masses.
Assuming that those corresponding to the $n=x$, $y$,
or $z$ directions are independent from one another,
they can be calculated in the Lipkin or Lipkin-Nogami approach, respectively,
as:
\begin{equation}
\label{eq:k2-trans-lipkin}
k_{2n}\equiv(2{\cal M}_{\text{PYn}})^{-1}=\frac{h(\phi_{2n})-k_{1n}p_{1n}(\phi_{2n})-h(0)}{p_{2n}(\phi_{2n})-n_2(0)} \,
\end{equation}
or
\begin{eqnarray}
\label{eq:k2-trans-LN}
k_{2n}&\equiv&(2{\cal M}_{\text{PYn}})^{-1} \nonumber \\
  &=&\frac{\langle\hat{H}\Delta\hat{P}_n^2\rangle\langle\Delta\hat{P}_n^2\rangle
          -\langle\hat{H}\Delta\hat{P}_n  \rangle\langle\Delta\hat{P}_n^3\rangle
          -\langle\hat{H}                 \rangle\langle\Delta\hat{P}_n^2\rangle^2}
          {\langle       \Delta\hat{P}_n^4\rangle\langle\Delta\hat{P}_n^2\rangle
          -\langle       \Delta\hat{P}_n^3\rangle^2
          -\langle       \Delta\hat{P}_n^2\rangle^3} \, ,  \nonumber \\
\end{eqnarray}
cf.~{\Eqs}~(\ref{eq:lambda2-lipkin}) or~(\ref{eq:lambda2-LN}).
Here $\Delta\hat{P}_n={\hat P}_n-P_{n0}$ are shifted momentum operators for $n=x$, $y$, or $z$
and $p_{in}(\phi_{2n})$ are their reduced kernels calculated at distances $\phi_{2n}$.

For conserved time reversal,
average values of all odd powers of momentum are equal to zero, and thus
{\Eq}~(\ref{eq:k2-trans-LN}) reduces to
\begin{eqnarray}
\label{eq:k2-trans-LN2}
k_{2n}&\equiv&(2{\cal M}_{\text{PYn}})^{-1}
   = \frac{\langle\hat{H}\hat{P}_n^2\rangle
          -\langle\hat{H}           \rangle\langle\hat{P}_n^2\rangle}
          {\langle       \hat{P}_n^4\rangle
          -\langle       \hat{P}_n^2\rangle^2} \, .
\end{eqnarray}
In addition, if we consider a spherical nucleus, where the Lipkin
coefficients corresponding to three directions $n=x$, $y$,
or $z$ are equal, and the Lipkin operator
takes the form ${\hat K}=k_2\hat{\bm{P}}^2$, we then have
\begin{eqnarray}
\label{eq:k2-trans-LN3}
k_2&\equiv&(2{\cal M}_{\text{PY}})^{-1}
   = \frac{\langle\hat{H}\hat{\bm{P}}^2\rangle
          -\langle\hat{H}           \rangle\langle\hat{\bm{P}}^2\rangle}
          {\langle       \hat{\bm{P}}^4\rangle
          -\langle       \hat{\bm{P}}^2\rangle^2} \, .
\end{eqnarray}
We see that equality of $k_{2n}$ and $k_2$ requires independence
of the three directions,
$\langle\hat{P}_n^2\hat{P}_m^2\rangle=\langle\hat{P}_n^2\rangle\langle\hat{P}_m^2\rangle$
for $n\neq{}m$, which was assumed when deriving {\Eqs}~(\ref{eq:k2-trans-LN}) and (\ref{eq:k2-trans-LN2}).
Further, within the Gaussian Overlap Approximation \cite{ring2000}, we have
$\langle\hat{\bm{P}}^4\rangle=3\langle\hat{\bm{P}}^2\rangle^2$, and {\Eq}~(\ref{eq:k2-trans-LN3})
simplifies to
\begin{eqnarray}
\label{eq:k2-trans-LN4}
2k_2&\equiv&({\cal M}_{\text{PY}})^{-1}
   = \frac{\langle\hat{H}\hat{\bm{P}}^2\rangle
          -\langle\hat{H}           \rangle\langle\hat{\bm{P}}^2\rangle}
          {\langle       \hat{\bm{P}}^2\rangle^2} \, .
\end{eqnarray}

\begin{figure}
\centering\includegraphics[width=0.40\columnwidth]{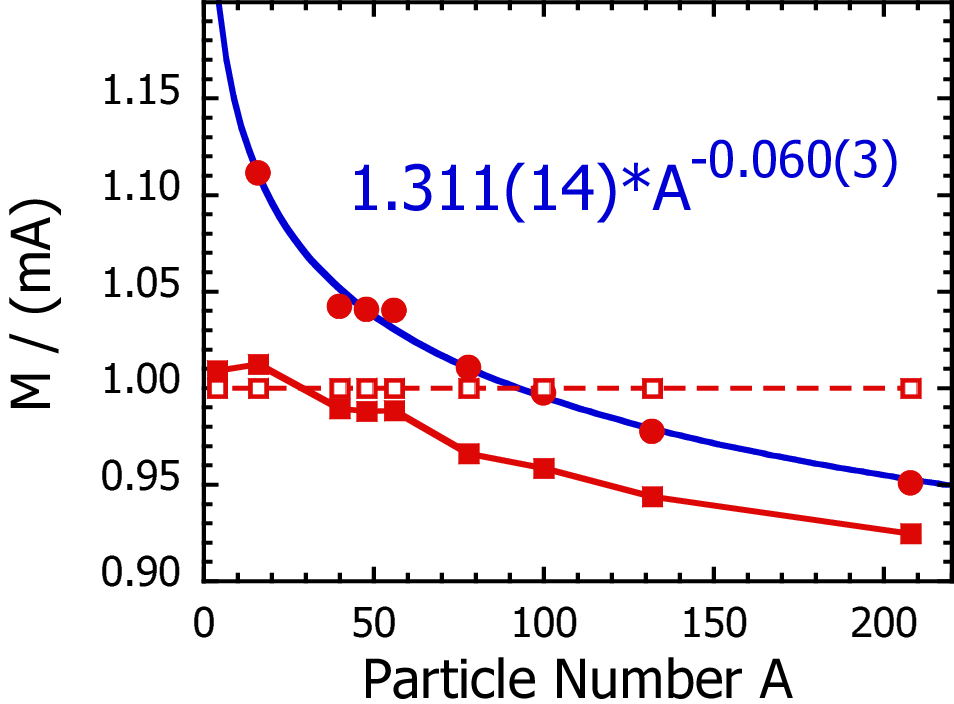}
\caption{\label{Dob09a}Exact masses ($M = mA$, open
squares) and the Peierls-Yoccoz masses $M={\cal M}_{\text{PY}}$, {\Eq}~(\ref{eq:k2-trans-LN3}),
calculated in doubly magic nuclei before (full circles) and after (full squares) including
the Lipkin operator.
Republished with permission of Institute of Physics, from \protect\citeasnoun{(Dob09a)};
permission conveyed through Copyright Clearance Center, Inc.}
\end{figure}

Figure~\ref{Dob09a} shows comparison of the Peierls-Yoccoz and
Thouless-Valatin (exact) masses calculated in doubly magic nuclei. We
see that the former are never equal to the latter, because they
represent different quantities. Indeed, the Peierls-Yoccoz masses
characterize the curvatures of energies projected from the
symmetry-breaking states at rest, whereas the Thouless-Valatin masses
characterize the increase of the energy when the symmetry-breaking
states are boosted to non-zero momenta. In addition, the figure shows
the Peierls-Yoccoz masses evaluated for energies minimized before and
after including the Lipkin operator. One clearly sees that the
self-consistent inclusion of the Lipkin correction does modify the
curvatures of projected energies. In practice, however, differences
between the Peierls-Yoccoz and Thouless-Valatin translational masses
do not exceed 10\% and vary smoothly with nuclear masses.
More discussion can be found in \citeasnoun{(Ben00g),(Dob09a)}.

Finally, let us mention here a popular way of implementing the so-called
center-of-mass correction in self-consistent calculations \cite{(Ben00g),(Dob09a)},
which amounts to combining the magic formula (\ref{eq:magic}) and the
second-order Lipkin operator (\ref{eq:KLipkin-trans2}) with the Thouless-Valatin
mass into an approximate expression for the projected energy,
\begin{equation}
E^{\bm{P}_{0}}\simeq\langle\Phi| {\hat H}|\Phi \rangle - \frac{\langle\Phi|\hat{\bm{P}}^2|\Phi \rangle}{2mA} .
\label{eq:KLipkin-trans2a}
\end{equation}
The Peierls-Yoccoz mass, which should appear in this formula is not evaluated but
approximated here by the true inertial mass of the system. Further approximations
of the average value $\langle\Phi|\hat{\bm{P}}^2|\Phi \rangle$ are also frequently
used  \cite{(Ben00g),(Dob09a)}.

\bigskip\noindent{\it Peierls-Yoccoz moment of inertia:}\newline
For the rotational symmetry, at second order, the
Lipkin operator reads
\begin{equation}
{\hat K}  = \bm{k}_1\cdot({\hat{\bm{J}}}-\bm{J}_{0})
          + \sum_{n=x,y,z} k_{2n} ({\hat J}_n^2-J_{n0}^2)~~~.
\label{eq:KLipkin-rot2}
\end{equation}
As already discussed in {\Sec}~\ref{sec:First-order}, the
first-order terms define the cranking model, wherein the nucleus
rotates in space with average angular momentum $\bm{J}_{0}$ and
frequency $\bm{\omega}\equiv\bm{k}_1$. This motion leads to an
increase of energy, which is (in deformed nuclei) approximately
quadratic in the angular momentum, and the
corresponding proportionality coefficient is called the
Thouless-Valatin moment of inertia (\ref{eq:Thouless-Valatin-MoI}).
On the other hand, the Lipkin coefficients $k_{2n}$ are meant to
flatten the spectrum of energies projected on angular-momentum
eigenstates. However, since the three components of the angular
momentum do not commute, the three first-order and three second-order
terms in the Lipkin operator (\ref{eq:KLipkin-rot2}) are not
independent from one another. Moreover, in axial nuclei, the
rotational symmetry is not fully broken, namely, the total angular
momenta are mixed, but the projections of the angular momentum on the
symmetry axis continue to be a good quantum numbers. In this case,
the second-order Lipkin operator in the form of
{\Eq}~(\ref{eq:KLipkin-axial}) is more appropriate.

Expressions for Peierls-Yoccoz moments of inertia can be obtained
in full analogy to those for Peierls-Yoccoz masses, {\Eqs}~(\ref{eq:k2-trans-LN})--(\ref{eq:k2-trans-LN4}).
In particular, for a one-dimensional rotation
of an axial nucleus about the $x$ axis perpendicular to the symmetry axis,
the analogue of {\Eq}~(\ref{eq:k2-trans-LN2}) reads
\begin{eqnarray}
\label{eq:k2-rot-LN2}
(2{\cal J}_{\text{PY}})^{-1}
   &=& \frac{\langle\hat{H}\hat{J}_x^2\rangle
          -\langle\hat{H}           \rangle\langle\hat{J}_x^2\rangle}
          {\langle       \hat{J}_x^4\rangle
          -\langle       \hat{J}_x^2\rangle^2} \, .
\end{eqnarray}

In analogy with approximate expressions used for translations
(\ref{eq:KLipkin-trans2a}), also the so-called rotational-motion
corrections \cite{(Fra69),(Ben03e)} are sometimes implemented along with various
approximations for the Peierls-Yoccoz moments of inertia.

\bigskip\noindent{\it Isospin:}\newline
Almost immediately after Nogami's work \cite{(Nog64)},
an analogous method was suggested in the
restoration of the isospin symmetry \cite{(Kis66),(Kis67a),(Gho75)},
whereupon the Lipkin operator,
\begin{equation}
{\hat K}  = \vec{k}_1\circ({\hat{\vec{T}}}-\vec{T}_{0})
          + k_{2} \left({\hat{\vec{T}}^2}-T_0(T_0+1)\right),
\label{eq:KLipkin-iso2}
\end{equation}
combines the isocranking term (\ref{eq:KLipkin-iso1}) with the second-order correction.
Such approach, however, was later not too often employed, because
the exact isospin restoration was early implemented \cite{(Cau80a),(Cau80b),(Cau82a)}
and is now efficiently used \cite{PhysRevLett.103.012502,PhysRevC.81.054310,(Sat16d)}.

\subsubsection{Higher-order terms}
\label{sec:Higher-order}

In the Lipkin approach or Kamlah expansion, higher-order terms were
studied for the particle-number restoration. \citeasnoun{(Rod05a)}
introduced the so-called reduced variation-after-projection
method, which aimed to improve second-order Lipkin-Nogami
or Kamlah results.
The Lipkin method up to sixth order was implemented in \citeasnoun{(Wan14c)}.
This work showed that away from semi-magic nuclei,
the Lipkin method converges already at 4th order, with the 6th order
corrections bringing almost no change. Near the semi-magic nuclei,
however, the non-analytical dependence of energy in function
of the particle number did not allow for obtaining well converged results.

\section{Projection methods in simple nuclear models and {\it ab initio} calculations}
\label{sec:simple}

The basic objective of the projection methods is to include the
many-body correlations beyond the mean-field
level, and in order to treat them accurately, the best is to perform them
in the $\bm{r}$-space \cite{(Bay84a)}. However, projection
in the $\bm{r}$-space leads to solution of non-local potential
problem, even with zero-range effective interaction. The non-local
potential problem is quite prohibitive to solve even with the modern
computing facilities. This has also to do with the fact that even
the bare three-dimensional  HFB problem is very difficult to solve
in $\bm{r}$-space. Most of the modern and efficient 3D methods in
$\bm{r}$-space are restricted the HF- or HF+BCS
calculations~\cite{(Rys19b)}, whereas full 3D-HFB
calculations are carried out in oscillator space~~\cite{(Egi95a)}.
Two-dimensional HFB equations were solved in $\bm{r}$-space using the
spline functions \cite{(Keg96a),(Pei08a)} and there were also
attempts to solve the three-dimensional HFB problem using the
multi-wavelet method \cite{(Pei14a)}. As evident from these studies,
the solution of the full HFB equations in the three-dimensional
$\bm{r}$-space requires, at present, a too high computational effort
to be applied in realistic applications.
For the above reason, most of the application of projection methods were
carried out by expanding the single-particle wave functions
in a finite basis. This leads to relatively simple matrix
operations. The basis states, normally chosen, are solutions of harmonic
oscillator or Woods-Saxon potentials. A very efficient method relies on
solving the HFB equations on the basis of eigenstates of the HF Hamiltonian
\cite{(Gal94a),(Ben03e)} that can be obtained in the $\bm{r}$-space.

For axial and triaxial mean-field calculations, the oscillator basis
has the advantage that these wave functions are separable in the
coordinates. This requires considerably less memory for the storage
of the basis states and, in addition, such a basis can be deformed.
Through a careful choice of the basis deformation parameters, this
allows us to reproduce, in a relatively small basis of roughly 20
major oscillator shells, the exact mean-field results of the system
with sufficient accuracy, i.e., with roughly 500 keV in the total
nuclear binding energies even for heavy nuclei \cite{(Dob02e)}. For
the nucleus $^{208}$Pb with an experimental binding energy of
1636.446 MeV this corresponds to an accuracy of 0.3~permille. We
also have to keep in mind, that the binding energy, being a very
sensitive difference between two large quantities, the repulsive
kinetic energy and the attractive potential energy, requires often
much more accuracy than other quantities that are mostly determined
by the properties of the valence shells, as for instance deformation
energies etc.

However, for the application of projection methods, the use of a deformed basis leads
to severe problems (see discussion in {\Sec}~\ref{EDF:notclosed}), because the basis violates
the corresponding symmetry from the beginning. By this reason, most of the applications of
angular-momentum projection were carried out in a spherical basis.

In most of the realistic nuclear models, for instance, in the DFT approach, the
projection methods were applied after variation and include correlations only partially
beyond the mean-field level. Projection calculations before variation become exceedingly difficult,
in particular, for the case of three-dimensional projection. Such studies were
performed with simpler model Hamiltonians using a few major oscillator shells around the
Fermi surface~\cite{McCullen1964_PR134-B515,(Law61),(Bar68_j)}.

The justification for using such a simpler approach, which considers only a restricted
set of active orbits around the Fermi surface (what is often referred to in the literature as
``the valence space"), is that we are mostly interested in properties associated to low energy excited
states.
Also, the pairing part of the effective interaction is dominated by single-particle states around the Fermi
surface. Even the correlations leading to deformation are dominated by contributions from a valence
space of a few major oscillator shells. It is known from several studies that
states that are far above or below the Fermi surface do not contribute to the correlations beyond the
mean-field level. This is clearly evident in the case of particle-number projection, where the states
far from the Fermi surface have occupations close to either one or zero and, therefore, do not contribute to
pairing correlations.

Such valence spaces are often too large for what is called ``full configuration-interaction"
calculations\footnote{The term refers to those calculations that consider as elements of the Fock state
all mean field states built by taking into account all possible distribution
of particles among the active orbits. The typical example being the interacting shell model \cite{(Cau05a)}.},
but still small
enough even for sophisticated projection techniques. A famous case is the Baranger-Kumar valence space~\cite{(Bar68_j)} containing the shells $N=3,4$ for protons and $N=4,5$ for neutrons and used for the description of the well
deformed Rare Earth region of the nuclear chart. Its configuration space contains 72 neutron and 50 proton levels.
This leads for mean-field calculations to matrices with dimension 72 and 50, which are easy to handle.
On the other side in a typical configuration-interaction calculation we have in the middle of the shells
roughly $5\times 10^{34}$ configurations.

Of course, the model interaction has to be adjusted carefully to the underlying configuration space.
If the interaction in such a valence space is properly chosen, much of the physics, in particular, the
interplay between collective degrees of freedom and single-particle degrees of freedom can be well
described in such a space, only global properties such as total binding energies, radii, or multipole moments
have to be
treated in the full space.

\subsection{The pairing-plus-quadrupole model} \label{sec:P+Q}

The applications are further simplified by the use of separable interactions.
From the configuration-interaction calculations \cite{(Duf96a),(Cau05a)}, one knows that an effective interaction in such a restricted space can be represented
as a sum of terms separable in the particle-hole and particle-particle channel, i.e.,
\begin{equation}
\hat{V}=\frac{1}{2}\sum_{\lambda \mu}\chi_{\lambda}\hat{Q}_{\lambda \mu}^{\dag}\hat
{Q}_{\lambda \mu}^{{}}+\sum_{\lambda \mu}G_{\lambda}\hat{P}_{\lambda \mu}^{\dag}\hat
{P}_{\lambda \mu}^{{}},
\label{VQQ}
\end{equation}
where,
\begin{equation}
~~\hat{Q}_{\lambda \mu}^{{}}=\sum_{nn^{\prime}} \langle n| r^{\lambda}{Y_{\lambda \mu}} |n^{\prime}\rangle a_{n^{{}}%
}^{\dag}a_{n^{\prime}},~~~~~\hat{P}_{\lambda \mu}%
=\frac{1}{2}\sum_{nn^{\prime}}\langle n| r^{\lambda}{Y_{\lambda \mu}} |n^{\prime}\rangle a_{n^{{}}}^{{}}a_{\bar{n}^{\prime}}^{{}%
},
\end{equation}
that is, $\hat{Q}_{\lambda \mu}$ and $\hat{P}_{\lambda \mu}$ are multipole operators
of multipolarity $\lambda$ in the particle-hole and particle-particle channel, respectively, with
$\chi_{\lambda}$ and $G_{\lambda}$ being the corresponding coupling constants.
The Hamiltonian of the model is
\begin{equation}
\hat{H}=\hat{H}_{0}+\hat{V},%
\end{equation}
where the doubly magic spherical core enters the calculations in terms
of the single-particle energies $\varepsilon_{n}$ in the operator
\begin{equation}
\hat{H}_{0}=\sum_{n}\varepsilon_{n}a_{n}^{\dag}a_{n}^{{}}~~~~~.%
\label{H0}
\end{equation}
The parameter of such models are the single-particle energies $\varepsilon
_{n}$ and the coupling constants $\chi_{\lambda}$ and $G_{\lambda}$.

In the configuration-interaction calculations with interactions derived from {\it ab initio} calculations
\cite{(Duf96a)}, it is found while deriving separable representations of these effective
forces that the quadrupole part with $\lambda=2$ plays the essential role in
the particle-hole channel and the monopole part with $\lambda=0$ dominates the particle-particle channel.
These two parts do not depend much on the underlying bare nucleon-nucleon force.
Therefore, it is easy to understand that the pairing plus quadrupole model, introduced
in the sixties \cite{(Bar68_j)}, was pretty successful to describe the bulk of the
important long-range correlations in nuclei in a very efficient way. Of course, it
can be improved without great difficulties through the inclusion of additional separable
terms, such as octupole ($\hat{Q}_{3}$) and hexadecapole   ($\hat{Q}_{4}$) operators or the
quadrupole pairing term $\hat{P}_{2}^{{}}$. Some of the advantages of
such models are:

\begin{enumerate}
\item[(i)] one works in a spherical basis which preserves, apart from
translational invariance, all the symmetries;

\item[(ii)] the basis and the corresponding single-particle matrices are
relatively small and, therefore, projection techniques can be applied easily;

\item[(iii)] if the parameters are carefully adjusted, one obtains excellent
results which can be compared with the experimental data and the basic physics can
be understood in this way.
\end{enumerate}

Of course, such model have restrictions as quantities influenced by
physics outside of the valence shells cannot be described. These
include, most importantly, the total binding energy of the nucleus
and also the phenomena driven by so-called intruder states coming
from the excluded higher shells or coming from the core at large
deformations. The fission process and superdeformed~\cite{Sheline1972_PLB41-115}
configurations belong to this category. There are also limitations
for phenomena driven by special forces not considered in the model
Hamiltonian, for instance, such parts of the tensor
force~\cite{(Ots20)} which are not included in the phenomenologically
adjusted single-particle energies.

\begin{figure}[t]
\centering
\includegraphics[width=3.0in]{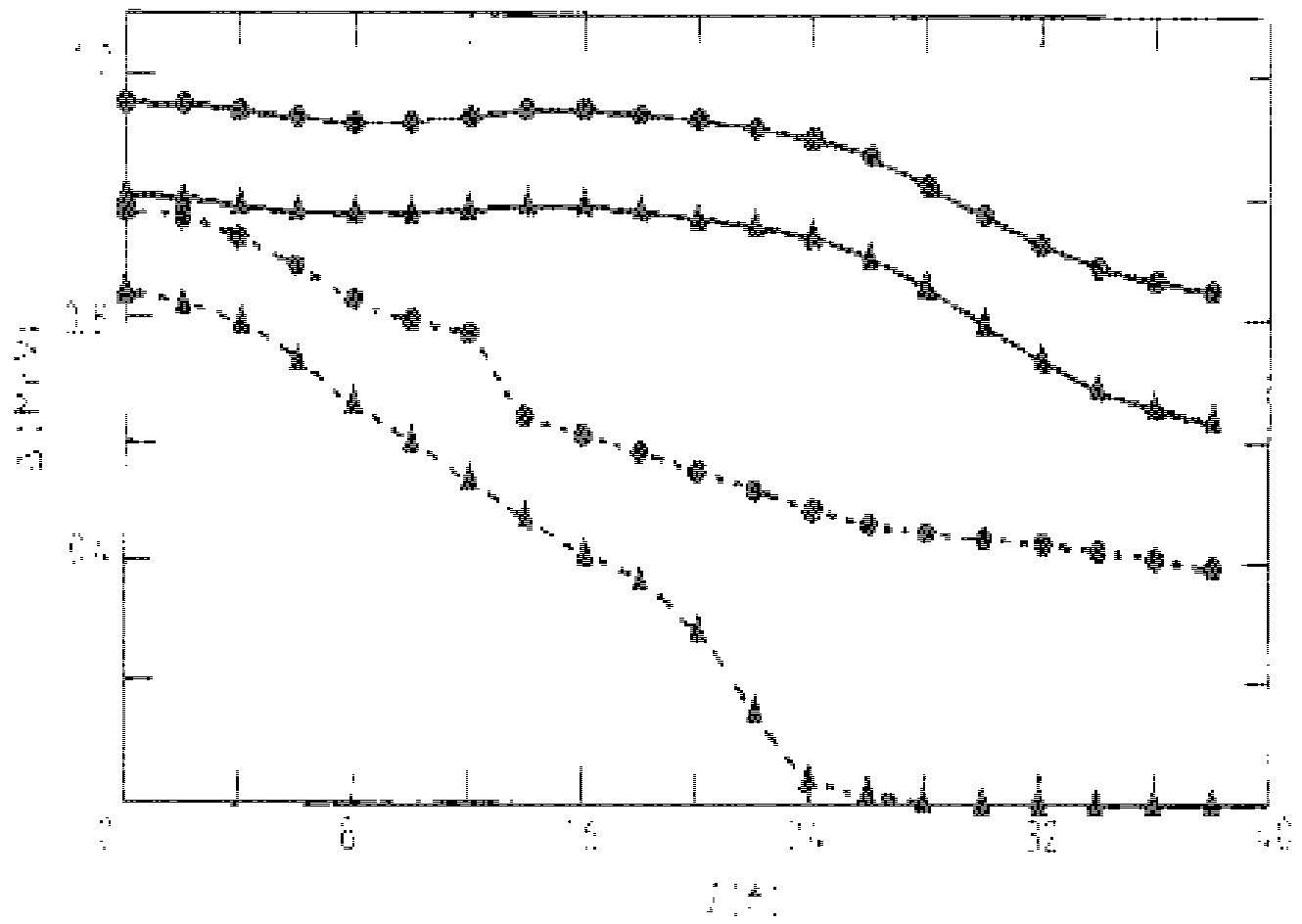}
\caption{Pairing correlations in
$^{168}$Hf as a function of angular momentum. Gap parameters for protons
(circles) and neutrons (triangles) obtained by a variation after number
projection (full curves) and by pure mean-field theory (dashed curves).
Republished with permission of Institute of Physics, from \protect\citeasnoun{(Mut84)};
permission conveyed through Copyright Clearance Center, Inc.}
\label{fig1}
\end{figure}
In {\Fig}~\ref{fig1}, we show self-consistent cranking results for
the pairing gap for protons and neutrons in the nucleus $^{168}$Hf as a
function of the average angular momentum, $I$ calculated with and without
particle-number projection before the variation. Without projection, we observe in the
neutron pairing collapse at spin $I=24\hbar$. The proton gap is also
somewhat quenched after $I=10\hbar$. On the other hand, for the case of number
projection the pairing correlations are reduced smoothly.

\begin{figure}[t]
\centering
\includegraphics[width=3.4in]{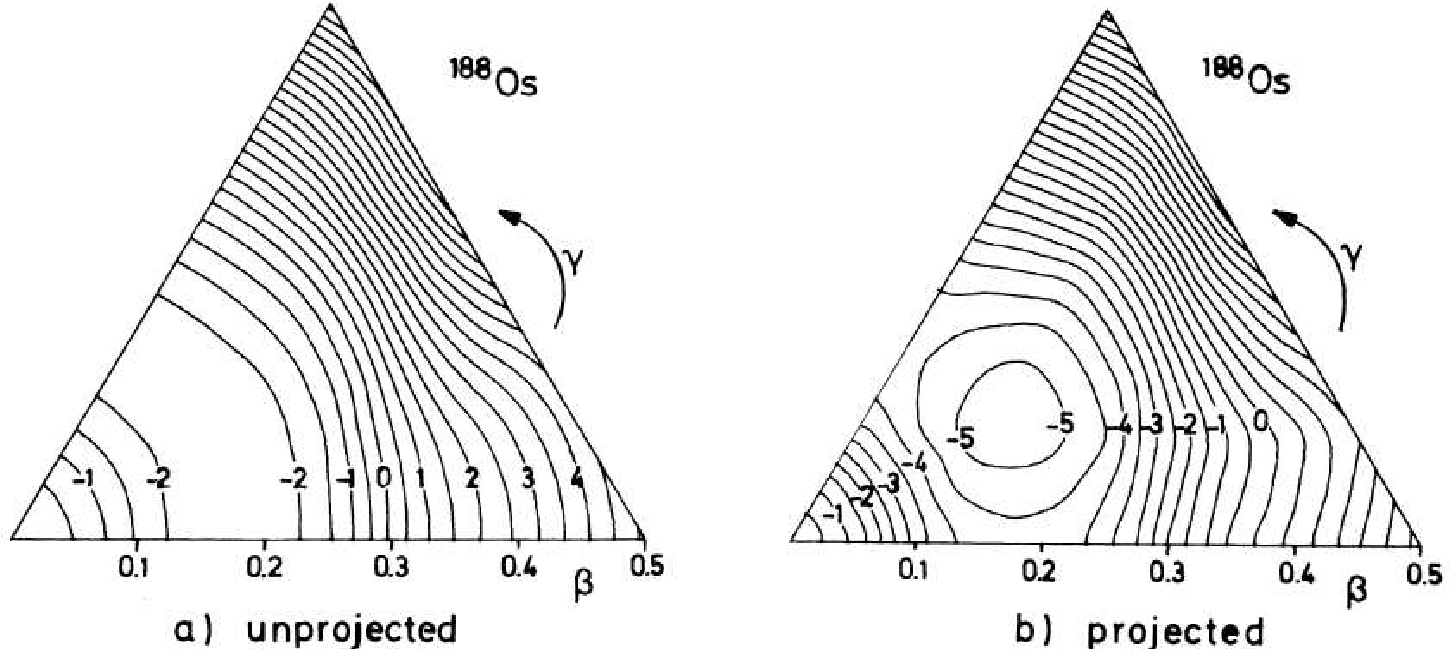}
\caption{Energy surface in
the $\beta$-$\gamma$ plane for the nucleus $^{188}$Os (a) without
angular-momentum projection and (b) with exact three-dimensional
angular-momentum projection. The units on the equipotential lines are in MeV.
Figure adapted with permission from \protect\citeasnoun{(Hay84)}.
Copyrighted by the American Physical Society.}
\label{fig2}
\end{figure}
Angular-momentum projection in triaxial systems is much more complicated. So
far there are very few results for variation after projection. In most
of the cases the variation of the angular-momentum projected energy is
restricted to a few external parameters as, for instance, deformation
parameters. In this case, one needs for each deformation only to evaluate the
projected energy in {\Eq}~(\ref{eq:60}) after a constrained mean-field
calculation. This is considerably simpler than the evaluation of the projected
gradient in each step of the iteration.

In {\Fig}~\ref{fig2}, we show as an example [taken from \citeasnoun{(Hay84)}], the energy surface of
the nucleus $^{188}$Os as a function of the Bohr quadrupole deformation
parameters $\beta$ and $\gamma$. The unprojected energy does not
depend on the triaxiality $\gamma$, and a mean-field calculation
without constraint would be relatively unstable. However, projection
on the angular momentum $I=0$ leads to a clear triaxially deformed
minimum with $\gamma\approx30^{\circ}$.

Several advanced nuclear models were developed based on the pairing
plus quadrupole-quadrupole interaction and the projection theory. Here,
we discuss the approach of triaxial projected shell
model (TPSM) \cite{(She99_j)}, which was extensively employed in recent years to
investigate the high-spin band structures in triaxial nuclei. In this
method, intrinsic basis states are obtained by solving the triaxial
Nilsson potential with the expected deformation values for the
system under investigation. Explicit three-dimensional angular-momentum
projection method is then employed to project out the states with good
angular-momentum. Apart from the projected vacuum
state, multi-quasiparticle states are also projected to the laboratory
frame of reference in this approach. In the most recent version of the
TPSM approach, for even-even systems, quasiparticle basis
states employed are:
\newcommand{\mydagger}{+}
\begin{equation}
\begin{array}{r}
\hat P^I_{MK}\ack\Phi\ket;\\
~~\hat P^I_{MK}~\alpha^\mydagger_{p_1} \alpha^\mydagger_{p_2} \ack\Phi\ket;\\
~~\hat P^I_{MK}~\alpha^\mydagger_{n_1} \alpha^\mydagger_{n_2} \ack\Phi\ket;\\
~~\hat P^I_{MK}~\alpha^\mydagger_{p_1} \alpha^\mydagger_{p_2}
                \alpha^\mydagger_{n_1} \alpha^\mydagger_{n_2} \ack\Phi\ket ;\\
~~\hat P^I_{MK}~\alpha^\mydagger_{n_1} \alpha^\mydagger_{n_2}
                \alpha^\mydagger_{n_3} \alpha^\mydagger_{n_4} \ack\Phi\ket ;\\
~~\hat P^I_{MK}~\alpha^\mydagger_{p_1} \alpha^\mydagger_{p_2}
                \alpha^\mydagger_{p_3} \alpha^\mydagger_{p_4} \ack\Phi\ket ,
\label{basis}
\end{array}
\end{equation}
where $\ack \Phi\ket$ is the vacuum state and
$\alpha^\mydagger_{n}~(\alpha^\mydagger_{p})$ are neutron (proton)
quasiparticle operators.

The projected basis of {\Eq}~(\ref{basis})
is then used to diagonalize the pairing-plus-quadrupole Hamiltonian, {\Eqs}~(\ref{VQQ})--(\ref{H0}),
consisting only of monopole ($\lambda=0$) and quadrupole ($\lambda=2$) terms
in the particle-hole and particle-particle chanel, respectively.
The $QQ$-force strength $\chi_2$ is
adjusted such that the physical quadrupole deformation $\beta$ is
obtained as a result of the self-consistent mean-field HFB
calculation \cite{(Har95a)}.

\begin{figure}[htb]
\centerline{\includegraphics[trim=0cm 0cm 0cm 0cm,width=\textwidth,clip]{{Fig_projRev_10.eps}}}
\caption{Comparison of the calculated (TPSM) and
measured excitation energies for various bands in $^{156}$Dy. The
comparison is made for the ground-state band (Eg), two-quasiparticle band (Es),
$\gamma$-band based on the ground-state (even-spin (Egae) and odd-spin (Egao) bands
are shown separately), and
$\gamma$-band built on the S-band, denoted by E17 and E20 for
even and odd spins, respectively. The $K=0$ two-quasiparticle excited band is
denoted by ESV. For more details on the labelling of various
bands the reader is referred to \citeasnoun{(Maj15a)}. \label{TPSM}}
\end{figure}

As an illustrative example, we present the TPSM results recently obtained in \citeasnoun{(Jeh17)}
for the high-spin band structures observed in  $^{156}$Dy. For this system, rich band structures
have been populated in \citeasnoun{(Maj15a)}. The TPSM calculated energies
for various band structures are compared with the experimental data in {\Fig}~\ref{TPSM}. It is
quite evident from the figure that TPSM results reproduce the data quite accurately. At first glance
it looks somewhat surprising that angular-momentum projection from a fixed mean-field employed
in the TPSM approach can reproduce the data so nicely up to quite high-spin. However, it is
to be reminded that in this approach, angular-momentum projection is performed not only from
the zero-quasiparticle, but also from multi-quasiparticle states. Further, configuration mixing
of these quasiparticle states is performed through diagonalization of the shell model
Hamiltonian \cite{(Har95a)}, which we consider takes into account minor perturbations
in the mean-field potential with increasing spin.

We would like to add that considering only mixing from two-quasiparticle states
without projection corresponds to the
Tamm-Dancoff approximation (TDA) that is known to show a number of disadvantages
as compared to the random phase approximation (RPA)~\cite{ring2000}, which can be justified as the
small amplitude limit of time-dependent-mean field theory. In order to take into
account ground state correlations beyond the projected two-quasiparticle TDA,
several versions of a symmetry conserving RPA were introduced
in \citeasnoun{(Fed85a),(San87a),(Kyo90a),Dukelsky2019_PLB795-537}.
Applications of these theories, however, were restricted,  so far, to very light systems or to
simple group-theoretical models.

\subsection{Projection in small configuration spaces}
\label{sec:small-conf}

Over the years many applications of various projection methods were
carried out in relatively small configuration spaces and for light nuclei.
This was done in the first place due to the fact that many of these
techniques require a considerable numerical effort and that it was or is
impossible to apply them in large spaces and for heavy nuclei. A second reason
was also the fact that exact configuration-interaction calculations can be
carried out in such small spaces, which allow a comparison of the projection
methods with the exact solution in the corresponding space.

A variation after projection on particle number and angular momentum was
carried out already in the eighties by the T\"{u}bingen
group~\cite{Schmid1984_NPA431-205}. This method was called VAMPIR
(variation after mean field projection in realistic model spaces). First it
was restricted to axially symmetric intrinsic HFB wavefunctions in the
valence space containing the 0s a 0d orbits (the sd-shell valence space)
and a phenomenological interactions adjusted to this space were
used. The results were compared with configuration-mixing calculations.
At low spins the results are reasonable. However, in this approach the
intrinsic wavefunctions obey time-reversal symmetry. This means that
they are not flexible enough as to contain all the physics at high spin  and the resulting moments of
inertia are too small. The missing physics can be considered by mixing to the
ground state band with projected two-quasiparticle configurations. This
model~\cite{Schmid1984_PRC29-291,Schmid1984_PRC29-308} was called MONSTER
(model handling many number- and spin-projected two-quasiparticle excitations
with realistic interactions and model spaces). In
\citeasnoun{Hammaren1985_NPA437-1} a similar method was used for odd-mass
nuclei in the mass $A=130$ region with great success.
These configuration-mixing of projected many-quasiparticle
configurations are based on a fixed HFB function for the ground state. Changes
in the deformation etc. cannot be included here. Therefore a new method was
developed by \citeasnoun{Schmid1986_NPA452-493}, the "excited VAMPIR". Here the
intrinsic wavefunction for the lowest energy configuration for each angular momentum (the yrast levels)
is determined in a first step. In
the next step, for each angular momentum, a set of excited states is
determined by new intrinsic wavefunctions, Schmidt-orthogonalized with respect
to all the earlier solutions at this spin. Thus one successively constructs an
optimal configuration space for the $A$-nucleon problem. Over the years, all
these methods were applied in relatively small configuration spaces with
realistic effective forces. They compare well with the corresponding exact
configuration-mixing calculations.  For reviews, see~\citeasnoun{Schmid1987_RPP50-731,(Sch04h)}. The problem is, as in the
case of the configuration-mixing calculations, the determination of the
effective interaction. It depends on the specific configuration space. In
many cases one starts with an effective interaction obtained by the Br\"uckner
method \cite{(Day67a),ring2000} and additional parameters are adjusted
to experimental data.

In \citeasnoun{GAO-ZC2015_PRC92-064310} calculations with variation after
projection on spin, isospin, and mass number were carried out in the
even-even nuclei in the sd-shell using the well known and successful USDB
interaction, which was adjusted to configuration-interaction calculations in
this space \cite{(Bro06)}. The binding energies turn out to be very close to
the exact configuration-interaction results. The differences are very small in
cases, where the number of parameters in the projected wave functions is close
or larger than the dimension of the configuration-interaction space. For the
opposite case one finds energy differences of up to 500 keV. Angular-momentum
projection is very important for these results. Calculations with
angular-momentum projection lead in most of the cases to triaxial deformations of the
intrinsic states. This is not the case for the calculation without
$\mathit{J}$-projection. This is in agreement with systematic mean-field
investigations with the Gogny-force without projection in large model spaces
\cite{Delaroche2010_PRC81-014303}.

Intrinsic wave functions with different deformations can be included in the
generator coordinate method. In \citeasnoun{GAO-ZC2009_PRC79-014311} this
method is extended by adding at each deformation not only the lowest
Hartree-Fock configuration, but in addition a large number $n$p-$n$h
configurations and projecting all these states on good angular momentum. This
is a projected configuration-interaction calculation. The results are compared
with exact solutions for nuclei in the sd- and pf-shell and it is found that
not only the energies, but also the quadrupole moments and the B(E2)
transition probabilities are in very good agreement with the exact
configuration-interaction calculations in the spherical basis.

It is an interesting question, whether it is possible to define an intrinsic
deformation for the exact wavefunctions in the laboratory frame obtained by
configuration-mixing calculations. In cases, where one finds good agreement
between the projected states and the exact solution not only for the energy,
but also for other operators, the projected mean field state is close to the
exact solution and in such a case it is possible to define the intrinsic
deformation of the exact wavefunction as the deformation of the intrinsic
state in the projected theory. It turns out that in some cases the above
mentioned calculations that the intrinsic deformation is not well determined,
because the projected energy surfaces are very flat and the resulting
projected wave functions are nearly identical, even if the intrinsic
deformations are different.

The Monte-Carlo Shell Model of the Tokyo group \cite{Otsuka2001_PPNP47-319}
(see also \citeasnoun{(Miz04)}) uses a fixed configuration space and it provides an
exact solution identical to the conventional calculations mixing spherical
configurations in an oscillator basis. Using the Hubbard-Stratonovich
transformation \cite{Hubbard59,(Str58a)} the
Monte-Carlo Shell Model uses a linear combination of angular-momentum and
parity projected intrinsic wavefunctions%
\begin{equation}\label{MCSM}
|\Psi_{M}^{I\pi}\rangle=\sum_{\sigma}c_{\sigma}\sum_{K}g^{\sigma}_K\hat{P}_{MK}^{I}\hat{P}^{\pi}|\Phi
(\sigma)\rangle~~~,
\end{equation}
where $\sigma$ runs over a large number of intrinsic Slater-determinants
$|\Phi(\sigma)\rangle$ which are determined, together with the coefficients $g_K^{\sigma}$, by a stochastic Monte-Carlo sampling based on the Hamiltonian $\hat{H}$. Finally the coefficients $c_{\sigma}$ are calculated
by the diagonalization of $\hat{H}$.
Each intrinsic state has a certain
deformation and the sum of the coefficients $|c_{\sigma}|^{2}$ with a specific
deformation determine the weight of the intrinsic deformation in the exact
state $|\Psi_{M}^{I\pi}\rangle$. In principle, one would not need the
projection operators, because the importance sampling would automatically lead
to eigenstates with good quantum numbers, i.e., would automatically carry out
the integration over the Euler angles and the summation over $K$. However, as it
turns out, the calculations are much more stable and faster by using the projection
operators. Technically, the 3-dimensional projection is possible because it is
applied here in relatively small configuration spaces, including only a few
oscillator shells as is usual in configuration-mixing calculations. For
recent reviews on modern extensions of this method to relatively large model
spaces and to {\it ab initio} applications, the reader is referred to
\citeasnoun{Shimizu2012_PTEP01A205,Shimizu2017_PS92-063001}.

\subsection{Projection methods in {\it ab initio} calculations}
\label{sec:ab-initio}

During the last few decades, {\it ab initio} calculations based on
effective Hamiltonians, derived without further approximations from
the bare nucleon-nucleon interaction adjusted to scattering
data~\cite{Wiringa1995_PRC51-38,(Ent03a)}, gained more
and more interest in nuclear structure physics. At short distances
these realistic potentials have a very strong repulsive core,
leading, in interaction mixing calculations, to the admixture of
configurations which very high momenta. Therefore such interaction
cannot be used for calculations in small model spaces. There are
several methods in the
literature~\cite{(Bog01),(Bog10a)} to
deduce, without phenomenological parameters, from these bare forces,
effective soft interactions, which can be used in small model spaces
and which lead, in principle, to the same results as calculations in
full space.

At the beginning these effective forces have been used for very small
spaces and therefore for very light systems. Here the effective
Hamiltonian were fully diagonalized by configuration
mixing~\cite{(Bar13a)} and the results were in good
agreement with exact numerical solutions of the these few-body
problems~\cite{Carlson2015_RMP87-1067}.

For heavier systems this is no longer possible, because for
calculations without core one needs a relatively large oscillator
basis. Here approximate methods, based on mean field calculations
with additional projection techniques, are used. The same applies to
heavy nuclei where {\it ab initio} calculations are possible in valence
shells in the neighborhood of magic configurations. However, already
for medium heavy nuclei these valence shells become relatively large,
such that full configuration-mixing calculations require
considerably numerical expense both in computer time and memory.
Here also mean field calculations together with projection methods
based on the corresponding effective interactions are useful.

No-core shell model calculations using chiral {\it ab initio} forces have
been carried out with symmetry adapted basis sets for the nuclei
$^6$Li, $^8$Be, and $^6$He in \citeasnoun{(Dyt13)}
and also for heavier nuclei in
\citeasnoun{(Dyt16),Launey2020_EPJST229-2429,(Lau21)}.
Such methods are also used in \citeasnoun{(Dyt15)}
for {\it ab initio} calculations of electron-scattering on the nucleus
$^6$Li. Within this symmetry-adapted framework one can achieve a
considerable reduction of the dimensions as compared to large
configuration-interaction calculations while retaining the accuracy of the results.

Broken symmetries and additional projection form also the basis of
the  Monte-Carlo Shell-Model discussed in {\Eq}~(\ref{MCSM}).  It has
been applied for no-core shell model calculations in light nuclei and
for valence shells in heavier nuclei. In recent years it has been
used for effective forces derived from {\it ab initio}
calculations~\cite{Shimizu2012_PTEP01A205}. Extensions have been
successfully applied in \citeasnoun{Shimizu2017_PS92-063001} for
heavier nuclei in the neighborhood of the magic shells with N = 28,
40, and 50. In most of these calculations this model uses a basis
with good particle number. Pairing correlations are included by the
additional correlations in the wave function (\ref{MCSM}). However,
for heavier nuclei where pairing correlations extend over a larger
region around the Fermi surface, the corresponding valence shells are
not large enough. Therefore, this method has been extended for
calculations based on the Bogoliubov quasiparticles and the corresponding
symmetry violation is restored by number projection in
\citeasnoun{(Shi21)}. Recently there have been also
{\it ab initio} investigations of cluster structures in nuclei based on using
projection techniques~\cite{Neff2008_EPJST156-69}.

Coupled Cluster methods provide a very successful tool in many-body
calculations in electronic systems. They are also applied in nuclear
physics for {\it ab initio} calculations~\cite{(Hag14a)}.
Modern version uses basis sets with symmetry violations and additional
projection, such as broken rotational symmetry~\cite{(Dug14a)} or
broken gauge
symmetry~\cite{(Sig15),(Dug16),(Qiu19)}

The generator-coordinate-method approaches together with symmetry projection in deformed mean-field
wavefunctions have also been used for {\it ab initio} calculations with
microscopic interactions derived by the Similarity Renormalization
Group method for even calcium and nickel isotopes
\cite{(Her14a)}. Recently this method has been
applied to the calculation of the nuclear matrix elements for
neutrinoless double beta ($0\nu\beta\beta$) decay
\cite{Yao2018,(Yao19)}. So far, this was possible only for the decay
from $^{48}$Ti to $^{48}$Ca, but this method opens up new ways towards
fully microscopic calculations in heavier isotopes, where this decay
is under discussion. This is an example of the importance of
reliable {\it ab initio} calculations in heavy nuclei not only for nuclear
physics, but also for general physics, in particular the physics
beyond the standard model. Earlier, the ($0\nu\beta\beta$) decay
matrix elements have been calculated only with phenomenological
forces of phenomenological density functionals.

\section{Projection methods and nuclear density functional theory}
\label{sec:functional}

The complexity of the bare
nucleon-nucleon force and strong in-medium effects characteristic of nuclear physics lead us to consider
phenomenological effective forces or relativistic Lagrangians to define the underlying intrinsic
mean-field, see the reviews in \citeasnoun{(Ben03e),(Vre05c),Robledo2019,(Sch19)}.
The associated HF or HFB equations are traditionally solved by
expanding  on a basis, as for instance the
harmonic oscillator or Woods-Saxon basis, or by using mesh techniques in
the coordinate representation (which is also equivalent to a basis expansion, see \citeasnoun{(Bay84a)}).
The presence of a phenomenological density-dependent term
containing the mean field density implies that such
effective forces do not lead to mean-field average values that can be obtained from a
Hamiltonian. Instead, they can be considered as a
special case of an energy density functional (EDF), where part of the
functional is obtained from a two-body interaction (and therefore is
quadratic in the densities) and the rest is purely phenomenological. Density dependent term of this kind
is also found in the Slater approximation to Coulomb exchange.
These peculiarities of the traditional nuclear EDF approach lead to difficulties
in the implementation of the methodology of symmetry restoration:

\begin{itemize}
        \item The harmonic oscillator or Woods-Saxon bases can break spatial symmetries like rotational
        and/or translational. This is also the case for the mesh representation
        of the wave functions. This symmetry breaking has to be taken
        into account in the formalism explicitly.
        \item The density-dependent term of the EDF is only well defined for mean values. For energy overlaps
         a prescription, satisfying consistency constraints, is required. It turns out that popular alternatives
         lead  to consider complex quantities that are often raised to non-integer powers, requiring
         additional considerations in the
         selection of the branch-cuts in the complex plane.
        \item Different interactions for the particle-hole and particle-particle
         channels are often used raising self-energy and violation of the Pauli principle issues.
\end{itemize}

The existence of these difficulties, and the fact that some of them
are still not satisfactorily addressed, slowed down the
application of symmetry restoration techniques in nuclear structure
with EDFs.  However, there are numerous indications that  symmetry restoration is required to
improve our qualitative and quantitative theoretical understanding of many nuclear
properties. The results obtained so far go in this direction and some of them will be
discussed below.

\subsection{Difficulties encountered in restoring symmetries with nuclear EDF}
\label{EDF:difficulties}

\subsubsection{Basis not closed under the symmetry action}
\label{EDF:notclosed}

Symmetry restoration requires
to consider the action of a symmetry operation (rotation, translation, etc) on an
intrinsic symmetry violating wave function. Often, the single-particle basis used for
the intrinsic state is not closed under the action of the symmetry operator and the ``rotated"
basis does not span the same subspace of the Hilbert space as the original basis.
In the rotational case, this happens, for instance,  when the
oscillator lengths along different spatial direction are not the same,
or when a Cartesian mesh is used in coordinate representation \citeasnoun{(Bay84a)}. In this situation,
the standard formalism to compute overlaps between rotated HFB states is not valid
and has to be generalized -- see below. A common alternative in the
applications of angular-momentum projection is to use spherical harmonic oscillator
bases with a sufficiently large number of complete major shells.
This strategy increases the computational cost, and becomes impractical in
some situations like fission, where the large variety of shapes involved in the dynamics would
require a huge rotationally invariant basis. In addition, it cannot be applied, e.g., to the case of spatial translations
as required in the restoration of translational and/or Galilean invariance
\cite{(Sch91b),Rod04a,Rod04b}.
To understand the reason, let us consider  a simple model in one dimension where the basis contains
just one single Gaussian state $\varphi_{0}(x)= e^{-x^{2}}$. After translation
$\varphi_{0}(x)\rightarrow \varphi_{0}(x-x_{0})=e^{x_{0}^{2}}e^{-x^{2}}
\sum_{k=0,\infty} \frac{(-2xx_{0})^{k}}{k!}$ an infinite number of Gaussian
wave functions $e^{-x^{2}}x^{k}$  is required to reconstruct the translated
wave function. The situation does not improve if another basis, like Woods-Saxon
or a discrete mesh in space \cite{(Bay84a)}, is used.

As mentioned above, the generalized Wick's theorem, cannot be used in this case as it is implicitly
assumed in its derivation (see, for instance, \citeasnoun{(Bal69a)})
that both HFB wave functions in the overlap are expanded in the same basis.
To circumvent this problem, the original finite  basis (denoted by 1) is formally expanded, by
using the orthogonal complement (denoted by 2), as
to span the whole Hilbert space \cite{(Bon90b),Rob94,(Val00a)}. The same procedure is applied to
the ``rotated" basis ($1^\prime$). The expanded $U$ and $V$ Bogoliubov
amplitudes have a block diagonal form
with the block corresponding to basis 2
having a simple form with an uniform occupancy of 0. With the introduction of the expanded basis,
the traditional generalized Wick's theorem can be applied, However, all quantities referring to basis 2 have to be
reexpressed \cite{Rob94,(Val00a)} in terms of quantities defined in basis 1, leading to an extended Wick's
theorem that differs from the traditional one by the presence of the non-unitary overlap matrix
between the two basis. Further details can be found in \citeasnoun{Rob94,(Val00a),(Mar19b),(Mar20)}.

\subsubsection{Self-energy and Pauli principle}
\label{EDF:Self-energy}

Reduced kernels of a two-body Hamiltonian between two HFB wave functions $|\Phi_{0}\rangle$
and $|\Phi_{1}\rangle$ can be expressed with the help
of the generalized Wick's theorem
in terms of the transition density matrix $\rho^{01}$ and
transition pairing tensors $\overline{\kappa}^{01*}$ and $\kappa^{01}$
as\footnote{Compare definitions in {\Eqs}~(\ref{eq:65})--(\ref{eq:67}), which were valid only for
the normalization specified in {\Eq}~(\ref{eq:61}).}
\begin{eqnarray}\hspace*{-2.2cm}
 \rho^{01}_{k_3 k_1} =
  \frac{\langle \Phi_{0} |  a^{\mydagger}_{k_1}  a_{k_3}           | \Phi_{1} \rangle}
       {\langle \Phi_{0} |                                         \Phi_{1} \rangle},
\quad\quad
  \overline{\kappa}^{01*}_{k_1 k_2} =
  \frac{\langle \Phi_{0} |  a^{\mydagger}_{k_1}  a^{\mydagger}_{k_2} | \Phi_{1} \rangle}
       {\langle \Phi_{0} |                                         \Phi_{1} \rangle},
\quad\quad
  \kappa^{01}_{k_3 k_4} =
  \frac{\langle \Phi_{0} |  a_{k_4}            a_{k_3}           | \Phi_{1} \rangle}
       {\langle \Phi_{0} |                                         \Phi_{1} \rangle},
        \label{eq:luistr}
\\ \hspace*{-2.2cm}
 \frac{\langle \Phi_{0} | a^{\mydagger}_{k_1}  a^{\mydagger}_{k_2} a_{k_4} a_{k_3} |\Phi_{1} \rangle }
       {\langle \Phi_{0} |                                                       \Phi_{1} \rangle }  =
\left[  \rho^{01}_{k_3k_1}
\rho^{01}_{k_4k_2} -  \rho^{01}_{k_4k_1} \rho^{01}_{k_3k_2} +
        \overline{\kappa}^{01*}_{k_1 k_2} \kappa^{01}_{k_3k_4} \right] ,
        \label{eq:luisme}
\end{eqnarray}
where the three right-hand-side terms of {\Eq}~(\ref{eq:luisme}) are referred to
as  direct, exchange, and pairing terms.
As the overlap $\langle \Phi_{0} | a^{\mydagger}_{k_1}  a^{\mydagger}_{k_2} a_{k_4} a_{k_3} |
\Phi_{1} \rangle $ is a finite quantity, the right hand side of {\Eq}~(\ref{eq:luisme})
must diverge  when the overlap $\langle \Phi_{0} | \Phi_{1} \rangle$
vanishes. The same argument applies to
$ \langle \Phi_{0} |  a^{\mydagger}_{k_1}  a_{k_3} | \Phi_{1} \rangle $, etc
and therefore $\rho^{01}_{k_3k_1}$, $\overline{\kappa}^{01*}_{k_1 k_2}$ and  $\kappa^{01}_{k_3k_4}$
are also divergent quantities when the overlap vanishes. The order of the pole present in  all
the quantities, including the Hamiltonian kernel, must be the same and equal to the order of the zero of the overlap.
Therefore, in the right hand side of {\Eq}~(\ref{eq:luisme}) there must be a cancellation
of  poles, so as
to reduce the order of the pole in the products of transition densities and pairing tensors.
The cancellation indeed takes place and is a direct consequence of the Pauli exclusion
principle \cite{(Str78a)}. The same
kind of cancellation also happens for the three-body and many-body operators.
This represents a serious problem if some of the contributions (typically exchange and/or pairing) for some interaction terms
are neglected. In \citeasnoun{Taj92} a somehow arbitrary ``regularization" scheme was introduced
to define ``regularized" contributions for the direct,
exchange and pairing terms. In \citeasnoun{(Don98)}, the problem was also
discussed, raising serious doubts about the validity of calculations with
the pairing plus quadrupole Hamiltonian (see {\Sec}~\ref{sec:simple}), where the exchange and pairing
parts of the quadrupole-quadrupole force, as well as the direct and
exchange parts of the pairing force, are all neglected.

In \citeasnoun{Ang01b} the same problem was discussed in the context of the
particle-number projection with the Gogny force,
see also \citeasnoun{Dobaczewski07,Duguet.09,Ben09,(Hup11b),(Dug15c)} for follow-up
studies with the Skyrme forces.
In traditional studies with the Gogny force, the Coulomb exchange is treated
in the Slater approximation, and Coulomb and spin-orbit pairing are neglected.
As a consequence, unphysical and inconsistent values of the projected energies were obtained.
The use of variation
after projection aggravated the problem, as the projected energy could get attractive contributions from
the unphysical contributions.  Due to the simple form of the overlaps
involved in the particle-number projection,
it was possible to trace back the problem to the existence of vanishing overlaps at
the gauge angle  $\varphi=\pi /2$, due to configurations
with occupancy $v_{k}^{2}=1/2$. The inclusion of all the missing contributions solved the problem.
The computational cost can increase by up to two orders of magnitude if the exchange and pairing
Coulomb contributions are required in a calculation with contact central potentials.

To illustrate the
problem, let us compute the general matrix element of {\Eq}~(\ref{eq:luisme}) with
$|\Phi_{1} \rangle=\exp (-i\varphi\hat N) |\Phi_{0}\rangle$
The transition density matrix and pairing tensor
take a very simple form in the canonical basis of the Bogoliubov
transformation  (see \citeasnoun{Man75,ring2000} for details).
If $c^{\mydagger}_{k}$ represent the creation operators in the canonical
basis of the HFB wave function $|\Phi_{0}\rangle$, we obtain
\begin{eqnarray}\hspace*{-2.2cm}
\frac{\langle \Phi_{0} |  c^{\mydagger}_{k_1}  c^{\mydagger}_{k_2} c_{k_4} c_{k_3}
\exp (-i\varphi \hat N)|\Phi_{0} \rangle}{   \langle \Phi_{0} | \exp (-i\varphi \hat N)|\Phi_{0} \rangle }
 &=&  \frac{v_{k_1}^2 e^{-2 i
\varphi}}{u_{k_1}^2 +v_{k_1}^2 e^{-2 i \varphi}} \cdot
\frac{v_{k_2}^2 e^{-2 i \varphi}}{u_{k_2}^2 +v_{k_2}^2 e^{-2 i \varphi}} \nonumber \\ &&\hspace*{4cm}\times
\left ( \delta_{k_3k_1} \delta_{k_2k_4} - \delta_{k_4k_1} \delta_{k_3k_2}
\right ) \nonumber \\\hspace*{-2.2cm}
 &+&   \frac{u_{k_1} v_{k_1}}{u_{k_1}^2 +v_{k_1}^2 e^{-2 i \varphi}}
\cdot \frac{u_{k_3} v_{k_3} e^{-2 i \varphi}}{u_{k_3}^2 +v_{k_3}^2
e^{-2 i \varphi}} \times \delta_{k_2 \bar{k}_1} \delta_{k_4 \bar{k}_3} .
\label{eq:overlap1}
\end{eqnarray}
When  $\varphi=\pi/2$ ($e^{-i2\varphi}=-1$) and the occupancy of one of the states $k$ is $u_k^2=v_k^2=1/2$, some of the
denominators in this expression go to  zero. In this case,
the overlap $\langle \Phi_{0} |\exp (-i\varphi \hat N)|\Phi_{0} \rangle  =
\prod_{k>0} (u_k^2 + v_k^2 e^{-2 i \varphi} ) $ also goes to zero. As long as there is a single pole,
or two poles but with $k_1 \neq k_2$ (or $k_3 \neq k_1$) the overlap in the
numerator of the l.h.s.\ of {\Eq}~(\ref{eq:overlap1}) remains finite.
The only  problematic case is
$\langle \Phi_{0} | c^{\mydagger}_{k} c^{\mydagger}_{\bar{k}} c_{\bar{k}} c_{k}
\exp (-i\varphi \hat N) | \Phi_{0} \ \rangle $ where the right-hand side of {\Eq}~(\ref{eq:overlap1})
apparently has a pole of order two that cannot be canceled out by the norm overlap.
Although each  of the two terms have an order two pole, the  sum turns out to be
a order one pole. When multiplying by the norm overlap, the sum gives a finite
contribution, namely \( v_{k}^2 \cdot e^{-2 i \varphi} \cdot
\prod_{m>0,m \ne k} (u_m^2 + v_m^2 e^{-2 i \varphi} )\).
As long as the direct, the exchange, and the pairing
terms of the Wick's factorization of
{\Eq}~(\ref{eq:luisme}) are taken into account, no divergences appear in the particle-number
projection formalism. This cancellation is also connected with the
so-called ``self interaction" problem \cite{(Per81a),Lac09}: the sum of the three terms is required
to give a zero two body energy for a Slater determinant containing just one particle.

Although the solution to the self-energy problem is simple, and it is routinely
used with Gogny forces (see {\Sec}~\ref{EDF:Applications}), it is not
possible to implement it for those nuclear EDFs where the particle-hole
and particle-particle interactions are independent of each other. In this case, by construction,
the particle-particle interaction do not contribute to the particle-hole channel and
vice versa and therefore the pole cancellation cannot take place.
The problem was extensively discussed in the literature
\cite{Dobaczewski07,Duguet.09,Ben09,(Hup11b),(Dug15c)} and
attempts to renormalize the divergences were
proposed in \citeasnoun{Taj92,Lac09,(Sat14f)}. Unfortunately, they only work for functionals depending on integer powers of the density
and a clear renormalization criteria is missing
in the most general cases. This difficulty is the major reason for the
recent interest in replacing EDFs by mean values of true Hamiltonians
\cite{(Ben14b),(Sad13d),(Ben17a)}.

Note that the arguments given above are rather
general, and thus they also apply to restoration of other symmetries like angular momentum \cite{(Zdu07c)}.
They do not depend on the kind of force,  or the kind of two
body operator considered in the mean value. Only for strictly density-independent forces (see {\Sec}~\ref{EDF:DenDep})
like Skyrme SV \cite{(Bei75b)} and no pairing, all symmetries, including the isospin symmetry,
can be consistently restored \cite{PhysRevC.81.054310}. Obviously, the projected
calculation of mean values of  one-body operators are never
affected by the inconsistencies mentioned above.

\subsubsection{Density-dependent prescription}
\label{EDF:DenDep}

Most of the EDF used nowadays include in one way or another a phenomenological
density dependence. In the non-relativistic case, it is introduced to
mock-up the saturation property of the nuclear interaction in a simple way.
Although the saturation could also be obtained without any density dependence \cite{(Bei75b)},
this inevitably leads to an unrealistic low value for the effective mass \cite{(Dav18)}.
In the relativistic case, the saturation property is a relativistic effect due to
the difference between the scalar density, the source of the attractive part of the force
and the baryon density, the source of the repulsive part of the
force. Here, the saturation
can also be obtained without any density dependence~\cite{Walecka1974_APNY83-491}. Nonetheless,
to obtain a realistic description of the nuclear incompressibility and surface properties,
an additional density dependence is used in the coupling constants.

In both the cases, density-dependent forces are implemented in a state-dependent way, so that
the density $\rho (\bm{r})=\langle \Phi | \hat \rho (\bm{r}) | \Phi \rangle$ is used
to evaluate the HFB energy associated with the HFB wave function $|\Phi\rangle$.
For most of the non-relativistic functionals, the density-dependent contribution
to the energy  is strongly repulsive,
it is usually proportional to $\delta (\bm{r}_{1}-\bm{r}_{2})$
(contact term) and depends on the intrinsic center of mass density
$\rho ((\bm{r}_{1}+\bm{r}_{2})/2)$ raised to some power $\alpha$ which is,
in most of the cases, a non integer number (usually 1/3, although 1/6 is also a common choice).
The typical form of the density-dependent part of functional reads
\begin{equation}
V_{DD}(\rho)= t_3 (1+x_0 \hat{P}_\sigma) \delta (\bm{r}_1 -\bm{r}_2) \rho^{\alpha}
(\frac{1}{2} (\bm{r}_1+\bm{r}_2)) , \label{eq:VDD}
\end{equation}
where $t_3$ and $x_3$ are parameters and $\hat{P}_\sigma$ is the spin-exchange operator,
see \citeasnoun{(Ben03e)} for details.

When the variational principle is used to derive the HF or HFB
equations, the density-dependent part has also to be varied, leading to the
so-called rearrangement term in the mean field, which in the non-relativistic case is given by
\begin{equation}
        \partial \Gamma_{ij} = \frac{1}{2}
                               \sum \langle k_{1} k_{2} |
                                     \frac{\delta V_{DD}}{\delta \rho_{ji}} |
                                     k_{3} k_{4} \rangle \rho_{k_{3}k_{1}}\rho_{k_{4}k_{2}}.
\end{equation}
Given the typical values of $t_{3}$, this rearrangement term
represents an important component of the HF or HFB mean field and it cannot be overlooked.
It enters in the expressions of many quantities like chemical potentials, quasiparticle excitation energies, etc..

In the evaluation of symmetry-restored energies based on HFB
intrinsic states, instead of the mean values of the HFB theory we
have to consider kernels of the Hamiltonian between different HFB
states. In the case of density-dependent forces some kind of
prescription has to be given for the evaluation of the Hamiltonian kernels.
This requirement also arises in the implementation of the generator
coordinate method.

There are a few fundamental requirements that must be used to guide the choice
of the prescription:

\begin{itemize}
        \item The projected energy has to be a real quantity.
        \item The projected energy has to remain invariant with respect to symmetry
         operations applied to the intrinsic state.
         \item In the strong deformation limit of the intrinsic state
         (see {\Sec}~\ref{sec:Kamlah}), the projected
         theory should reduce to the traditional mean-field approach \cite{robledo07p}.
\end{itemize}

Among the various proposals made over the years \cite{(Bon90b),(Val00a),(Rod02a),Duguet03,(Sch04h)},
we only comment on the two that are in vogue:

\vspace{1ex}\noindent\underline{Mixed-density prescription:}\newline
The ``mixed" or ``transition" density prescription, proposed already in
early nineties \cite{(Bon90b)} within the framework of the generator coordinate method, and
extended to the symmetry restoration case in \citeasnoun{(Val00a),(Rod02a)},
employs the transition density,
\begin{equation}
\rho_{q, q'} =  \langle\Phi (q)| \hat{\rho}
|\Phi(q')\rangle  / \langle\Phi (q) |\Phi (q') \rangle ,
\label{kernel1}
\end{equation}
where $\hat{\rho}$ is the one-body density operator.
Then, $\rho_{q, q'}$ is used in the evaluation of the Hamiltonian kernel
\begin{equation}
h(q,q')= \langle\Phi (q)| \hat{H}\{\rho_{q, q'}\} |\Phi (q')\rangle
/ \langle\Phi (q) |\Phi (q') \rangle,
\label{kernel2}
\end{equation}
where $\hat{H}\{\rho\}$ denotes any density-dependent Hamiltonian
containing terms similar to $V_{DD}$ of {\Eq}~(\ref{eq:VDD}).
This prescription is inspired by
contractions used in the generalized Wick's theorem when it is applied
to three-body forces. It
satisfies all the consistency requirements mentioned above
\cite{(Rod02a),(Egi04b)}. However, since the transition density is in
general a complex quantity, this prescription requires additional rules
when it is implemented for the density raised to a non-integer power.

\vspace{1ex}\noindent\underline{Projected-density prescription:}\newline
The density to be used in the Hamiltonian kernel (\ref{kernel2}) is given by
\begin{equation}
\rho^P_{q, q'} = \frac{\langle\Phi (q)|\hat{P} \hat{\rho} \hat{P} |\Phi (q')\rangle}
                      {\langle\Phi (q)|\hat{P}                    |\Phi (q')\rangle} ,
\end{equation}
where $\hat{P}$ represents generic projection operator (\ref{eq:51a}).
This prescription produces a density-dependent term which is invariant
under the broken symmetries.
For instance, in the case of the rotational symmetry breaking,
this prescription leads to density $\rho^P_{q, q'}$ that is spherically
symmetric, and because of this, it was advocated by
\citeasnoun{(Sch04h)}. This prescription satisfies the first two
requirements mentioned above, but the third one is more problematic.

\begin{figure}
\begin{center}
\includegraphics[width=0.5\textwidth]{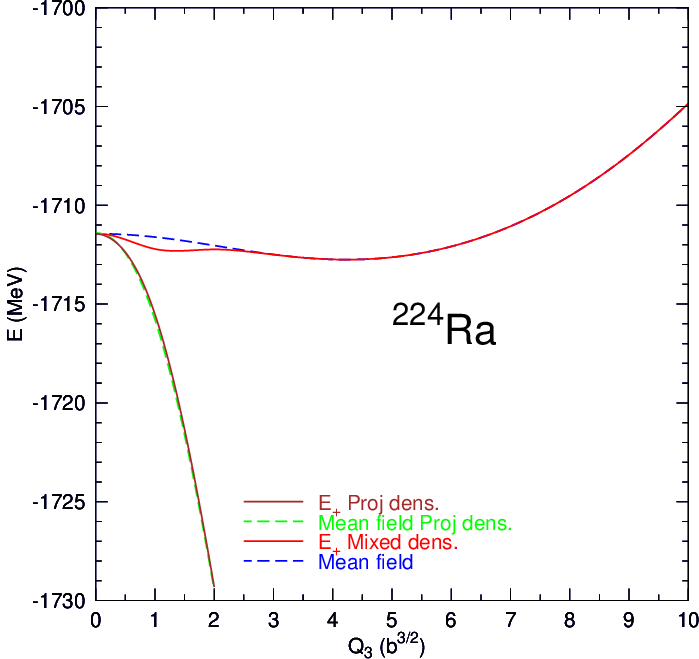}
\caption{Various positive-parity projected energies corresponding to
different prescriptions for the density in the density-dependent term of
the Gogny force are plotted as functions of the axial octupole moment
for the nucleus $^{224}$Ra. The brown downsloping curve corresponds to
the parity-projected density prescription, whereas the red curve corresponds
to the mixed-density prescription. The mean-field energies corresponding
to the two prescriptions are also plotted with dashed curves: blue for the
mixed-density prescription and green (on top of the brown curve) for the
projected-density prescription.
Republished with permission of Institute of Physics, from \protect\citeasnoun{Robledo10};
permission conveyed through Copyright Clearance Center, Inc.}
\label{Fig:Robledo10}
\end{center}
\end{figure}

The projected density prescription is highly inconsistent with the underlying
intrinsic mean field \cite{Robledo10} when spatial symmetries like the parity
(see discussion in {\Sec}~\ref{DSPW}) or
rotational invariance are under consideration. An example of the catastrophic
consequences can be seen in {\Fig}~\ref{Fig:Robledo10}, where the parity-projected
energy for the nucleus $^{224}$Ra is plotted as a function
of the octupole moment $Q_{30}$ for the two prescriptions discussed here.
For the projected-density prescription, with increasing octupole moment, the projected energy continuously
decreases as a consequence of the mismatch
between the projected density and the transition density.

We note here, that in the case of the particle-number projection, the
use of the projected-density prescription produces reasonable
results. This is related to the fact that the particle-symmetry
restoration affects only the occupation numbers of single-particle
canonical states and does not lead to a spatial mismatch between the
left and right states for which the Hamiltonian kernel
(\ref{kernel2}) is evaluated.  The Madrid group routinely uses a
hybrid prescription, that is, the mixed-density prescription is used
for all symmetries apart from the particle-number symmetry
restoration when the projected-density prescription is used. In
addition, as it was shown in \citeasnoun{(Val00a)}, the Lipkin-Nogami
approximation (see {\Sec}~\ref{sec:LN}) leads to results that are very
similar to those obtained using the mixed-density prescription.
We also note that the random phase approximation applied to the
generator coordinate method \cite{Jancovici64} can also be used as an
argument in favor of the transition-density prescription \cite{robledo07p}.

\subsubsection{Non-integer powers of the density}
\label{EDF:NonInt}

The transition-density prescription requires to consider the transition density raised
to some non-integer power. However, the transition density is a complex
quantity, in general, and we are confronted with the problem of how to consider the
evaluation of a non-integer power of a complex number.
The choice of branch cut in this case can have an enormous impact on the value
of the matrix element as a consequence of the large positive value of the strength
of the density-dependent term.

In addition, in the symmetry restoration case, where integrals over the
parameters of the symmetry group have to be carried out, the presence
of branch cuts associated with the density-dependent part break the analyticity of the
integrand and leads to spurious
dependencies on the integration path in the complex plane of the symmetry
group parameters. In a simple case of the particle-number
restoration, this issue was discussed in great
detail by \citeasnoun{Dobaczewski07,Duguet.09}. Similar problems are expected for the restoration of other
symmetries. Treatment of the density dependence with non-integer powers of the density
is still an unsolved problem and represents a serious limitation
in the applicability of the symmetry restoration and configuration-mixing
methodology within nuclear EDF.

\subsubsection{Future implementations}
\label{EDF:Ahead}

In recent years, the idea to give up phenomenological density-dependent terms
in favor of real (multibody) operators is becoming increasingly
popular in non-relativistic density functional theory \cite{(Ben14b),(Sad13d),(Ben17a)}.
The idea is to find a three-body interaction (not excluding higher order terms) that
is able to mimic the saturation property induced by the density-dependent interaction and
at the same time is not spoiling pairing properties.
Plain use of the generalized Wick's theorem would be the only thing required to compute
the kernel overlap. The difficulty encountered with this approach is the large
number of possible terms with their associated free parameters. In order
to reduce the number of terms, several assumptions were made, but
so far, the three-body interactions proposed are not very accurate in
describing nuclear properties.

On the other hand, the influence of the poles in the projected energy surfaces
seem to be relatively narrow and in many applications, in particular
in a projection after variation, they can be regularized by using a wider mesh in the
integration over the various angles. In this way, the results depend on
the choice of the discretization mesh,  but at the end the numerical errors
connected with such procedures do not play an essential role, in particular for
heavy nuclei. In any case, one finds in the literature many very successful applications
of generator coordinate method and projection after variation (see below), where
the potential impact of such poles in the physical observables do not spoil the physical
interpretation.

\subsection{Applications of symmetry restoration with nuclear EDFs}
\label{EDF:Applications}

The importance of  symmetry restoration in nuclear structure was
noticed very early in nuclear physics \cite{(Bay60),Die64}.
Applications of this method to simple models were already discussed in
{\Secs}~\ref{sec-PNPN} and~\ref{sec:simple}.
Other approaches used shell-model-like effective interactions defined
in a restricted configuration space involving a limited number of orbits.
Among them we can mention the work developed by the  T\"ubingen
group using very sophisticated many body techniques involving mixing
of configurations through projecting out mean-field intrinsic states
obtained in a variation after projection framework \cite{(Sch04h)}.
We will not dwell on these approaches as we are more concerned with
realistic nuclear energy density functionals.

\subsubsection{Non relativistic EDFs}
\label{Non-rel}

The first application of symmetry restoration with an EDF possibly dates
back to the work of Caurier and Grammaticos \cite{(Cau77a)} where rotational bands
of light nuclei were computed using several flavors of the
Skyrme interaction and angular-momentum projection.
A few years later, the parity-projected excited intrinsic configurations of $^{20}$Ne
were studied \cite{Mar83} using the BKN interaction \cite{(Bon76)}. In that work, the center of mass correction was
also computed using symmetry restoration techniques.
Reflection symmetry restoration was also used
along with the Gogny D1S interaction in \citeasnoun{Egi91} to describe the
physics of parity doublets and octupole deformation.  Parity restoration was also applied to
cranking wave functions in order to study the emergence of octupole
deformation at high spins
\cite{Gar97,(Gar98c)}. Parity projection of non axial intrinsic states was
performed in several mercury and lead isotopes in a generator coordinate context
\cite{Ska93}. Systematic calculations of the excitation energies, E3
transition strengths and ground state octupole correlation energies
were performed in a combined generator coordinate method and parity projection framework
\cite{Rob11a,Rob15a} with several flavors of the Gogny force, depicting
the importance of the dynamical octupole correlations in nuclear structure.

To describe collective negative parity states, intrinsic wave functions
breaking reflection symmetry are important ingredients and therefore octupole deformation
becomes important in this case. At the mean field level
only a few  nuclei in the actinide and
rare earth regions are octupole deformed. The number of nuclei accessible increases
substantially if the intrinsic state is determined in the variation after parity projection scheme,
irrespective of whether it is implemented exactly or in a restricted variational space
\cite{Rob11a,Rob15}. Octupole correlations are better described using
the generator coordinate method with the octupole moment  as
one of the collective coordinates \cite{Ber16a,(Buc17),(Ber17)}.

Particle-number restoration is another important application of
symmetry restoration in nuclear structure as pairing correlations are
known to be rather weak in atomic nuclei and therefore the use of an
intrinsic mean field wave function (BCS or  HFB wave function) is not
easy to justify. Fluctuations of the order parameter associated with
pairing correlations (for instance, the fluctuation of the particle number
$\langle \Delta N^{2}\rangle$) as well as the corresponding gauge angle
associated with particle-number restoration are important ingredients for
a proper description of pairing correlations in the weak-pairing regime of
nuclear physics.
Most of the applications use intrinsic states obtained from a HFB calculation, supplemented with
the Lipkin-Nogami  procedure (see {\Sec}~\ref{sec:LN}) generalized for density dependent forces \cite{(Val97a)}.
There are many examples of full variation after
particle-number projection calculations with the Gogny force,
mostly in the framework of a particle-number projected generator
coordinate method with restoration of additional symmetries. There are
also early examples  \cite{Ang01b} aimed to understand the effect of
particle-number projection on the moment of inertia of rotational bands or in
the ground state correlation energy \cite{Ang02}.  Particle-number projection
in the variation after
projection scheme increases pairing correlations and, as a consequence, decreases the moment of inertia, increasing thereby the
excitation energy of rotational $2^{+}$ states.

The structure of some Sr
\cite{Hee93} or Pb isotopes \cite{(Hee01)} was analyzed with Skyrme EDF and the particle-number projected generator coordinate method.
In \citeasnoun{(Sto03a)}, a mass table from proton to neutron dripline was
generated with the SLy4 EDF and using volume pairing and implementing
Lipkin-Nogami method followed by a full particle-number projection. The
procedure was implemented in a computer code that is publicly
available \cite{(Sto05b)}. Similar calculations were performed
by the Brussels group in their quest for an accurate mass model \cite{(Sam04a)}.

A full variation after particle-number
projection calculation with a Skyrme functional plus a zero range
pairing force was carried out in \citeasnoun{(Sto07e)} using the formulation
of the particle-number projection method of  \citeasnoun{(She00c)}, solely involving functions of the standard density
and the abnormal pairing tensor, see Sect \ref{sec_srgf}. The results were
compared to the ones obtained with the Lipkin-Nogami method followed by
a subsequent projection on particle number and a substantial improvement
is observed, specially for magic or near magic nuclei.

Spontaneous symmetry breaking of rotational invariance is a defining
characteristic of the nuclear interaction. It leads to the fruitful
concept of nuclear deformation \cite{Casimir1935} that allows us to explain a variety of
phenomenology like, for instance, the ubiquitous  existence of rotational bands in
nuclear spectra \cite{(Boh53a),(Boh55a)}.
Most of nuclei in the Segr\'e chart are thought to exhibit
rotational symmetry breaking in some of their quantum states, either the
ground or excited states. Although it is possible to extract a lot of
information out of the intrinsic deformed states by using the strong
deformation limit [see \citeasnoun{Man75,ring2000} and {\Sec}~\ref{sec:Kamlah}],
the existence of weakly deformed states and/or the
coexistence of different types of deformations in a limited range of energies
requires the explicit restoration of the rotational symmetry. This
is also the case for calculations of electromagnetic transition strengths
in weakly-deformed nuclei \cite{Rob12}.

\begin{figure}
\begin{center}
\includegraphics[width=0.48\textwidth]{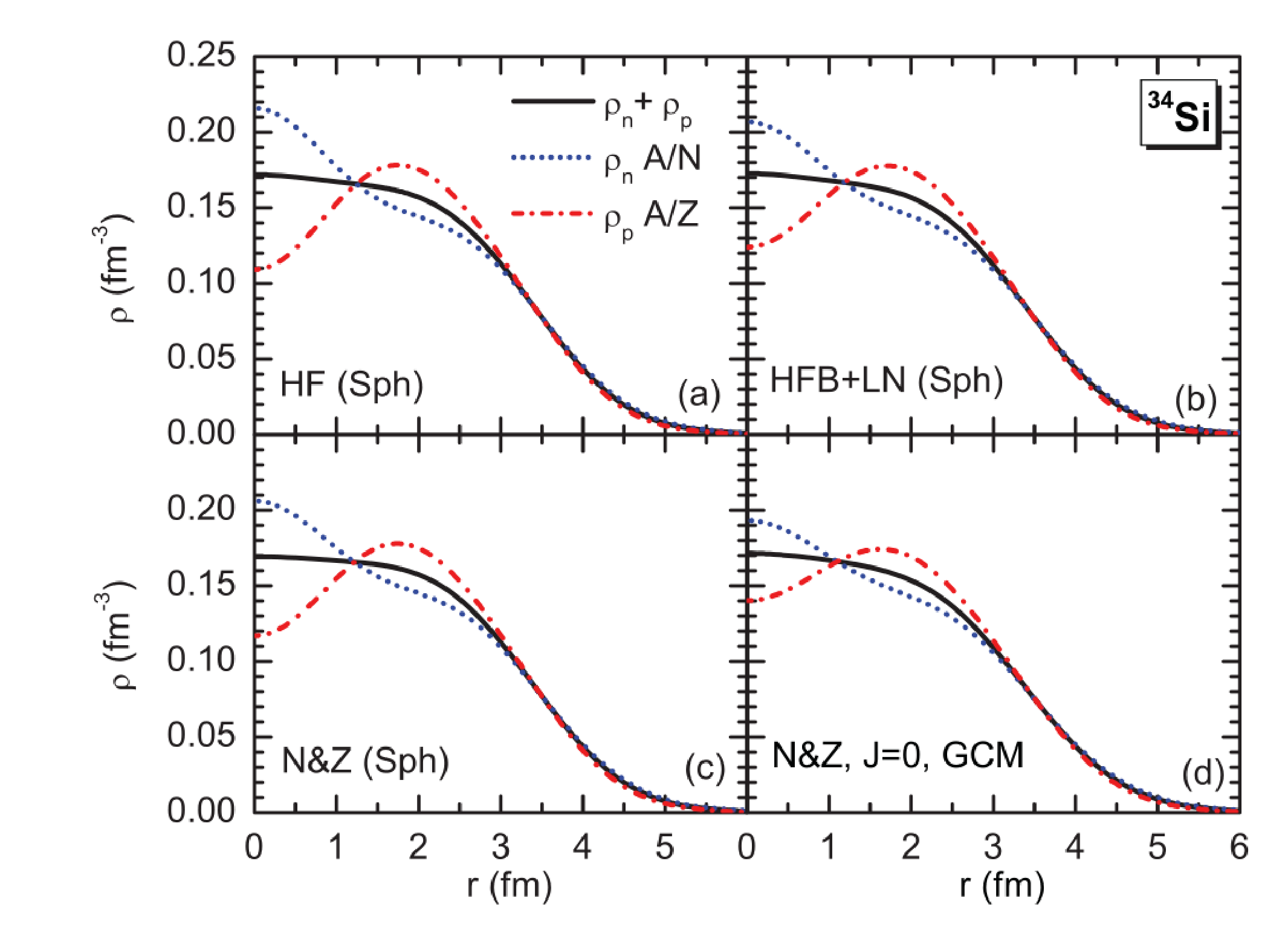} \\
\includegraphics[width=0.48\textwidth]{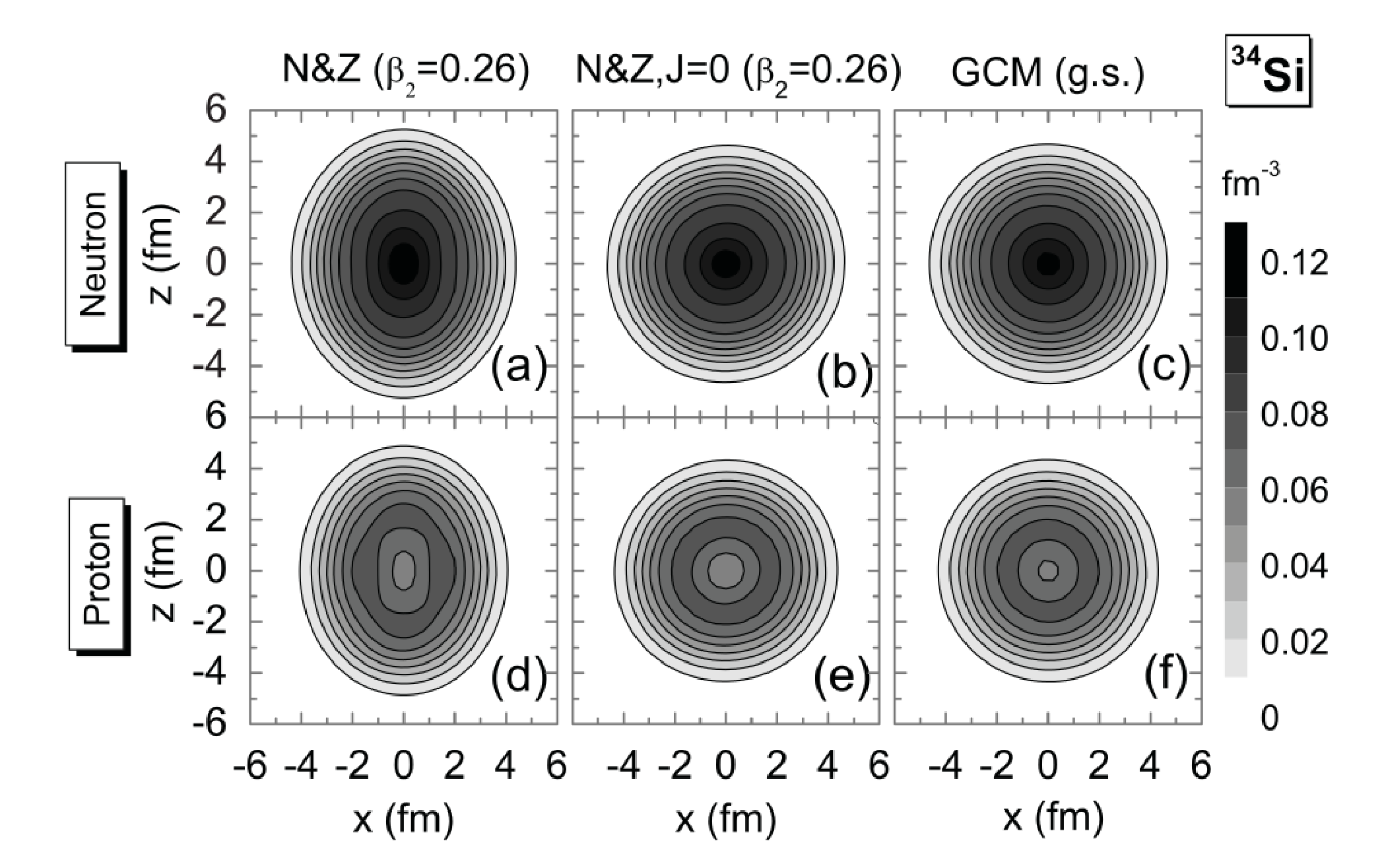}
\caption{Upper part: the proton, neutron and
total radial densities in $^{34}$Si are plotted as a function of the radius $r$ for
four different types of calculations: a) pure HF imposing spherical symmetry
b) HFB plus Lipkin-Nogami also imposing spherical symmetry c)
Particle-number projected density d) Particle-number and angular-momentum ($I=0$)
projected density obtained after  generator coordinate method calculation
using the quadrupole moment as generating coordinate.
Lower part: the contour plots of neutron and proton densities obtained
using particle-number projected wave functions obtained from a
quadrupole deformed ($\beta_2=0.26$) intrinsic state (panels a) and d));
Panels b) and e), same as panels a) and d) but for particle-number
and angular-momentum ($I=0$) projected wave functions;
the densities in panels c) and f) are obtained from generator coordinate
method wave functions projected to good number of particles and angular momentum.
Figures reprinted with permission from \protect\citeasnoun{Yao2012}.
Copyrighted by the American Physical Society.}
\label{Fig:Yao2012}
\end{center}
\end{figure}

One of the difficulties of the restoration of angular momentum is
the three-dimensional integration over the Euler angles. Often, the assumption
of axial symmetry is made to reduce the number of integrals to just one, facilitating the
application of the method, see {\Sec}~\ref{sec_rot3D}.
Axial angular-momentum projection along with particle-number projection
was first carried out in  \citeasnoun{(Val00a)} with several
parameterizations of the Skyrme EDF. This paper, together with
\citeasnoun{Hee93}, contains a detailed description of the evaluation of
operator matrix elements and different prescriptions for the density-dependent
part of the interaction. Further applications include the study of
the impact of the rotational energy correction in fission barriers \cite{Bender2004}
or the analysis of collective low lying structures in Kr  \cite{Bender2006}
or Pb isotopes \cite{(Ben04d)}. Other applications include the study
of bubble structures in $^{34}$Si \cite{Yao2012} (see {\Fig}~\ref{Fig:Yao2012})
that shows the relevance of angular-momentum projection for the description of
matter density. The intrinsic density shows a bubble structure at the center that
is much less pronounced for the density obtained from the projected ground-state
wave function.

Applications with the Gogny
D1S energy density functional include the work of \citeasnoun{(Rod00a)} where the erosion
of the $N=20$ magic number in the magnesium isotopic chain was addressed.
The physics of super-deformation in sulfur isotopes
was also analyzed \cite{RoG00k} in this framework. The calculations
were further extended  to consider the generator coordinate method with the
quadrupole degree of freedom as generating coordinate to explain more
quantitative features of the deformation of Mg isotopes \cite{(Rod00c),(Rod02a)},
the erosion of the $N=28$ shell closure
\cite{(Rod02c)} and
triple shape coexistence of neutron deficient lead isotopes \cite{(Rod04e)}.

\begin{figure}
\begin{center}
\includegraphics[width=0.58\textwidth]{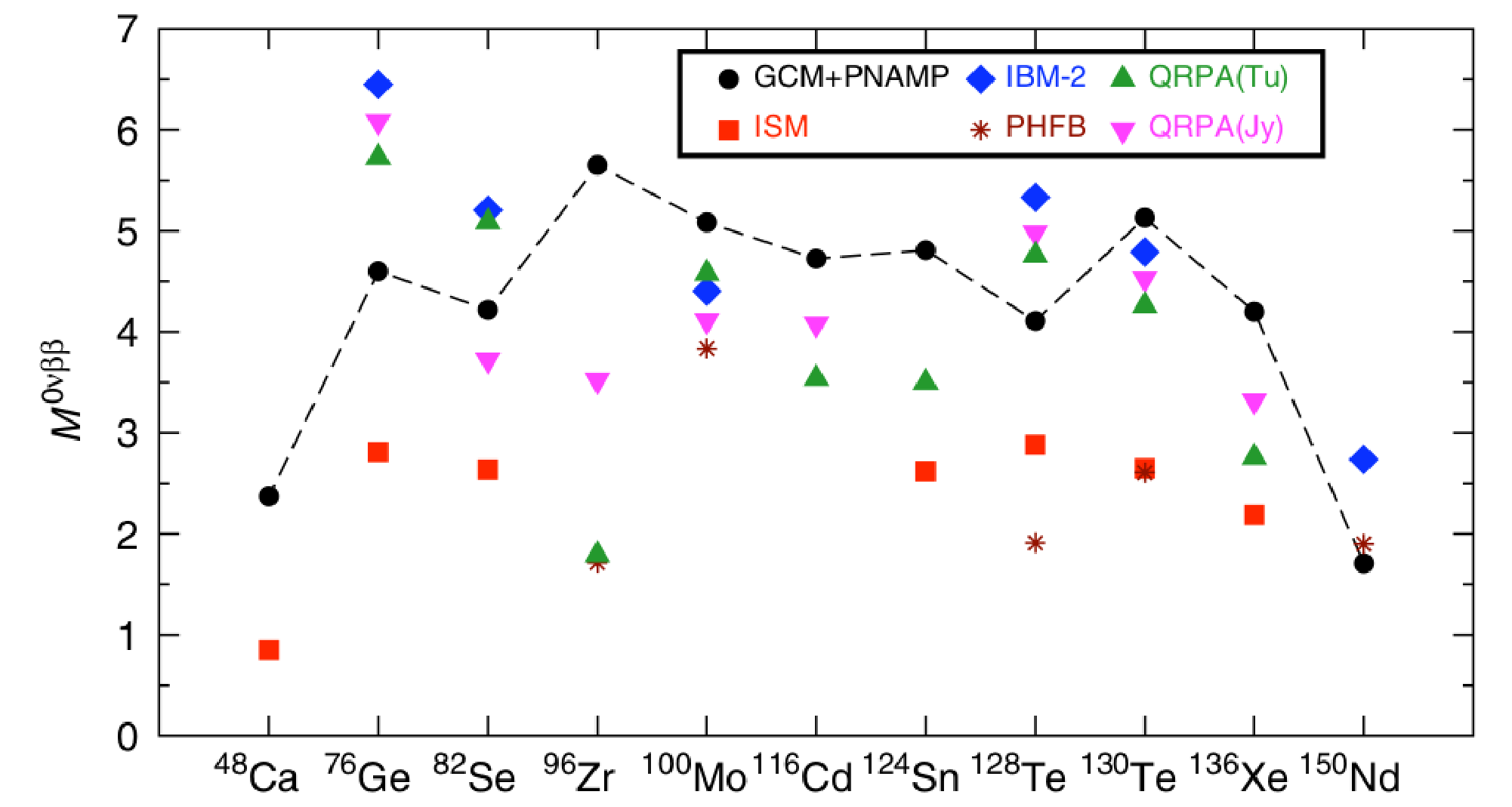}
\caption{ Neutrinoless double beta decay matrix element for various
nuclei of interest is computed with an angular-momentum plus particle-number
projected generator coordinate method calculation with the Gogny
D1S force. The results are compared with those of other approaches, showing
a large uncertainty in the theoretical predictions.
Figure reprinted with permission from \protect\citeasnoun{Rod10}.
Copyrighted by the American Physical Society.}
\label{Fig:Rod10}
\end{center}
\end{figure}

Calculation
of mass tables of even-even nuclei including angular-momentum projection
restricted to axial symmetry plus particle-number projection and including  generator coordinate method
for the axial quadrupole degree of freedom with Skyrme \cite{(Ben05),Bender2006a,Bender2008} or Gogny
forces \cite{Rodriguez2015} were presented in the literature. Other large scale studies include the
excitation energy of the collective $2^{+}$ state and its B(E2)
transition strength to the ground state \cite{Sabbey2007,Rodriguez2015}. Applications
of this methodology to the study of neutrino-less double beta decay
are essential to extract relevant nuclear matrix elements in medium mass
and heavy nuclei \cite{Rod10}, see results presented in {\Fig}~\ref{Fig:Rod10}.

Angular-momentum projection of triaxial intrinsic states was first
carried out in an early work by Caurier and Grammaticos
\cite{(Cau77a)} with the Skyrme SIII interaction. A restricted
variation after projection was used with the radii and quadrupole moments as variational
quantities.  A very small configuration space was used and
therefore the applications were restricted to very light nuclei. Quite
a few years later, \citeasnoun{BONCHE1991149} used conveniently
chosen combinations of intrinsic states oriented along different axes
to carry out an approximate projection of triaxial intrinsic
states on $I=0^+$. A full triaxial angular-momentum projection was carried out
with Skyrme interactions in \citeasnoun{(Bay84a)}, but restricted to HF wave functions.
The HF wave functions were represented on a mesh of nodes in coordinate representation
and special care was taken to define the rotation operator in that case.
In \citeasnoun{(Ben08a)} triaxial angular-momentum plus particle-number projection of HFB
states was performed with Skyrme SLy4 for light nuclei.  The main conclusion was that the inclusion of
triaxiality improves the description of rotational bands as compared
to axial results. The calculations
were subsequently extended to heavier systems in \citeasnoun{Yao2013}. Using the
finite range Gogny forces, the technology to project triaxial states was
developed in \citeasnoun{Rodriguez2010} and applied to the study of $^{44}$Si
in \citeasnoun{Rodriguez2011} and to the waiting point nucleus $^{80}$Zr in \citeasnoun{Rodriguez2011a}.

Most of the angular-momentum projected calculations so far were of the projection after variation type with intrinsic states
preserving time-reversal invariance. As a consequence, the rotational
bands obtained were stretched with respect to experiment by a typical
factor of 1.4 \cite{(Rod02c),LI-Zhipan2012_PRC86-034334}, a consequence of implicitly using the Peierls-Yoccoz
moment of inertia instead of the Thouless-Valatin one, see {\Sec}~\ref{sec:First-order}. In order to
overcome this difficulty, the use of time-reversal breaking intrinsic
states is required.

\begin{figure}
\begin{center}
\includegraphics[width=0.7\textwidth]{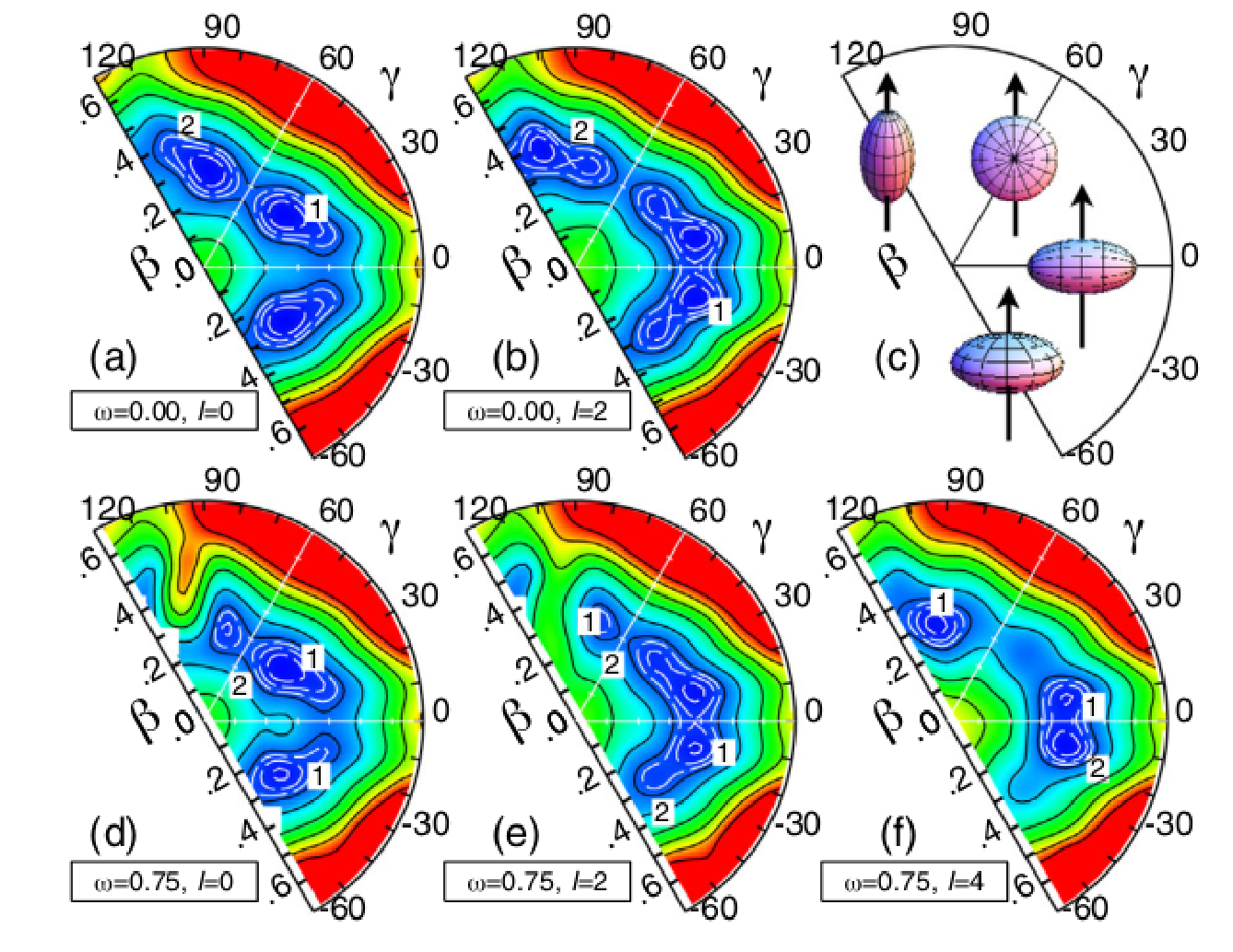}
\caption{Potential energy surfaces in the ($\beta,\gamma$) plane obtained
by computing the projected energy of each of the intrinsic states obtained
in a constrained cranked HFB calculation for different ($\beta,\gamma$)
and $\omega$ values. Those states are used in a generator coordinate
method calculation to obtain ground state and excited rotational bands in
$^{44}$S with the Gogny D1S force. The use of the cranking frequency $\omega$
as a generator coordinate improves the agreement with experiment substantially.
Figure reprinted with permission from \protect\citeasnoun{Egido2016}.
Copyrighted by the American Physical Society.}
\label{Fig:Egido2016}
\end{center}
\end{figure}

Cranked HF states were employed as early as in  \citeasnoun{(Bay84a)}
with a simple Skyrme like interaction (BKN+Coulomb).
They were also used as intrinsic states in a full three-dimensional
angular-momentum projection with the Skyrme SLy4
force to analyze (i) band termination in nuclei around
$A=44$ \cite{(Zdu07c)} and (ii) angular-momentum projection in $^{156}$Gd and $^{155}$Er
\cite{(Zdu07d)}. Later, calculations with HFB cranking intrinsic states
were carried out with the Gogny force
in \citeasnoun{Borrajo2015,Egido2016} [see also \citeasnoun{Rodriguez2016} for an
analysis of the different moments of inertia obtained in the different approaches].
In \citeasnoun{Egido2016}, collective and single-particle degrees of freedom
were studied in the nucleus $^{44}$S. Intrinsic wave functions $|\Phi (\beta,\gamma,\omega)\rangle$
of the HFB type with quadrupole deformation parameters $\beta$ and $\gamma$,
and obtained at different angular frequencies $\omega$ were used in a configuration
mixing calculation including projection on good particle number and
angular momentum (see {\Fig}~\ref{Fig:Egido2016}). The consideration of cranking states improves notably
the description of moments of inertia, whereas inclusion of states with
deformation $\gamma$ in the full interval of
$-60^\circ\leq\gamma\leq120^\circ$ allows for considering
two-quasiparticle excitation like those that are present in the
experimental spectrum of $^{44}$S.

The use of the cranking frequency $\omega$ as a generator coordinate
was considered in \citeasnoun{Shimada2015,Shimada2016}
without particle-number projection. In these calculations, the intrinsic
cranking state
was rotated along a single axis. In contrast, in \citeasnoun{Tagami2016}
intrinsic states were generated by using infinitesimal cranking frequencies
$\omega_i$ along three different cranking axes. This technique was well suited
to describe $\gamma$ bands and was applied to study these structures in
$^{164}$Er. Applications to wobbling motion in odd-$A$ nuclei and to chiral doublet bands were presented
in \citeasnoun{Shimada2018} and \citeasnoun{Shimada2018_PRC97-024319}, respectively.

\begin{figure}
\begin{center}
\includegraphics[width=0.48\textwidth]{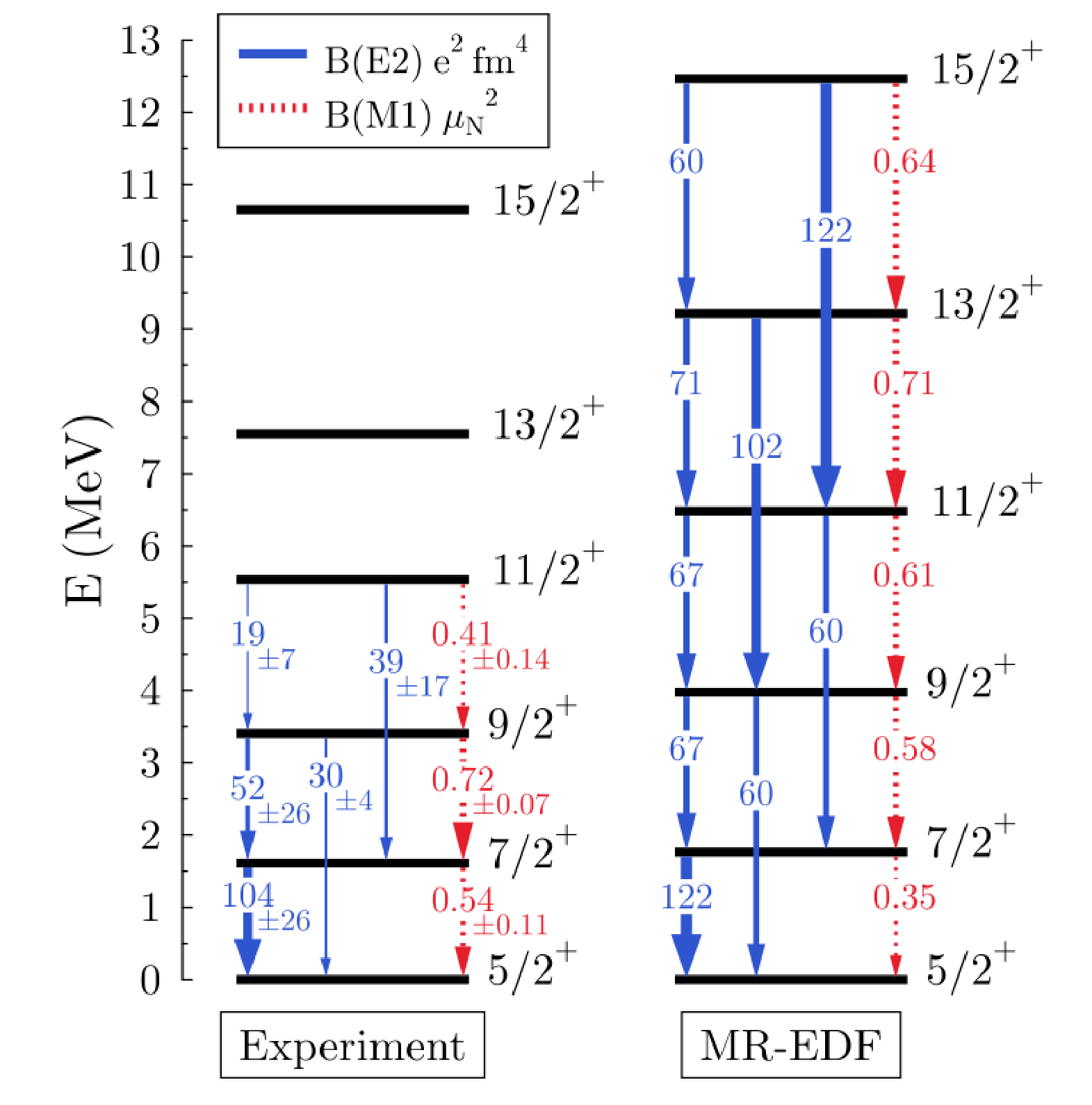}
\caption{Excitation energies and transition strengths
between the members of the ground state band of $^{25}$Mg are compared
with the experimental spectrum. The theory results are obtained from a calculation including
angular-momentum and particle-number projection of intrinsic states breaking
time reversal invariance. The density-independent SLyMR0 functional is used both
for the particle-hole and the particle-particle channels to avoid all the
difficulties associated to symmetry restoration and described in the text.
Figure reprinted with permission from \protect\citeasnoun{(Bal14d)}.
Copyrighted by the American Physical Society.}
\label{Fig:Bally2014}
\end{center}
\end{figure}

Odd mass nuclei also require  time-reversal breaking of intrinsic
states for their analysis.  In this case, the existence of zeros in the overlaps between
rotated wave functions seem to be more likely  than when time-reversal
is preserved \cite{(Oi05a),(Zdu07d)}. It is therefore to be expected
that the problems concerning self-energies and ill-behaved density-dependent forces should be
more relevant for odd mass nuclei. This consideration led \citeasnoun{(Bal14d)}
to consider for their first projected calculation of an odd mass nucleus a
Skyrme interaction which is fully derived from a Hamiltonian \cite{(Sad13d)}. The results for
$^{25}$Mg look reasonable, see {\Fig}~\ref{Fig:Bally2014}. Surprisingly, the
implementation with the density-dependent Gogny force \cite{Borrajo2015,Borrajo2018}
does not lead to any apparent inconsistency in the Mg isotopes considered. Some
results in heavier systems and for high spin states were discussed in \citeasnoun{Shimada2018}
and magnetic moments in $^{45}$Sc were determined in \citeasnoun{(deG20)}.

Isospin symmetry is explicitly broken in the atomic nucleus due to the
Coulomb interaction and the tiny differences observed in the different isospin
channels of the nucleon-nucleon interaction originating from the different mass and
charge of the $u$ and $d$ quarks. In addition, when working at the
mean field level, isospin symmetry is also spontaneously broken by the mean
field wave functions \cite{(Eng70a),(Bri70)}.  Independently
of the origin of the broken symmetry, working in a basis preserving
isospin quantum numbers is advantageous to understand the impact of the
different sources of isospin symmetry breaking in the nuclear wave
function. The first application of isospin projection in a variation-after-projection framework
was carried out in \citeasnoun{(Cau82a)} using a density-independent
interaction with a Brink-Boecker central potential and applied
to the study of Coulomb displacement energies.
Later, the formalism was applied to Skyrme functionals in
\citeasnoun{PhysRevLett.103.012502,PhysRevC.81.054310,Satula2012}
for the case of HF wave functions not mixing protons and neutrons at
the single-particle level.

The proper treatment of isospin related effects often involves
the use of EDF mixing proton and neutron densities as well as cranking
techniques in the isospin space (isocranking) \cite{(Sat13e),(She14b)}.

The traditional applications of symmetry restoration for EDF are combined
with the use of the generator coordinate method with collective continuous coordinates (deformation
parameters, pairing gaps, etc) so as to describe low-energy collective states.
An alternative is to use multi-quasiparticle excitations as discrete generating
coordinates (like in the Projected Shell Model  discussed in {\Sec}~\ref{sec:P+Q}) to gain flexibility
in generating the correlated wave function.  Both the projected shell model  and the Shell Model employ restricted
configuration spaces that require the introduction of a core as well as
effective charges. Recently, a No Core Configuration Interaction (NCCI)
method was implemented along with the density-independent SV Skyrme
interaction \cite{(Sat16d)}. The method  uses the
full configuration space, removing the need for a core and/or effective
charges. The present implementation of NCCI used p-h excitations of
Slater determinants projected to good angular momentum and isospin and was
employed to study excitation spectra of several $N\approx Z$ nuclei
as well as $\beta$-decay and exotic-processes matrix elements \cite{Konieczka2016,Konieczka2018}.

\subsubsection{Relativistic EDFs}
\label{Relativistic_EDFs}

The relativistic mean-field approach, also called Covariant Density Functional Theory,
represents an alternative approach to
describe the structure of the nucleus. Its main ingredient is the Dirac
equation, which is used to determine the nucleon orbits. The
potentials entering Dirac's equation are deduced in different ways
depending on the version of the relativistic model
used. In most of these models, the potentials experienced by nucleons
depend upon several meson fields \cite{Rein89,RING1996193}
which are determined through classical inhomogeneous Klein-Gordon
equations, where the sources are given in terms of the nucleon
densities and currents. The simplest version of this model
\cite{Walecka1974_APNY83-491} cannot reproduce the right incompressibility
of nuclear matter. Therefore a density dependence was introduced
by non-linear meson couplings \cite{Boguta1977_NPA292-413}
or by an explicit density dependence of the meson-nucleon
coupling constants \cite{Lalazissis2005_PRC71-024312}.

In deformed nuclei the classical meson fields $\phi_i({\bm r})$ are deformed. In order to
restore symmetries  the meson fields have to be quantized using bosonic creation and
annihilation operators, $b_i^\dag({\bm{r}})$ and $b_i({\bm{r}})$. The total wave function
$|\tilde{\Phi}\rangle$
of the system becomes the product of a Slater determinant $|\Phi\rangle$ in Fermion space
and a coherent state in boson space, that is,
\begin{equation}
|\tilde{\Phi}\rangle \propto |\Phi\rangle \exp\left(\sum_i\int {\rm d}^3\bm{r} \phi_i({\bm{r}})
b_i^\dag({\bm{r}})\right)|0\rangle.
\end{equation}
A variation of the corresponding energy with respect to the single-particle wave functions of
the fermions and with respect to the meson field leads to the classical
relativistic mean-field equations. For the angular-momentum projection discussed in
{\Sec}~\ref{sec:grouptheory}, one has to evaluate integrals of norm and Hamiltonian overlap kernels,
{\Eqs}~(\ref{eq:60a}) and~(\ref{eq:62}), not only in the fermion space,
but also in the boson space. Expressions pertaining to the boson space
can be found in \citeasnoun{(Bal69a)}. However, because of the numerical complexity of
the Hamiltonian matrix elements with finite-range interactions of Yukawa-type, such calculations were
not carried out so far.

A simple way to bypass these problems is to use the relativistic point coupling  models.
Here the meson propagators with the large meson masses are expanded in momentum
space up to second order in $q/m_i$, where $q$ is the momentum transfer and $m_i$
are meson masses, and one ends up with a Lagrangian without mesons, but
containing zero-range fermion interactions and zero-range derivative terms, in full
analogy to the non-relativistic Skyrme functionals \cite{(Bur02a)}. The large repulsive
density-dependent contact term of the non-relativistic case
is not needed here. However, for a good description of nuclear matter properties
a density dependence  is
introduced in the Lagrangian by either three- and four-body contact terms
\cite{(Bur02a),(Nik08a),ZHAO-PW2010_PRC82-054319} or by density-dependent coupling
constants of the two-body contact terms \cite{(Nik08a)}. This leads to coupling
constants depending on integer powers of density only, and therefore the
difficulties mentioned in {\Sec}~\ref{EDF:NonInt} related to a
non-analytical dependence on complex densities are not present.
The self-energy and self-pairing problem of {\Sec}~\ref{EDF:Self-energy} is however
present because, in practice, the Fock terms, which
are also of zero range, are usually neglected.

A much
more serious problem is, however, related to the pairing channel. Because of
the extremely large relativistic scalar and vector fields, one cannot use the same force
in the HF and in the pairing channel~\cite{Kucharek1991_ZPA339-23}. As a result,
in the relativistic point coupling models it
is a common practice to use
pairing interaction different from that in the particle-hole channel.
Therefore, the self-energy
problems associated with the violation of Pauli principle are present in these
calculations and cannot be easily avoided. The conclusion is that, as in
most of the non-relativistic cases, the results obtained by symmetry
restoration can contain some spurious contamination and their stability
with respect to the parameters of the calculations should always be carefully checked.

Point coupling models with many-body contact terms and without explicit density dependence were
used in the mean-field + BCS approximation for beyond mean-field calculations with
symmetry restoration (and configuration mixing) in \citeasnoun{(Nik06c)}, with angular-momentum
projection of axially symmetric intrinsic states, and also in \citeasnoun{(Nik06d)} with
simultaneous angular-momentum and particle-number projection.
Angular-momentum projection was carried out after variation.
The particle number was treated in the intrinsic state with the Lipkin-Nogami method (see {\Sec}~\ref{sec:LN})
and after that an exact particle-number projection was carried out. In this way it was possible to
provide a microscopic description of the $X(5)$ quantum-phase transition, which
was introduced in a group-theory model by \citeasnoun{Iachello2001_PRL87-052502}.
It describes a transition from spherical to axially symmetric deformed intrinsic shapes and
it is realized, e.g., in the chain of Nd-isotopes between the spherical nucleus $^{142}$Nd
and the axially deformed nucleus $^{152}$Nd. At $^{150}$Nd, a first order phase transition
occurs and this nucleus has a very characteristic spectrum (see {\Fig}~\ref{fig:X5}), which can be described
in the $X(5)$ model with only two phenomenological parameters.

In \citeasnoun{(Nik07)}, collective states in $^{152}$Nd were
determined within a fully microscopic relativistic mean-field calculation with subsequent
angular-momentum and particle-number projection. The entire calculated spectrum was subsequently scaled
to match the experimental $2^+$ energy, see {\Sec}~\ref{Non-rel}. The
resulting spectrum is even in a better agreement with experiment than the group-theory
spectrum, which indicates that the nucleus $^{150}$Nd is not exactly at the transition point
of the $X(5)$ model.
The transition rates were calculated in the full configuration
space without effective charges and show excellent agreement with
the experimental data.
\begin{figure}
\centering
\includegraphics[width=0.7\textwidth]{Fig_projRev_16.eps}
\caption{The particle-number-projected generator-coordinate-method spectrum of
$^{150}$Nd (left), compared with experiment (middle), and the
$X(5)$-symmetry predictions (right).
Figure reprinted with permission from \protect\citeasnoun{(Nik07)}.
Copyrighted by the American Physical Society.}
\label{fig:X5}
\end{figure}

\begin{figure}
\begin{center}
\includegraphics[width=0.7\textwidth]{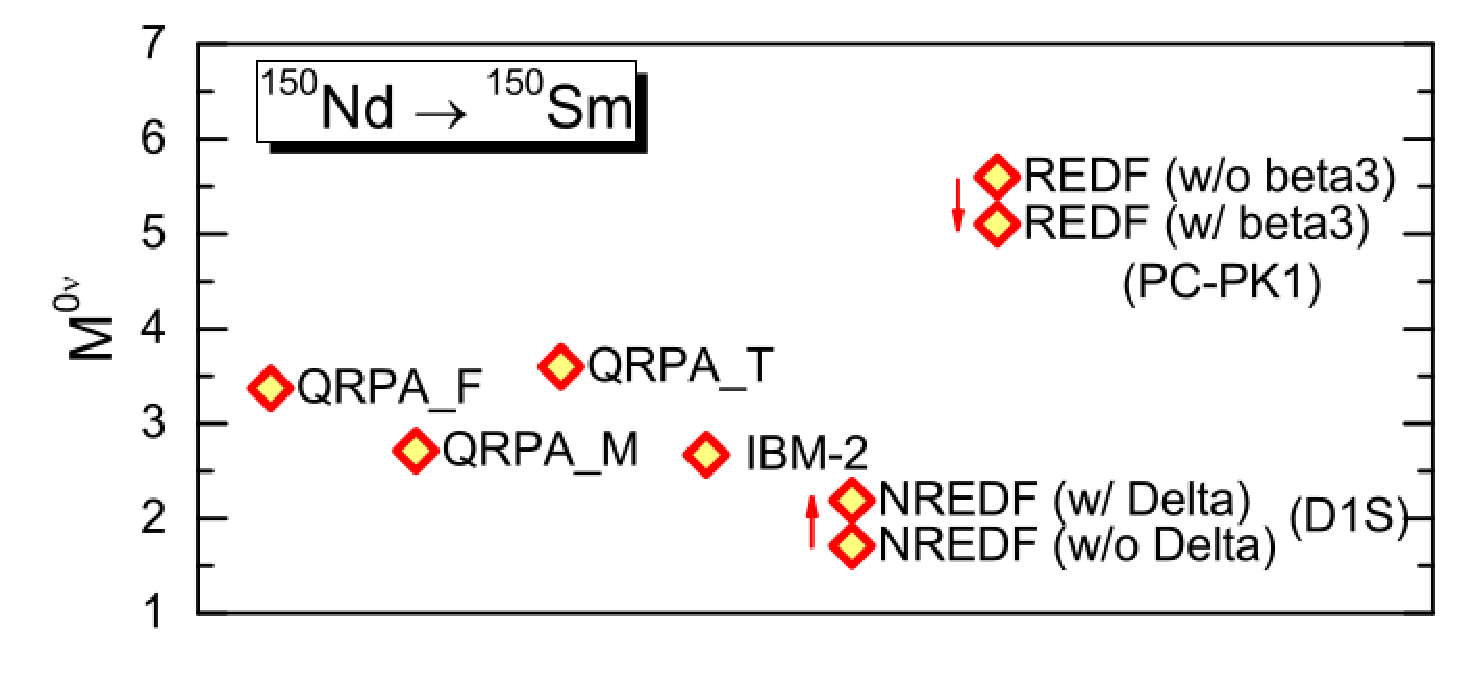}
\caption{Nuclear matrix element of the $0\nu\beta\beta$ decay of
the ground state of $^{150}$Nd into $^{150}$Sm obtained by different types of
calculations. The inclusion of octupole correlations in the relativistic-EDF
calculation has little impact on the results.
Figure reprinted with permission from \protect\citeasnoun{Yao16}.
Copyrighted by the American Physical Society.}
\label{Fig:Yao2016b}
\end{center}
\end{figure}

The model based on Covariant Density Functional Theory was further extended to
describe (i) triaxial intrinsic states \cite{Yao09,Yao10,Yao11},
(ii) reflection asymmetric states in \citeasnoun{Yao15,ZHOU-EF2016_PLB753-227,Yao2016_PRC94-011303},
(iii) admixtures of projected two-quasiparticle configurations
\cite{ZHAO-PW2016_PRC94-041301} important for
band-crossing phenomena in rotating nuclei, and
(iv) nuclear matrix elements for $0\nu\beta\beta${} decay
\cite{SONG-LS2014_PRC90-054309,Yao2015_PRC91-024316}. In \citeasnoun{Yao16}, the $0\nu\beta\beta$
decay of $^{150}$Nd to $^{150}$Sm was studied by including octupole correlations
in the description of the ground and lowest-lying $0^{+}$ collective excited
states. In {\Fig}~\ref{Fig:Yao2016b}, the result obtained for the nuclear matrix
element corresponding to the $0^{+}_{1} \rightarrow 0^{+}_{1}$ $0\nu\beta\beta${}
transition is compared with predictions including octupole correlations
and with those of other similar calculations \cite{Rod10}.
The inclusion of octupole correlations in the ground states of both
mother and daughter nuclei reduces the value of $M^{0\nu}$ by
7\%, which, however, is not enough to reduce discrepancies with
non-relativistic calculations using a similar framework, or with other
calculations using the quasiparticle random phase approximation or the interacting boson model.

\subsection{Approximate projection for nuclear EDF}
\label{EDF:approx}

\begin{figure}[htb]
\begin{center}
\includegraphics[width=0.8\textwidth]{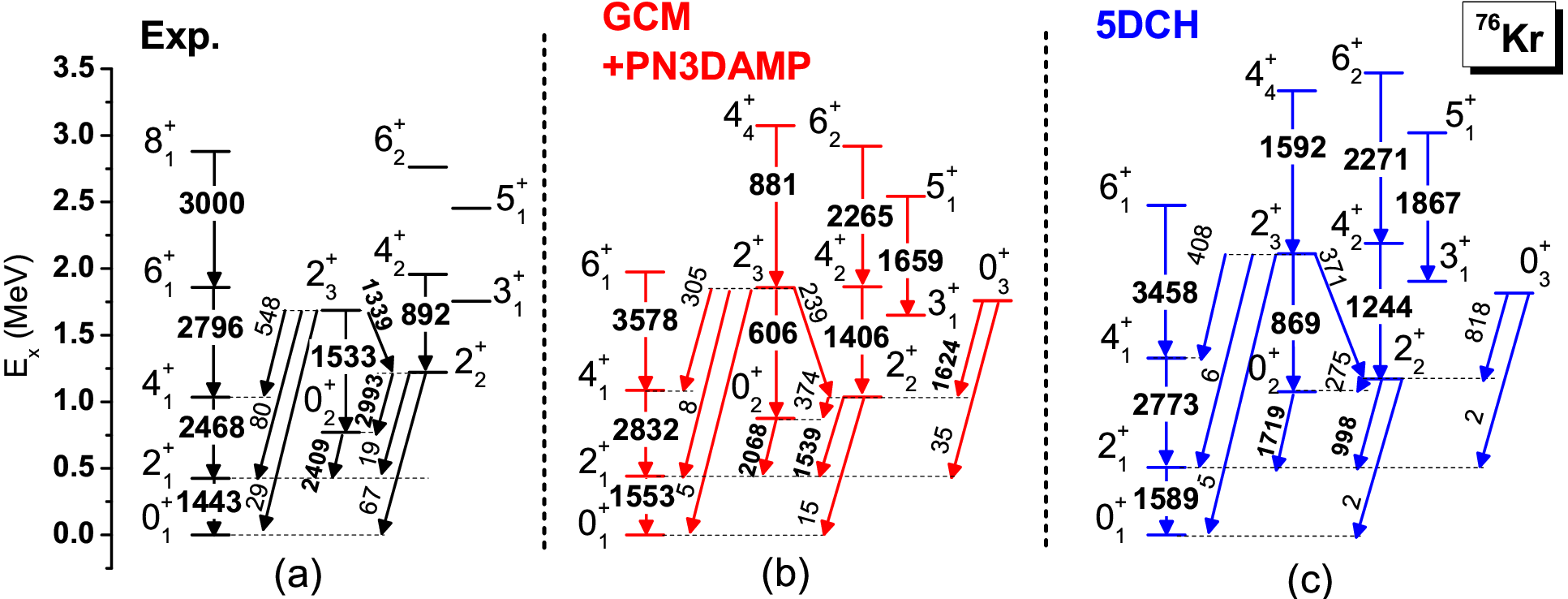}
\caption{Low-lying spectra and B(E2) values (in e$^2$ fm$^4$) of
$^{76}$Kr. Experimental data (a) are compared with the full
relativistic generator-coordinate-method calculations with particle-number and
three-dimensional angular-momentum-projection (b) and with five-dimensional Bohr Hamiltonian
results (c).
Figure reprinted with permission from \protect\citeasnoun{Yao2014_PRC89-054306}.
Copyrighted by the American Physical Society.}
\label{Fig:76Kr}
\end{center}
\end{figure}

As we discussed in {\Sec}~\ref{EDF:Applications}, practical applications of projection
and configuration-mixing methods within nuclear-EDF approaches are quite successful.
However, large configuration spaces make applications of the variation after
three-dimensional angular-momentum projection in heavy nuclei quite difficult. Many of the applications are
therefore restricted to light nuclei, where one has in principle also other methods
such as configuration-interaction calculations or coupled-cluster methods. Methods based on the mean-field
approximation are assumed to work better in heavy systems, where other methods
cannot be applied. Therefore, for heavy nuclei, approximate methods were developed for nuclear EDFs.
They are based on the fact, that the overlap
$\langle\Phi (q) |\Phi (q') \rangle$ and Hamiltonian
$\langle\Phi (q)| \hat{H}\{\rho_{q, q'}\} |\Phi (q')\rangle$
kernels between two different HFB wave functions, cf.~{\Eq}~(\ref{kernel2}),
are sharply peaked at $q=q'$.

For heavy systems, the Gaussian Overlap Approximation \cite{ring2000} is well justified.
It was shown \cite{Haff1972_PRC7-951,Giraud1974_NPA233-373}
that under this approximation one can derive a collective Hamiltonian in collective variables $q$.
It contains a potential energy $V(q)=\langle \Phi (q)| \hat{H}\{\rho_{q, q}\} |\Phi (q)\rangle$,
a kinetic term with
microscopically derived inertia parameters and zero-point corrections [for details see
\citeasnoun{Libert1999_PRC60-054301}].
In the case of three-dimensional angular-momentum projection,
one ends up with the rigid-rotor Bohr Hamiltonian \cite{Une1976_PTP55-498},
where angular momentum is automatically preserved.
For the generator-coordinate-method ansatz that includes quadrupole-deformation parameters
$\beta$ and $\gamma$, one finds in this approximation the five-dimensional rotation-vibration
Bohr Hamiltonian \cite{(Pro09)}. A similar
collective Hamiltonian can also be derived within the adiabatic
time-dependent HF theory \cite{(Bar78b)}.

The advantage of above approximation is that one only has to solve
the constrained mean-field equations on the energy surface
characterized by the parameters $q$ and to determine the expectation
values of certain operators, e.g., $\langle \Phi (q)| \hat{H}\{\rho_{q, q}\} |\Phi (q)\rangle$ or
$\langle \Phi (q)| \hat{H}\{\rho_{q, q}\} \hat{J^2}|\Phi (q)\rangle$. One avoids the complicated
matrix elements and the problem of singularities connected with
those.

As an example, in {\Fig}~\ref{Fig:76Kr}, we show the results of  benchmark calculations
by \citeasnoun{Yao2014_PRC89-054306}, where full three-dimensional angular-momentum
and particle-number projected generator-coordinate-method calculations
are compared with experiment and with the results of the corresponding
five-dimensional Bohr Hamiltonian (see also another example in
\citeasnoun{Delaroche2010_PRC81-014303}).
The agreement between the
two calculations for this complicated spectrum in the transitional
nucleus $^{74}$Kr is excellent. Having in mind, that the generator-coordinate-method
calculations for this spectrum required 200 CPU hours
with one processor, it is easy to understand that nowadays one can find
many applications based on this approximation. Unfortunately, as discussed in \citeasnoun{Rodriguez2015},
the method also has its downside related
to negative values of zero-point-energy corrections.

\section{Projection methods in other mesoscopic systems beyond atomic nuclei}
\label{sec:mesoscopic}

The last few decades witnessed extraordinary advances in experimental techniques leading to
the fabrication of mesoscopic and nanoscopic many-body systems with unparalleled control and
diversity over the finite number of constituent particles, temperature, interparticle interactions,
dimensionality, particle density, statistics (fermions versus bosons), and spin
\cite{yann06,hans07,(Ser11a),joch12,joch15,grei17,(Win87a),haef17,nogu14}. Such manmade
systems can be viewed as artificial atoms and molecules, and they offer unprecedented opportunities
for generating and observing novel and exotic many-body states and phenomena, as well as for testing
fundamental aspects of quantum physics that are beyond the reach of the natural chemical and
condensed-matter systems. These nanosystems include two-dimensional semiconductor
\cite{hans07,yann07} and
graphene \cite{yann09} quantum dots confining electrons and ultracold traps confining neutral atoms
\cite{(Ser11a),joch12,joch15,phil11} or ions \cite{(Win87a),haef17,nogu14} in a variety of trap
shapes. Among the rich physics studied in these systems, one can mention Wigner molecules
(which extend
Wigner crystals to the quantum regime), the connection to the fractional quantum Hall effect for
high magnetic fields, Aharonov-Bohm phenomena and quantum space-time crystals in ring-shaped
devices, wave function entanglement\footnote[1]{
A pure quantum state describing two or more particles is entangled
if it is unfactorizable. A mixed state is entangled if it cannot be written as a mixture of
factorizable pure states \cite{woot98}; see further \citeasnoun{(Eck02a),aspe04}.
},
Schr\"{o}dinger-cat-state superpositions in strings of ultracold ions in
linear traps, and the elucidation of the nature of correlations in assemblies of strongly repelling
electrons (long-range Coulomb interaction) or strongly interacting neutral atoms with both an
attractive or repulsive contact interaction. Areas of potential applications include quantum
information\footnote{
For increasing interest in the intersection between nuclear physics and quantum information science,
see also \citeasnoun{cloe19}} and computing, improved electronic and photonic devices,
atomic clocks, metrology, etc.

This section provides an outline of the attainments of symmetry-restoration methods in the area of
mesoscopic systems beyond atomic nuclei. For an extensive background exposition, we note an earlier
review in \citeasnoun{yann07}.

\begin{figure}[t]
\centering\includegraphics[width=7.8cm]{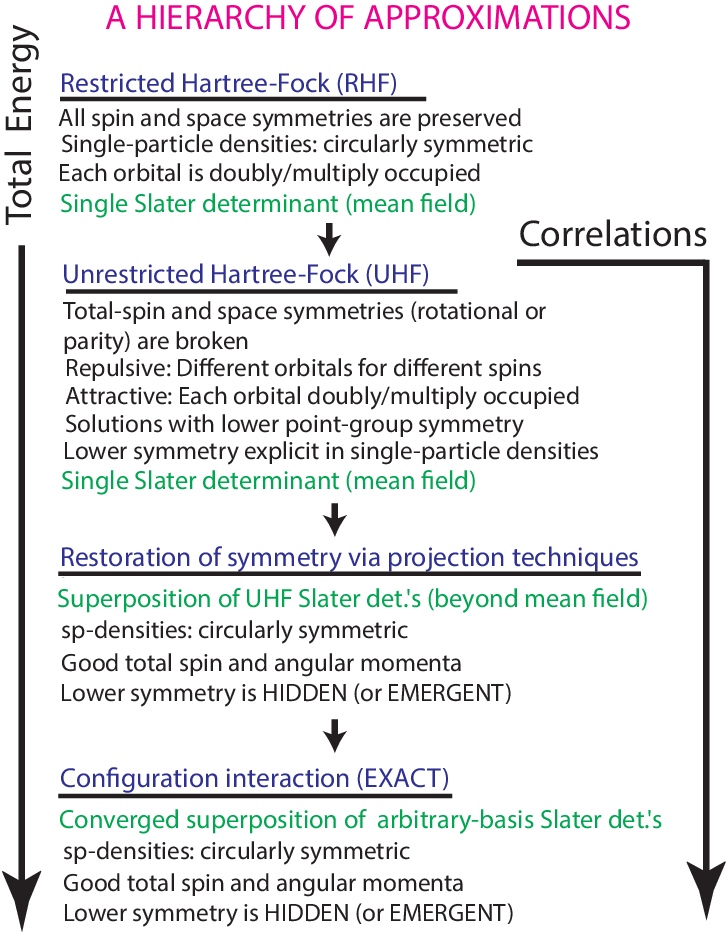}  
\caption{
Synopsis of the method of hierarchical approximations, illustrating that symmetry breaking at
the mean-field level (single Slater determinant) must be accompanied by a subsequent post-Hartree-Fock
step of symmetry restoration yielding a linear superposition of UHF Slater determinants. The downward
arrow on the left emphasizes that the total energy of the finite system is lowered with each successive step,
approaching from above the exact configuration-interaction total energy; see {\Fig}~\protect\ref{succ} below
for a simple example. The arrow on the right emphasizes that the steps beyond the Restricted Hartree-Fock
introduce correlations.
Republished with permission of Institute of Physics, from \protect\citeasnoun{yann07};
permission conveyed through Copyright Clearance Center, Inc.}
\label{hiera}
\end{figure}

\subsection{The general hierarchical methodology.}
\label{hier}

Symmetry restoration in electronic and atomic systems is a particular step in
a multilevel hierarchical scheme, which produces a lower total energy at each step;
see {\Fig}~\ref{hiera} that describes the successive levels in this
hierarchy. One starts with the restricted Hartree Fock (RHF) whose
wave function [a single Slater determinant for fermions or permanent
for bosons \cite{roma04,roma06})]
preserves all the symmetries of the many-body Hamiltonian; in particular, it imposes the
spatial symmetries of the many-Hamiltonian on each individual HF orbital.
The next level is an unrestricted Hartree Fock (UHF) whose single determinant allows for the
breaking of some (or all) of the Hamiltonian symmetries in an appropriate range of the two-body
interaction. In this case, the UHF total energy is lower compared to the RHF one, but the UHF space
orbitals do not reflect the space symmetries of the many-body Hamiltonian, a behavior that is often
referred to as "L\"{o}wdin's symmetry dilemma" \cite{lyko63}.

In a subsequent step, the broken symmetry in the UHF solutions is restored and the symmetry dilemma
is resolved. This level produces a multi-determinantal (or multi-permanent) wave function by applying
the projection-operator technique on the UHF single determinant.
This level, which is depicted as a single item in {\Fig}~\ref{hiera}, consists of two substeps, namely the step
of variation before projection and the step of variation after projection, see {\Sec}~\ref{sec_srgf}.
The variation-after-projection step produces lower energies in general, while the wave functions retain
the same multi-determinantal structure as in the variation-before-projection step. The energy difference
between these two projection variants decreases as the symmetry breaking becomes stronger. In the context
of this section, an example of the variation-after-projection step is offered by
\citeasnoun{roma04,roma06} where the localized-particle orbitals
[displaced Gaussians of {\Eq}~(\ref{uorb}) with variational parameters]
were used to build an approximate UHF Slater determinant for fermions (or
permanent for bosons).

The final level corresponds to a configuration-interaction treatment
which in principle provides the exact many-body energies and wave
functions. The RHF and UHF are mean-field approximations; the
restoration of symmetry and the configuration interaction are often
referred to as beyond-mean-field approaches.

We note here that the symmetry-restoration approaches discussed earlier
within the context of nuclear physics imply a similar hierarchy as the
one shown in {\Fig}\ \ref{hiera}.

\subsection{Quantum dots}
\label{sec:qdots_main}

Advances in nanolithography and growth techniques  enabled the fabrication of small
semiconductor devices with dimensions in the nanoscale range; they are known in the
literature as quantum dots and they play a central role in the modern field of nanotechnology.
Here we focus on two-dimensional electrostatically controlled quantum dots \cite{kouw97}.
Quantum dots are often referred to as "artificial atoms" \cite{kast93,marc98} due to
their having a discrete single-particle spectrum arising from their finite size. Such a
terminology invokes a 2D analogue of the physics of 3D electronic shells (whether closed or
open) which is associated with the Mendeleev periodic table of natural elements \cite{marc98}.

However, it was rather early realized
\cite{yann99,yann00.2,(Yan01),(Yan02b),(Yan02a)}, through UHF
calculations that in 2D quantum dots, the process of symmetry
breaking is highly operative [see also \citeasnoun{koon96}]. This is unlike
the case of natural atoms, where due to the overwhelming Coulomb
attraction from the central nucleus, the extent of spherical-symmetry
breaking is minimal \cite{fert00}. As a result, the physics of 2D
quantum dots overlaps in several ways \cite{yann07} with the nuclear
many-body problem, transposed however in the milli-eV (meV) energy
range, instead of the mega-eV (MeV) range of atomic nuclei.

\subsubsection{The microscopic many-body Hamiltonian.}
\label{mbh}

Before proceeding with the description of the many-body Hamiltonian,
a brief discussion concerning the spin-orbit coupling is informative.
As usual in atomic and molecular physics, the $L$-$S$ coupling
scheme, where $L$ stands for orbital angular momentum and $S$ for
spin, is also used for condensed-matter nanosystems and trapped
ultracold atoms. This allows that the restorations of the orbital angular
momentum and spin can be carried out independently of each other,
whereas the restoration of the combined (orbital and spin) total
angular momentum ${\bm J}={\bm L} + {\bm S}$ ($J$-$J$ coupling) is
pertinent in nuclei due to the strong spin-orbit interaction. In
contrast, the spin-orbit coupling in atomic, molecular, and
electronic systems considered here is often weak compared to the
corresponding coupling in nuclei.

The spin-orbit in condensed-matter systems (like quantum dots) is treated in two varieties: (i) the Rashba
type \cite{(Byc84a)} and (ii) the Dresselhaus type \cite{dres55}. The Rashba or Dresselhaus couplings can be
included following the steps of restoration of the spin and angular-momentum broken symmetries. For
an example of incorporating the Rashba and Dresselhaus spin-orbit couplings in the context of
two-dimensional (2D)
quantum dots, see the configuration-interaction calculations in \citeasnoun{szaf09}.

The many-body Hamiltonian describing $N$ fermions or bosons
interacting via a two-body potential $U({\bm r}_i - {\bm r}_j)$ is given by
\begin{equation}
\hat{{\cal H}} =\sum_{i=1}^N \hat{H}_{\text{sp}}({\bm r}_i,{\bm p}_i) +
\sum_{i=1}^N \sum_{j>i}^N U({\bm r}_i - {\bm r}_j),
\label{mbhn}
\end{equation}
where $\hat{H}_{\text{sp}}({\bm r}_i,{\bm p}_i)$ denotes the single-particle Hamiltonian,
which depends on the position ${\bm r}_i$ and momentum ${\bm p}_i$
of the $i$th particle. For electrons and ultracold ions, the Coulomb repulsion
is pertinent as the two-body interaction in {\Eq}~(\ref{mbhn}), namely
\begin{equation}
U({\bm r}_i - {\bm r}_j)=\frac{e^2}{\kappa |{\bm r}_i - {\bm r}_j|},
\label{clmb}
\end{equation}
where $e$ is the elementary charge (we assume single-ionized ions) and $\kappa$ is
the dielectric constant of the material in the case of semiconductor
quantum dots; for trapped ultracold ions, $\kappa=1$. Besides the familiar electrons, examples of
trapped ultracold ions are: Be$^+$, Ca$^+$, and Yb$^+$.

We assume that, in practice, the system is two dimensional and thus, for ${\bm r}=(x,y,z)$, it is
confined to the $z=0$ plane.
The single-particle Hamiltonian in a perpendicular external field ${\bm B}=(0,0,B)$ is given by
\begin{equation}
\hat{H}_{\text{sp}}({\bm r},{\bm p})=\frac{\left({\bm p}- \eta {\bm A}({\bm r})\right)^2}{2m} + V(x,y),
\label{hsp}
\end{equation}
where $m$ and ${\bm p}$ denote mass and momentum of the particle, respectively,
the external-confinement potential acting in the $z=0$ plane is denoted by $V(x,y)$,
and in the symmetric gauge, the vector potential ${\bm A}({\bm r})$ is given by
\begin{equation}
{\bm A}({\bm r})=\frac{1}{2}{\bm B} \times {\bm r} =\frac{1}{2}(-By,Bx,0).
\label{vectp}
\end{equation}
In the case of charged particles, ${\bm B}$ coincides with the natural magnetic field and $\eta=e/c$.
For electrons confined within a quantum dot,
the mass $m$ should be replaced by the effective mass $m^*$.

Unlike in nuclear physics, in finite two-dimensional systems
like quantum dots, the magnetic field plays an important role because of their relatively large spatial size.
This allows the full range of
orbital magnetic effects to be explored for magnetic fields that are readily attained in
the laboratory (less than 40 T). In contrast, for natural atoms and molecules, magnetic fields
of extremely large strength (i.e., larger than $10^5$ T) are needed to produce novel phenomena
related to orbital magnetism (beyond the perturbative regime). Such strong fields are known
to occur only in astrophysical environments (e.g., on the surface of neutron stars) \cite{ruder}.
A main orbital effect is the progressive spatial shrinking of the single-particle orbitals as
the magnetic field increases; this behavior can be directly visualized from the analytic width
$\lambda$ [{\Eq}~(\ref{lamb})] of the displaced Gaussian wave function given in {\Eq}~(\ref{uorb}).
Another orbital effect is the acquisition of a Peierls phase factor [see again {\Eq}~(\ref{uorb})].
These orbital effects are prerequisites behind the appearance of celebrated magnetic-field-dependent
phenomena, like the Aharonov-Bohm effect and the formation of quantized Landau levels supporting integer
and fractional quantum Hall effects.

To model a single circular or elliptic quantum dot, or ultracold confining trap,
or a molecule-like double well, the external-confinement potential $V(x,y)$
can assume various parametrizations. In the case of an elliptic confinement in a harmonic-oscillator
potential, one has
\begin{equation}
V(x,y) = \frac{1}{2} m (\omega_x^2 x^2 + \omega_y^2 y^2),
\label{vxy}
\end{equation}
where $\omega_x$ ($\omega_y$) is the
oscillator frequency in the $x$ ($y$) direction.
When $\omega_x=\omega_y=\omega_0$, the elliptic confinement reduces to the circular (parabolic) one.
The appropriate parametrization of $V(x,y)$ in the case of a double potential well is
more complicated. Often a parametrization based on a 2D version of a two-center oscillator
with a smooth necking is used. Details of the double-potential-well parametrization are described in
\citeasnoun{(Yan02b),yann09.2}.

\subsubsection{Mean-field equations for electrons: UHF wave functions using the
Pople-Nesbet equations or Slater determinants with displaced Gaussians.}
\label{popnes}

The UHF many-body wave function for $N$ fermions is a single Slater determinant,
\begin{equation}
\Phi_{\rm UHF}(1,\ldots,N) = \frac{1}{\sqrt{N!}}
\det[\chi_k({\bm x}_j)],
\label{psiuhf}
\end{equation}
where $\chi_k({\bm x})$ stands for one of the $k=1,\ldots,{N}$ spin orbitals, with the symbol
${\bm x}_j$ denoting both the space and spin coordinates of the $j$th particle.

We stress here that the spin orbitals are characterized by conserved
values of the projection of spin on the $z$ axis. Therefore, they are
expressed through spatial orbitals, $\varphi^\alpha_n({\bm r})$ and
$\varphi^\beta_{\bar{n}}({\bm r})$, and spinors, $\alpha$ (up) and $\beta$ (down), as
$\chi_{k=n                 }({\bm x})=\varphi^\alpha_n      ({\bm r})\alpha, n=      1,\ldots,N^\alpha$
for spin-up fermions and
$\chi_{k=\bar{n}+N^{\alpha}}({\bm x})=\varphi^\beta_{\bar{n}}({\bm r})\beta, \bar{n}=1,\ldots,N^\beta $
for spin-down fermions. Thus the UHF Slater determinants (\ref{psiuhf}) are eigenstates of the
projection $\hat{S}_z$ of the total spin on the $z$ axis with
eigenvalue $S_z=(N^\alpha-N^\beta)/2$, where $N^{\alpha}$
($N^{\beta}$) denotes the number of spin up (down) fermions. However,
except for the fully spin-polarized case of only spin-up
($N^{\alpha}=N$) or only spin-down ($N^\beta=N$) orbitals being occupied,
these determinants are not eigenstates of the square of the total
spin, $\hat{\bm S}^2$.

To specify the spin orbitals entering in the UHF Slater determinant
(\ref{psiuhf}), one usually solves the self-consistent Pople-Nesbet
equations, which are described in Chapter 3.8 of \citeasnoun{so89};
see also \citeasnoun{yann07}. To derive them, one minimizes the total
energy $\langle \Phi_{\rm UHF}(1,\ldots,N) | \hat{{\cal H}} | \Phi_{\rm UHF}(1,\ldots,N)
\rangle$ by varying the two sets of spatial orbitals
$\{\varphi^\alpha_n({\bm r})\}$ and $\{\varphi^\beta_{\bar{n}}({\bm r})\}$
under the constraint that both sets consist of orthonormal functions.
Because these two sets of spatial orbitals are allowed to be different,
the Pople-Nesbet equations are also referred to as the approach of
``different orbitals for different spins.''
We note that each UHF spatial orbital (the output of the Pople-Nesbet
equations) is allowed to break the rotational symmetry. On the other hand, for
$N^{\alpha}=N^{\beta}=N/2$ and for rotational-symmetry-conserving
spatial orbitals that are pairwise identical, $\varphi^\alpha_n({\bm
r}) = \varphi^\beta_{\bar{n}}({\bm r})\equiv\varphi_k({\bm r})$, the
Slater determinant of {\Eq}~(\ref{psiuhf}) corresponds to the RHF
approximation, see the hierarchy of approximations displayed in
{\Fig}~\ref{hiera} and Chapter 3.4 in \citeasnoun{so89}.

\begin{figure}[b]
\centering\includegraphics[width=7.3cm]{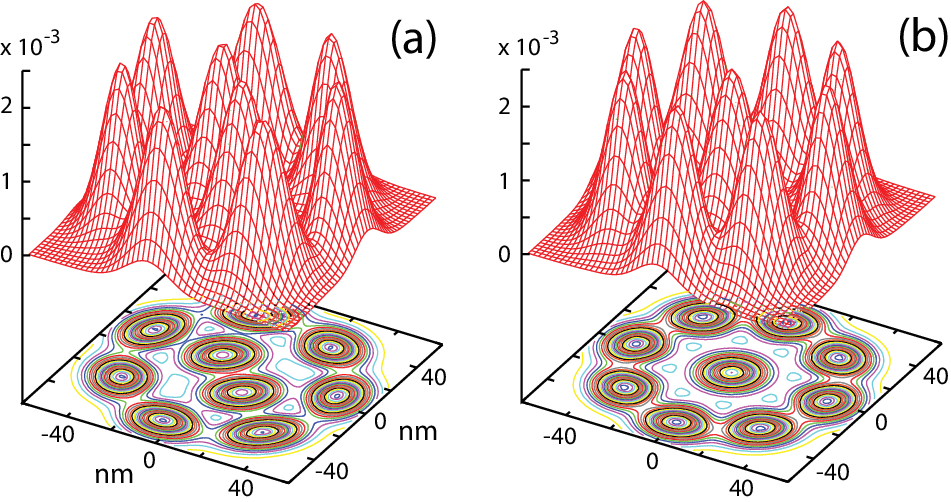}
\caption{UHF electron densities for two isomers in a parabolic quantum dot with $N=9$ electrons and $S_z=9/2$,
exhibiting breaking of the circular symmetry
at $R_W=6.365$\,nm [see {\Eq}~(\ref{rw})]
and $B=0$. (a) The (2,7) ground-state isomer with total energy 570.0093\,meV.
(b) The (1,8) first-excited isomer with total energy 570.2371\,meV.
The choice of the remaining parameters is: parabolic confinement $\hbar \omega_0=5$\,meV,
dielectric constant $\kappa=3$, and effective mass of electron $m^*=0.067 m_e$.
Distances are in nanometers and the electron density in nm$^{-2}$.
}
\label{uhfn9}
\end{figure}

An illustrative example of broken-symmetry UHF solutions is given in {\Fig}~\ref{uhfn9} for the case of $N=9$
electrons in a parabolic quantum dot at $B=0$. In the case of repulsive interactions [but also for high
magnetic fields \cite{yann07}], the symmetry breaking results in particle
localization and a lowering of the continuous
rotational symmetry to a point-group one. The localized humps in the UHF densities in {\Fig}~\ref{uhfn9} result
from the tendency of the particles to avoid each other due to their strong mutual repulsion. For high magnetic
fields, a similar localization effect is related to the shrinking of the space orbitals, as mentioned above.

Two UHF isomers of localized electrons, in the notation introduced
in {\Sec}~\ref{exh2} denoted as (2,7) and (1,8), are displayed in {\Fig}~\ref{uhfn9}.
Such nested polygonal-ring isomers are denoted in general as $(n_1,n_2,\ldots,n_r)$, with $n_r$ being the
number of localized electrons in the $r$th ring. They may compete with each other in a similar way to the
prolate and oblate nuclear shape deformations.

The localization of individual particles, revealed by using the
self-consistent Pople-Nesbet equations \cite{yann99}, suggests a
convenient and physically transparent approximation for the
broken-symmetry UHF mean-field solution. Namely, one can use a Slater
determinant $\Phi_{\rm UHF}^{\text{app}}(1,\ldots,N)$ made out of non-orthogonal
spatial orbitals having the form of displaced Gaussian functions localized at
positions ${\bm R}_j$ \cite{yann06.3}, i.e.,
\begin{eqnarray}
u({\bm r}, {\bm R}_j) =\frac{1}{\sqrt{\pi} \lambda} \exp \biggl( -\frac{({\bm r}-{\bm R}_j)^2}{2\lambda^2}
-i \psi({\bm r},{\bm R}_j; B) \biggr),
\label{uorb}
\end{eqnarray}
where $\lambda$ can be used as a variational parameter. However, for strong magnetic fields, one can fix it as
\begin{eqnarray}
  \lambda=\sqrt{\hbar/m\widetilde{\omega}} \quad\mbox{for}\quad
  \widetilde{\omega}=\sqrt{\omega_0^2+\omega_c^2/4},
\label{lamb}
\end{eqnarray}
where $\omega_c=\eta B/m$ is the cyclotron frequency, familiar for electron systems under the
influence of a magnetic field. The augmentation of the effective trap frequency from $\omega_0$ to
$\widetilde{\omega}$ expresses the associated diamagnetic behavior, which is operative due to
the large size of the quantum dot (compared to natural atoms). The phase in {\Eq}~(\ref{uorb}) is due
to the gauge invariance of magnetic translations \cite{(Pei33a),(lond37),(ditch74),(paus20)} and is given by
$\psi({\bm r},{\bm R}_j; B)= (xY_j-yX_j)/(2l_B^2)$,
with $l_B=\sqrt{\hbar /\eta B}=\sqrt{\hbar/(m\omega_c)}$ being the magnetic length.
This approximation proved to be accurate and, in addition, it bypasses the numerical effort involved in
solving the self-consistent Pople-Nesbet equations.

At zero magnetic field and for both the cases of a contact potential and a Coulomb interaction, the
resulting energy gain from symmetry breaking becomes larger for stronger repulsion.
Controlling this energy gain (the strength of correlations) is the ratio $R_\delta$ (for a contact
potential) and $R_W$ (for a Coulomb interaction) between the strength of the repulsive potential
and the zero-point kinetic energy. Specifically, for a 2D trap, one has
\begin{eqnarray}
R_\delta = gm/(2\pi\hbar^2) \;\;\; {\rm and} \;\;\; R_W=Z^2e^2/(\hbar \omega_0 l_0),
\label{rw}
\end{eqnarray}
with $l_0=\sqrt{\hbar/(m\omega_0)}$ being the characteristic harmonic-oscillator
length. [The subscript $W$ stands for ``Wigner'' \cite{wign34}.]

\begin{figure*}[t]
\centering\includegraphics[width=14.5cm]{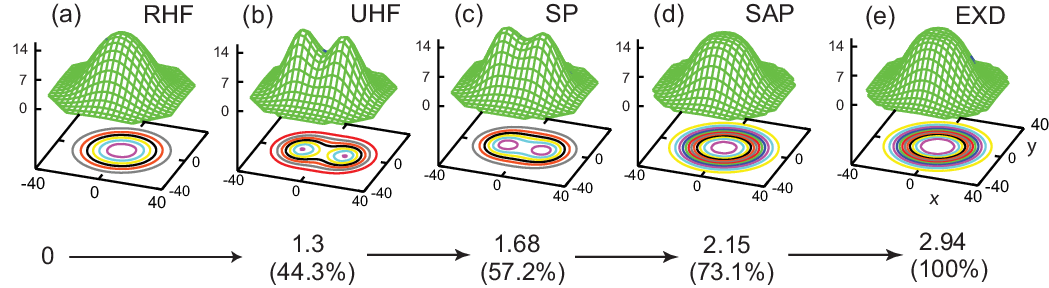}  
\caption{
Various successive approximation levels for the lowest singlet state of a field-free
($B=0$) two-electron circular quantum dot with $R_W=2.40$. The corresponding gain in
correlation energy (in meV and percentage wise) are shown at the bottom of the figure.
(a): Electron density of the RHF solution, exhibiting the circular symmetry (due to the
imposed symmetry restriction upon the HF wave functions). The correlation energy,
$E_{\rm corr} = 2.94$ meV, is defined as the difference between the energy of the RHF state
(with energy 22.74 meV) and the exact (EXD) solution [with a lower energy of 19.80 meV,
shown in panel (e)]. (b): Electron density of the symmetry-broken ``singlet''
UHF solution exhibiting non-circular shape. The energy
of the UHF solution shows a gain of 44.3\% of the correlation energy.
(c): Electron density of the spin-projected
singlet (SP) showing broken spatial symmetry, but with an additional gain
in correlation energy. (d): the spin-and-angular-momentum projected state
(SAP) exhibiting restored circular symmetry with a 73.1\% gain of the
correlation energy. The choice of parameters is: dielectric constant
$\kappa = 8$, parabolic confinement $\hbar \omega_0 = 5$ meV, and effective
mass $m^* = 0.067m_e$. Distances are in nanometers and the densities in
$10^{-4}$ nm$^{-2}$.
Republished with permission of Institute of Physics, from \protect\citeasnoun{yann07};
permission conveyed through Copyright Clearance Center, Inc.}
\label{succ}
\end{figure*}

\subsubsection{Combining spin and total-angular-momentum restorations.}
\label{resc}

When the fermions are not fully polarized, the symmetry-broken
Pople-Nesbet UHF determinantal solutions do have the total spin projection $S_z$ as good quantum numbers.
However, the total spin $\hat{\bm S}=\sum_{i=1}^N \hat{\bm s}_i$ is not preserved. A simple example is the
UHF Slater determinant which describes the $S_z=0$ ground state of two electrons in a parabolic quantum dot
for $R_W=2.40$ (and $B=0$). {\Fig}~\ref{succ}(a) displays the azimuthally symmetric RHF electron density,
which contrasts with the symmetry-broken UHF one displayed in {\Fig}~\ref{succ}(b).
The associated UHF determinant is given by {\Eq}~(\ref{det}) in {\Sec}~\ref{exh2}.

The next step in restoring the total spin is described in detail in
{\Sec}~\ref{exh2}, resulting in a $s=0$, singlet state given by {\Eq}~(\ref{gvb});
it is a superposition of two UHF Slater determinants. We note that the
spatial reflection symmetry (parity) is automatically restored along
with the spin symmetry \cite{fuku81}.

For the exact singlet state of the circular confinement, one can
generate approximate projected wave functions with good total angular
momentum, exhibiting the required azimuthally-uniform electron
densities, by applying the product operator,
\begin{equation}
{\cal O} \equiv \hat{P}_L \hat{P}^s_{\rm spin}~.
\label{cpr}
\end{equation}
The singlet spin-projection operator $\hat{P}^s_{\rm spin}$ (\ref{prjp2}) produces the anisotropic singlet
wave function of {\Eq}~(\ref{gvb}), and then the total-angular-momentum projection operator,
\begin{eqnarray}
\hat{P}_L = \frac{1}{2\pi} \int_0^{2\pi} e^{i \gamma (L-\hat{L})} d\gamma,
\label{prjop}
\end{eqnarray}
acts upon this two-determinant wave function to restore the total 2D angular momentum $L$.\footnote{
The description of the general projection-operator formalism regarding symmetry restoration is
presented in {\Sec}~\ref{sec_srgf}. See also Tables~\ref{tab:table1} and~\ref{tab:table2},
where the nuclear-style notation $\hat{J}_z$ is used for the 2D
angular momentum $\hat{L}$, commonly used in the condensed-matter and atomic-physics literature. Here we
repeat the definition in {\Eq}~(\ref{prjop}) for clarity and for the convenience of the reader. Note that "1D"
in Tables~\ref{tab:table1} and~\ref{tab:table2} refers to the dimension of the integral,
and not to the geometry of the physical system.
For a full exposition of the adaptation of the general formalism to the 2D electronic and atomic systems,
see \citeasnoun{yann07}.}
In {\Eq}~(\ref{prjop}), $\hat{L}=\sum_{i=1}^N \hat{l}_i$, $i=1,2,\ldots,N$, $\hbar \hat{L}$
is the two-dimensional total angular-momentum operator, and $\gamma$ is the azimuthal angle.
This double projection describes all the lowest-energy states [yrast band \cite{yann00}]
with good total angular momentum $L=0, 2, 4,\ldots$. (The yrast-band states with odd values,
$L=1, 3, 5,\ldots$, are generated via a projection of the fully-polarized UHF state.)

The evolution of the ground-state $(L=0)$ electron densities
according to the successive approximations, RHF, UHF, spin projection
(SP), and combined spin-and-angular-momentum projection (SAP)
is illustrated in {\Fig}~\ref{succ}. The exact wave functions for two
electrons in a parabolic confinement are available \cite{yann00}, and
the corresponding ground-state electron density is plotted in
{\Fig}~\ref{succ}(e). The successive lowering of the ground-state total
energies is also displayed.

\subsubsection{More on spin restoration.}
\label{sec:moron}

The literature of spin restoration in systems other than nuclei has a
more complicated history compared to that of 2D angular momentum.
L\"{o}wdin introduced a spin projection operator through the expression \cite{(Low55a)}
\begin{equation}
\hat{P}_{\rm spin}(S) \equiv \prod_{s^\prime \neq S}
\frac{\hat{\bm S}^2 - s^\prime(s^\prime + 1)}
{S(S+1) - s^\prime(s^\prime + 1)},
\label{prjp}
\end{equation}
where the index $s^\prime$ runs over the quantum numbers associated with
the eigenvalues $s^\prime(s^\prime+1)$ of $\hat{\bm S}^2$
(in units of $\hbar^2$), with $\hat{\bm S}$ being the total spin operator.
When applying $\hat{\bm S}^2$ on a Hartree-Fock determinant, one uses:
\begin{equation}
\hat{\bm S}^2 \Phi_{\rm UHF}=\hbar^2
\left[ (N_\alpha-N_\beta)^2/4+N/2+\sum_{i<j}\varpi_{ij} \right] \Phi_{\rm UHF},
\label{s2hf}
\end{equation}
where $\varpi_{ij}$ interchanges the spins of electrons $i$ and $j$ provided that they are different;
$N_\alpha$ and $N_\beta$ denote the number of spin-up and spin-down electrons, respectively, while
$N=N_\alpha+ N_\beta$.

For a large number of electrons,
a computationally practical implementation of L\"{o}wdin's spin projection formalism
was recently discussed in \citeasnoun{vive18}.

The operator $\hat{P}_{\rm spin}(S)$ was used \cite{(Yan02b),cava07} to describe the energy spectra
and wave functions for electrons in 2D quantum dots. We note that, for $N \geq 3$, there
are multiple spin eigenfunctions for a given value $S$ of the total spin, and this multiplicity
is important for obtaining a complete set of excited states (e.g., the yrast band), in addition to the
ground state. The spin multiplicities are tabulated in the so-called branching diagram
\cite{yann09.2,paun00,yann16,(salm74)}. For example, for $N=3$ fermions,
there are two spin eigenfunctions with $S=1/2$. For a spin projection $S_z=1/2$ and using the
notation ${\cal S}(S,S_z;i)$ (where the index $i$ is employed for the degeneracies),
a pair of basis spin eigenfunctions that spans the associated two-dimensional spin space is given by
\begin{equation}
\sqrt{6} {\cal S} (\mbox{$\frac12$},\mbox{$\frac12$};1) =
2 | \uparrow
 \downarrow
 \uparrow \;\rangle
- | \uparrow
 \uparrow
 \downarrow \;\rangle
- | \downarrow
 \uparrow
 \uparrow \;\rangle,
\label{wf3e12121}
\end{equation}

\begin{equation}
\sqrt{2} {\cal S} (\mbox{$\frac12$},\mbox{$\frac12$};2) =
| \uparrow
 \uparrow
 \downarrow \;\rangle
-| \downarrow
 \uparrow
 \uparrow \;\rangle.
\label{wf3e12122}
\end{equation}

For $N>3$ electrons, the complete set of spin eigenfunctions can be specified by solving a Heisenberg
Hamiltonian $H_H=\sum_{i,j} J_{ij} \hat{\bm S}_i \cdot \hat{\bm S}_j$ \cite{yann09.2,yann16}, whose
purely-spin solutions are given as a superposition of spin primitives
$| \sigma_1 \sigma_2 \ldots \sigma_N \rangle$,
where $\sigma$ stands either for $\alpha$ (spin-up) or $\beta$ (spin-down). Then using the fact that
each electron is associated with a localized space orbital, one can generate a corresponding Slater
determinant of spin orbitals from each spin primitive, thus generalizing \cite{jain07,yang07} the
two-electron Heitler-London-type expression of {\Eq}~(\ref{gvb}).

Naturally, the total-spin restoration\footnote{In the context of a generalized UHF with a broader
set of unrestrictions, restoration of both the total spin $\hat{\bm S}$ and its projection
$S_z$ (in the case that the mean-field wave functions break both of these symmetries) has
also been discussed in \citeasnoun{fuku81}.}
can also be performed \cite{yann07,fuku81,hash82,igaw95} by using the
projection-operator formulas that restore the three-dimensional (3D) total angular
momenta, see {\Sec}~\ref{sec_rot3D}.

\subsubsection{Molecular symmetries of the UHF wave functions and magic angular momenta.}

The 2D projected wave functions have good total angular momentum $L$, and the
corresponding single-particle density is circular and azimuthally uniform. Thus any association of the
projected wave functions to a point-group symmetry
(which corresponds to a single-particle density that is not azimuthally uniform) is counterintuitive.
Despite this expectation, the projected trial wave functions do embody and reflect hidden (or
emergent) molecular point-group symmetries similar to the case of natural molecules. Specifically, the $C_N$
point-group symmetry of the ``classical'' crystalline configuration, which is accounted
by the symmetry-broken mean-field determinants $\Phi_{\rm UHF}(1,\ldots,N)$, {\Eq}~(\ref{psiuhf}), is
reflected in the fact that the symmetry-restored wave functions $\Psi^{\rm PRJ}_L$ are identically zero except for a
subset of {\it magic\/} angular momenta $L_m$.
Of course this vanishing of wave functions is not present for the exact configuration-interaction
ones with non-magic angular momenta. In the context of the symmetry-restoration approach, correspondence
with the full set of configuration-interaction wave functions is established by considering the vibrations
of the molecular configurations. This additional step was described in \cite{yann10,yann11}; see also
\cite{yann00}. An analogous situation in nuclear theory is provided by the generator-coordinate
method, see {\Sec}~\ref{GCM}, which is a generalization of the strict symmetry-restoration approach via
projection techniques.

\begin{figure}[t]
\centering\includegraphics[width=7.5cm]{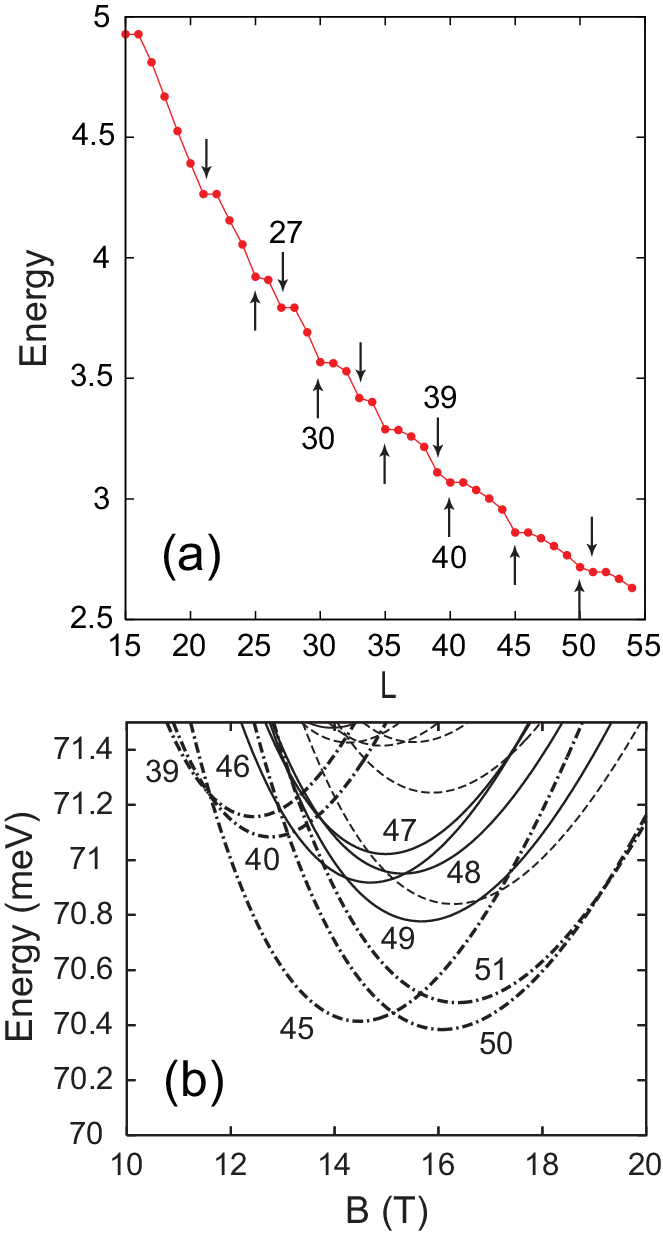}  
\caption{
(a) Exact-diagonalization yrast-band energies for $N=6$ lowest-Landau-level electrons as a function
of the total angular momentum (with $15 \leq L \leq 55$). Only the Coulomb term was retained
in the many-body Hamiltonian. The cusp states are marked by arrows;
they correspond to either a (1,5) (upward arrows) or to a (0,6) (downward arrows) Wigner-molecule
ring \cite{yann03.2,yann07}. Energy in units of $e^2/{\kappa l_B}$.
(b) The associated global spectrum (when the effect of the external confinement, with
$\hbar \omega_0 = 3.6$ meV, is included) as a function of the magnetic field $B$ (in units of Tesla).
The cusp states in (a) with $L=39$ (0,6), 40 (1,5), 45 (1,5), 50 (1,5), 51 (0,6) are associated with the
energy curves (marked by the $L$ values) in thick dashed-dotted lines.
Parameters: $\kappa=13.1$ and $m^*=0.067 m_e$.
Figure adapted with permission from \protect\cite{yann11}.
Copyrighted by the American Physical Society.}
\label{cusp}
\end{figure}

For the simpler case of $N$ repelling fermions with parallel spins on a ring [i.e., with $S_z=N/2$ and a
$(0,N)$ configuration], the magic total
angular momenta can be determined by considering the point-group symmetry operator
$\hat{R}(2\pi/N) \equiv \exp (-i 2\pi {\hat L} /N)$ that
rotates on the ring simultaneously the localized particles by an angle $2\pi/N$.
In connection to the state $\Psi^{\rm PRJ}_L$, the operator $\hat{R}(2\pi/N)$ can be invoked in two different
ways, namely, either by applying it on the ``intrinsic'' (symmetry-broken) wave function
$\Phi_{\rm UHF}(1,\ldots,N)$ or on the phase factor $\exp(i \gamma L)$ appearing in the ``laboratory'' (symmetry-restored)
wave function $\Psi^{\rm PRJ}_L$ of {\Eq}~(\ref{prjop}), see Chapter 4-2c in \citeasnoun{bomo98}.
One gets
\begin{equation}
\hat{R}(2\pi/N) \Psi^{\rm PRJ}_L = (-1)^{N-1} \Psi^{\rm PRJ}_L,
\label{rot1}
\end{equation}
from the first alternative and
\begin{equation}
\hat{R}(2\pi/N) \Psi^{\rm PRJ}_L = \exp(-2\pi L i/N) \Psi^{\rm PRJ}_L,
\label{rot2}
\end{equation}
from the second alternative. The $(-1)^{N-1}$ factor in {\Eq}~(\ref{rot1}) results from the fact that the
$2\pi/N$ rotation is equivalent to exchanging $N-1$ rows in the determinant $\Phi_{\rm UHF}(1,\ldots,N)$.
Now, if
$\Psi^{\rm PRJ}_L \neq 0$, the only way that {\Eqs}~(\ref{rot1}) and~(\ref{rot2}) can be simultaneously
valid
is if the condition $\exp (2\pi L i/N)=(-1)^{N-1}$ is fulfilled. This leads to the following sequence of magic
angular momenta,
\begin{equation}
L_m = k N; \;\;\; k=0,\pm 1, \pm 2, \pm 3, \ldots,
\label{mag1}
\end{equation}
for $N$ odd, and
\begin{equation}
L_m = (k + \tfrac{1}{2}) N; \;\;\; k=0, \pm 1,\pm 2,\pm 3, \ldots,
\label{mag2}
\end{equation}
for $N$ even.

The physics associated with magic-angular-momentum yrast states was extensively explored in the
literature of 2D quantum dots \cite{yann07,yann06.4,yann03.2,ruan95,maks96,seki96,maks00}.
An important property is the enhanced energy stabilization (compared to the rest of the spectrum as described
by configuration-interaction calculations) that they acquire in their neighborhood in the regime of
strong interactions (i.e., for large $R_W$, $R_\delta$, see {\Eq}~(\ref{rw}), or for large magnetic fields).

In the case of strong magnetic fields, when the many-body Hilbert space can be restricted to the lowest Landau
level, it is customary to calculate the configuration-interaction energy spectra keeping only the interaction
term in the many-body Hamiltonian. In such partial spectra, the magic angular momenta are associated with the
so-called ``cusp'' states \cite{yann03.2,yann04,jainbook} [see {\Fig}\ \ref{cusp}(a)], which are precursors of
the fractional quantum-Hall-effect bulk states \cite{yann03.2,yann04,jainbook}.
When the single-particle part of the Hamiltonian is also included (i.e., kinetic energy plus external
confinement), one obtains global energy spectra as a function of the magnetic field $B$; see
{\Fig}~\ref{cusp}(b). In these global spectra, all the ground states correspond to cusp states,
illustrating the
role played by the magic angular momenta in enhancing energy stabilization.

For magnetic-field-free systems, this energy stabilization leads to a separation of energy scales between the
rotational the vibrational motions (formation of a near-rigid rotor), which is a familiar prerequisite in the
formulation of nuclear effective field theories \cite{pape15}. An example of such a separation of energy scales
is portrayed in {\Fig}~1
of \citeasnoun{yann00} for the case of two electrons in a parabolic 2D quantum dot.

In the above derivation, we considered fully polarized fermions only, that is cases when $S=S_z=N/2$, where $S$
is the total spin and $S_z$ is its projection. Consideration with this methodology of the other spin values
$S_z < N/2$ is straightforward; it requires, however, restoration of both the total spin ${\bm S}^2$
and the total angular momentum. An explicit example for $N=3$ fermions is discussed in \citeasnoun{yann03}.

\subsubsection{A tour of the literature and Wigner molecules.}
\label{sec:dots2}

Naturally, there are several key differences between the physics of 2D quantum dots and that
of atomic nuclei, which arise from the fact that the inter-particle interaction in quantum dots
is repulsive, instead of attractive as in nuclei,
and that quantum dots consist of one kind of fermions (electrons), instead of two kind of particles
(protons and neutrons).
As a result, symmetry breaking in quantum dots is associated with individual-electron localization
in space in the
intrinsic frame (leading to formation of mean-field crystalline configurations), rather than the familiar shape
deformations of the
nuclear central mean-field confining potential.

Such mean-field crystalline configurations in quantum dots
[see, e.g., {\Fig}~\ref{uhfn9}] are referred to as "Wigner molecules" \cite{yann99}. After restoration of the
angular-momentum symmetries, they are often referred to \cite{yann04.2}
as "rotating Wigner molecules"\footnote{
The rotating-Wigner-molecule wave functions are stationary, exhibiting a time-independent single-particle
density. The term "rotating" here refers to these wave functions having good quantal total
angular momenta, unlike a symmetry-broken crystalline UHF wave function. Time-evolution of
wave packets formed through the superposition of several rotating Wigner molecules \cite{yann17} reintroduces
(beyond mean field) symmetry breaking and leads to the concept of a quantum space-time crystal \cite{wilc12,li12,yann17}
and to phenomena of quantum-mechanical revival \cite{seid99,yann17,kavo18}; see also \citeasnoun{manz18,manz19} for analogous
superpositions that break and restore symmetries, experimentally achieved in natural molecules and atoms.},
exhibiting two special cases of "rotating electron molecules" \cite{yann03.2} or "rotating boson molecules"
\cite{roma06}. Localized corpuscular patterns or cluster structures arise also in symmetry-broken mean-field single-particle densities
of lighter nuclei \cite{rein12,schu13,ebra13,(Ebr17)}; they are, however, associated with $\alpha$-particle
multi-nucleon clustering.

Starting with the early 2000's, the two-step methodology that
combines symmetry breaking with subsequent symmetry restoration was
employed extensively to investigate the physics of quantum dots. In
particular, using the L\"{o}wdin projection for restoring the total
spin, \citeasnoun{(Yan01)} investigated the coupling and dissociation
of two electrons in a double-well confinement (artificial H$_2$
molecule), see {\Sec}~\ref{exh2}.
In addition to the spin restoration, the formation of a two-electron
rotating Wigner molecule was described in \citeasnoun{(Yan02b)} by restoring
simultaneously the total angular momentum in the
case of two electrons confined in a parabolic (circularly symmetric)
single-well quantum dot.

For the case of zero or low magnetic fields and using the two-step
method, subsequent literature studied a larger number of electrons in
parabolic quantum dots (in the range of $3 \leq N \leq 10$). An explicit
demonstration that the projected ground-state wave function has a
lower energy compared to the UHF one was given in \citeasnoun{mikh02}
for $N=2-8$ fully spin-polarized electrons. For $N=3$ electrons, a
detailed analysis of the lower point-group symmetries of the UHF
broken-symmetry molecular solutions and their influence upon the
angular-momentum-restored wave functions was also carried out
\cite{yann03}. The richness of the physics embodied in the projected
wave functions was illustrated in \citeasnoun{yann04.2}, where it was
shown that the rotating Wigner molecule can attain two opposite
limits depending on the
parameters of the system. Namely the limit of a rigid 2D rotor is
reached for strong Coulomb repulsion (e.g., $R_W = 200 >> 1$) in the
absence of an applied magnetic field; the rotational spectrum (yrast
band) in this case exhibits energy levels $\propto L^2$. An opposite
limit of a hyper floppy rotor is reached for smaller $R_W \sim 10$,
but very high magnetic field (the lowest-Landau-level regime); in this case the
rotational energies (yrast band) have an ${\cal A} L +{\cal
B}/\sqrt{L}$ dependence on the total angular momentum $L$. The limit
of a 2D rigid rotor for $R_W \rightarrow \infty$ and low magnetic
field was also demonstrated for the case of $N=9$ and $N=8$ ultracold
ions confined in a 2D ring-shaped trap \cite{yann17}. The limit of
the 2D rigid-rotor rotational spectrum extracted in the papers above
is reminiscent of the projected-energy Kamlah expansion\footnote{The Kamlah
expansion needs to be used in conjunction with the $C_N$
Wigner-molecule lower symmetry; otherwise \cite{koon96} the
multifaceted effects originating from the magic angular momenta are
missed. Moreover in the lowest-Landau-level regime, use \cite{koon96} of the Kamlah
expansion cannot reproduce the $1/\sqrt{L}$ energy component
characteristic of the hyper-soft rotor \cite{yann04.2,yann06.3}.} in
integer powers of $L$ for strong symmetry breaking in rotating nuclei
[see {\Sec}~\ref{sec:approximate} and Chapter 11.4.4 of \citeasnoun{ring2000}]; in the present cases,
however, only the dominant term $\propto L^2$ survives for $R_W
\rightarrow \infty$.

Using broken-symmetry UHF solutions and following the L\"{o}wdin
prescription for the total-spin projection [see {\Eq}~(\ref{prjp})] in
connection with the construction of spin eigenfunctions presented in
\citeasnoun{smith64}, the combined restoration of both total-spin and
angular-momentum approach was applied in a systematic investigation
\cite{cava07,cava08} at zero and low magnetic field $B$ of the
properties of 2D parabolic quantum dots with up to $N=12$ electrons.
In particular for $B=0$, it was confirmed that Hund's rules apply for
weaker interaction with $R_W \leq 2$; for stronger interaction ($R_W
> 4$), Hund's rules are violated signaling the dominance of a strong
Wigner molecule \cite{yann99}.

For completeness, we mention that collective modes associated with
the spurious RPA states were used
to restore the broken rotational symmetry of UHF solutions in
parabolic quantum dots. The case of $N=2$ electrons was
systematically studied \cite{serr03,birm13}. This RPA-based approach,
however, becomes computationally prohibitive for larger $N$, due to
the increasing number of RPA modes that are required.

The symmetry-restoration methodology
was also successfully used to describe aspects of the many-body
physics of few electrons in the lowest Landau level. This level forms at
very large magnetic fields $B \rightarrow \infty$, and it consists
exclusively of all single-particle levels $\propto r^\ell e^{i\ell\phi}
e^{-r^2/2\lambda_c^2}$ with zero radial nodes and arbitrary single-particle angular
momentum $\ell$. These levels are degenerate with energy of $\hbar
\omega_c/2$, where $\omega_c$ is the cyclotron frequency $\omega_c=eB/m^*c$, and
$\lambda_c = l_B\sqrt{2} = \sqrt{2\hbar/(m^* \omega_c)} $;
see, e.g., the Appendix in \citeasnoun{yann07}.

By constructing a Slater determinant out of the
displaced Gaussian orbitals in {\Eq}~(\ref{uorb}) (with
$\lambda=\lambda_c$), and projecting out the good total angular
momentum $L$, one can derive analytic
expressions for the rotating electron molecule \cite{yann02.2} for
any number $N$ of fully spin-polarized electrons (i.e., with
$S=S_z=N/2$) and any $L$. Note that for large magnetic fields, the electrons
in the ground state are fully spin-polarized. Analytic expressions
were derived for both the cases of rotating electron molecules with $(0,N)$
\cite{yann02.2} and $(1,N-1)$ \cite{yann03.2} ring configurations. A
numerical investigation of lowest-Landau-level
rotating electron molecules exhibiting a configuration where
the electrons are arranged in a configuration consisting of $r$
concentric regular polygons ($n_1,n_2,\ldots,n_r$, $N=\sum_i^r n_i$)
was also presented \cite{yann04}. Corresponding analytic expressions
for rotating bosonic molecules for $N$ spinless bosons in the lowest Landau level in
a double-ring configuration, $(n_1,n_2)$ with $n_1+n_2=N$, were
subsequently derived \cite{yann10}.

Going beyond the rotating-electron or rotating-boson
molecular states (which describe pure
vibrationless rotations), a class of trial wave functions portraying
combined rotations and vibrations of Wigner molecules associated with
concentric polygonal rings was further introduced \cite{yann10,yann11}.
These trial functions, referred to as rovibrational molecular
functions, are valid for both bosons and fermions and provide a
correlated basis that spans the translationally invariant part of the
lowest-Landau-level spectra for both the yrast and excited lowest-Landau-level
states, and for both low and high angular momenta \cite{yann10}. As a result, the
restoration of broken symmetry approach can describe the totality of
the lowest-Landau-level states and not only the cusp states which are associated with
ground states exhibiting magic angular momenta that are precursors of
the fractional quantum Hall effect states characterized by fractions $\nu$.

A major subject in the lowest-Landau-level physics was the emergence of actual
broken-symmetry Wigner-solid crystal states. Such Wigner-solid
crystals were expected to appear for smaller fractions $\nu \leq
1/5$. It was thus surprising that a Wigner-crystal regime was
experimentally observed \cite{tsui10} in the neighborhood of
$\nu=1/3$ in the case of very clean samples. An interpretation of
these observations was achieved using linear superpositions (wave
packets) of angular-momentum-restored wave functions (specifically
the analytic ones of the rotating electron molecules).
These superpositions involve summation over
several cusp states with different magic angular momenta; they
naturally break the rotational symmetry to exhibit explicitly the
crystalline structure, without necessarily reverting back to the UHF
level. The triggering agent for the pinning of the rotating Wigner
molecule and the enforcing of symmetry breaking is the presence of
residual impurities and disorder in the sample.

For non-fully spin polarized electrons, the symmetry restoration in
the lowest Landau level must involve both the total spin $S$ and the angular momentum
$L$. Such combined $S$ and $L$ projection leading to spin-dependent
rotating electron molecules with $S < N/2$ was
performed for $N=4-5$ localized
electrons in the lowest Landau level \cite{jain07,yang07}. The combined spin and
space projection was also demonstrated for $N$ lowest-Landau-level electrons
confined in a ring geometry \cite{yang08.2}.

Of interest is the property that the edge states at zero-magnetic
field in a circular graphene dot with a zig-zag termination form a
collection mimicking the lowest-Landau-level manifold; these edge states appear due to
the existence of two valleys in the single-particle spectrum of the
zero-mass Weyl-Dirac graphene electron. The formation of rotating
Wigner molecules in this novel lowest-Landau-level manifold was investigated using
both configuration-interaction and projection techniques \cite{guin08,yann09}.

\subsection{Trapped ultracold neutral atomic gases and ions}
\label{sec:ultracold}

Ultracold trapped neutral atoms interact via a Dirac-delta contact potential,
namely, in {\Eq}\ (\ref{mbhn}) one takes
\begin{equation}
U({\bm r}_i - {\bm r}_j)= g \delta( {\bm r}_i - {\bm r}_j ).
\label{delt}
\end{equation}
In {\Eq}\ (\ref{delt}) above, the strength parameter $g$ can take both negative (attractive interaction)
and positive (repulsive interaction) values. Experimentally, this parameter can be varied
continuously from the attractive to the repulsive regime; see, e.g., \citeasnoun{joch12,bran15}.
For ultracold neutral atoms, the magnetic field ${\bm B}$ in {\Eq}\ (\ref{mbhn}) can be mimicked with
artificial synthetic fields \cite{spie14} or the rotational frequency $\Omega$ of a rotating
harmonic trap \cite{roma06} [with an appropriate modification of the parameter $\eta$ in {\Eq}\ (\ref{mbhn})].
$^6$Li (fermionic) and $^{87}$Rb (bosonic) are examples of trapped neutral atoms.

In the case of ultracold atoms and molecules, dipole-dipole two-body interactions were also
experimentally realized. However, no symmetry-restoration investigations with dipolar
interactions were reported as yet.

The restoration of angular momentum was employed \cite{roma04}
to investigate systems with a finite number $N$ of spinless neutral
and charged bosons (ions) confined in a 2D harmonic trap. The
broken-symmetry UHF-type orbitals were approximated as in {\Eq}~(\ref{uorb}),
treating the positions and the widths of the displaced
Gaussians as variational parameters (which corresponds to a
variation-after-projection scheme). Wigner molecules were described
for both neutral and charged bosons in the regime of strong
interparticle repulsion. For the case of neutral bosons, the
Wigner-molecule regime corresponds to a process of 2D fermionization,
when the strong repulsion keeps the particles away from each other
overtaking the propensity of bosons to bunch together due to
statistics. This fermionization behavior is well known for strongly
repelling strictly 1D bosons \cite{gira60}; in two dimensions, it has
also been recently further verified via exact numerical calculations
for two interacting bosons \cite{poll18}.
\begin{figure}[t]
\centering\includegraphics[width=6.0cm]{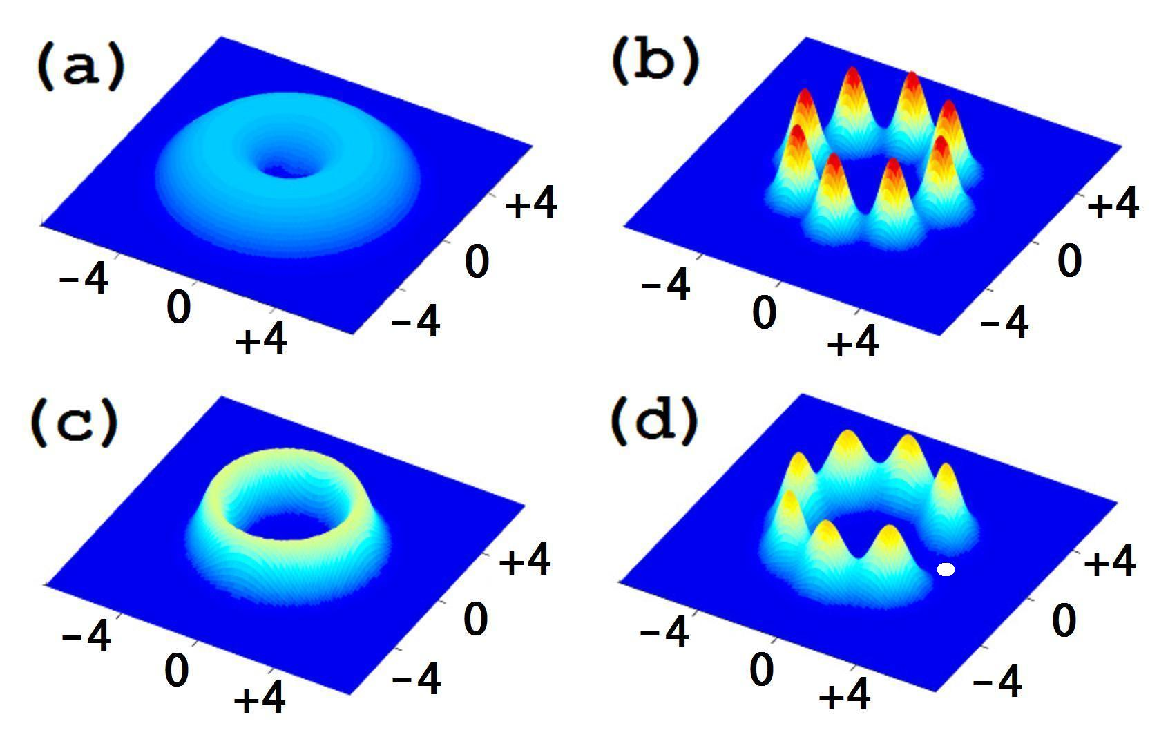}
\caption{
Single-particle densities and conditional probability distributions
for $N=8$ bosons in a rotating toroidal trap, see \citeasnoun{roma06} for details.
(a) Gross-Pitaevskii single-particle density. (b) UBHF single-particle density exhibiting
breaking of the circular symmetry. (c) Rotating-boson-molecule single-particle density
exhibiting circular symmetry. (d) Conditional probability distribution
for the rotating-boson-molecule wave function
(PRJ wave function, symmetry restored) revealing the hidden point-group symmetry in the
intrinsic frame of reference. The fixed (observation) point is denoted by a white dot.
The rotating-boson-molecule ground-state angular momentum is $L=16$.
Figure reprinted with permission from \protect\citeasnoun{roma06}.
Copyrighted by the American Physical Society.}
\label{rbm}
\end{figure}

The system of $N$ spinless bosons in {\it rotating\/} harmonic and
toroidal traps was also studied \cite{roma06} using
angular-momentum restoration techniques. Figure~\ref{rbm} illustrates
the patterns in the single-particle and two-body conditional
probabilities as the successive steps sketched in {\Fig}~\ref{hiera}
are applied. The conditional probability is defined [see {\Sec}~1.5 in \citeasnoun{yann07}]
as the probability for finding a particle at position ${\bm r}$ given that another
particle is located (fixed) at a reference point ${\bm r}_f$.
To be noted is the fact, see \cite{roma06}, that the ground state has zero
angular momentum only for small values of the rotational frequency
$\Omega$ (or equivalently for small magnetic field $B$); for larger values
of $\Omega$, the ground-state angular momentum increases in steps of
$N$, which is the hallmark of the emergence of magic angular momenta.

An interesting application \cite{roma08} of the methodology of
projection techniques is the restoration of the rotational symmetry
starting from broken-symmetry Gross-Pitaevskii solutions that
describe vortices in rotating traps. In this case, the
symmetry-restored many-body wave functions can be characterized as
rotating vortex clusters. The presence of vortices is not visible in the
single-particle densities of the rotating vortex clusters, which are homogeneous;
to reveal the hidden vortices, one needs to employ the conditional
probability distributions. The Gross-Pitaevskii vortex states are shown to be wave
packets composed of such rotating-vortex-cluster states.

Finally, we mention that the restoration of angular-momentum (under
the name of "continuous configuration-interaction") was applied
\cite{sakm04} in the case of an attractive Bose gas on a ring. For
strong attraction, the full-configuration-interaction method converges very slowly with the
increasing dimension of the employed discrete single-particle basis,
and thus the restoration of broken symmetry is advantageous, yielding
lower total energies compared to the nonconverged configuration-interaction ones.

\subsection{Spin-projected UHF, Hubbard systems, and natural molecules}
\label{sec:spin}

In addition to the description of novel strongly-correlated many-body
phases and phenomena \cite{yann07,yann06.4} for electrons in quantum
dots and trapped ultracold neutral atoms and ions, the two-step
method of symmetry breaking and symmetry restoration beyond nuclear
physics is also being developed in the direction of a powerful
computational approach that can rival in effectiveness the Kohn-Sham
density-functional computational technique. This computational
direction is mainly associated with the spin-projected UHF
(restoration only of spin) in the context of condensed-matter Hubbard
systems and natural molecules traditionally belonging to the field of
chemistry. The specific approach used to restore the total spin is
similar to that used for the 3D total angular-momentum projection in
nuclear physics \cite{ring2000}. This has definite practical
advantages \cite{voor15} for large-scale computations compared to the
prescription suggested by L\"{o}wdin \cite{(Low55a)}; however, see
\citeasnoun{vive18}. In this respect,
we mention that, unlike the symmetry-restoration techniques, the
single-determinantal Kohn-Sham density-functional formalism fails to
provide a complete description of magnetic states \cite{kapl18}, as
well as to describe properly the regime of strong static correlation
[when there are several competing degenerate states, as in
dissociation processes and the formation of Wigner molecules; see,
e.g., \citeasnoun{(Coh08),yann07,yann06.4}].

The Hubbard model \cite{hubb63} is widely used to describe strongly
interacting electrons in periodic potentials (associated with natural
ionic lattices in condensed-matter systems) and most recently
ultracold atoms trapped in artificial optical lattices. The Hubbard
model Hamiltonian for $N$ fermions is given by
\begin{eqnarray}
H=-t\sum_{<i,j>,\sigma}(a_{i,\sigma}^\mydagger a_{j,\sigma}+
a_{j,\sigma}^\mydagger a_{i,\sigma})+U\sum_{i=1}^N n_{i\uparrow}n_{i\downarrow},
\end{eqnarray}
\noindent
where $<i,j>$ denotes summation over nearest-neighbor sites and $\sigma$ sums over the up ($\uparrow$) and down
($\downarrow$) spins, with $n_{i\sigma} = a_{i,\sigma}^\mydagger a_{i,\sigma}$. The parameters $t$ and $U$ are the
hopping parameter and the on-site repulsion, respectively.

In particular, following an early publication \cite{igaw95}, the
method of restoration of spin symmetries was applied with
increasing numerical sophistication in the case of one-dimensional
Hubbard chains \cite{tomi04,tomi09,schm05,scus13}. Partially
two-dimensional Hubbard geometries (e.g., 2$\times$2 and 2$\times$4
plaquettes) were also studied \cite{scus12} using the
spin-projected HF approach. Interestingly, it was shown
that the exact ground-state in a four-site Hubbard plaquette can be
recovered by the spin-projected wave function irrespective of the
interaction strength \cite{fres14}. A combination of spin projection
with Gutzwiller-type double-occupancy screening was also
applied to 2D Hubbard lattices \cite{scus16}.

The restoration of the total-spin symmetry was employed further
to describe the ground-state correlations and dissociation profiles
of natural molecules \cite{scus11,scus12.2,scus13.2} familiar from
quantum chemistry. Examples of investigated systems were O$_2$, N$_2$,
H$_2$O, [Cu$_2$O$_2$]$^{2+}$ core, etc. A projected coupled cluster
theory is being also developed for both natural molecules and the
Hubbard model. It was shown \cite{scus17} that spin restoration via
projection techniques significantly improves unrestricted
coupled-cluster-method results while reimposing the required good
quantum numbers.

\subsection{Other electronic systems}
\label{sec:other}

In the early 1980's, it was discovered that the doubly-excited states
of the Helium atom exhibit rovibrational spectra that reflect the
formation of a highly nonrigid linear symmetric XYX "molecular"
structure \cite{kell80,berr89}, where the X's stand for the two
excited electrons and Y for the He nucleus. In addition to other
methods \cite{berr89}, these rovibrational spectra and corresponding
underlying collective wave functions were studied using the approach
of 3D angular-momentum restoration \cite{iwai89}. Such molecular
structures in highly excited atoms indicate that physical processes
associated with symmetry breaking cannot be dismissed even in the
case of natural atoms; they provide a bridge \cite{yann07,yann06.4}
to the regime of Wigner molecules in two-dimensional quantum dots.

Another notable application of projection techniques beyond nuclear
physics is the use of a number-projection method,
see {\Sec}~\ref{sec:restoration-at-finite} and \citeasnoun{ross95}, to
calculate the canonical-ensemble,
temperature-dependent free energy
of metal clusters for describing temperature effects in electronic-shell
energy contributions \cite{(Fra96b)}, and in particular for describing
temperature attenuation in ionization
potentials, electron affinities, and fission fragmentation of poly-cationic
and poly-anionic clusters \cite{yann97,yann00.1,yann02}.

Particle-number projection techniques were also used to
investigate the properties of superconducting metal grains
\cite{egid03,egid05}.

Finally worth mentioning is the use of projection-operator techniques
to describe the dynamic Jahn-Teller effect in natural molecules in
the case of tunneling between equivalently distorted energy-minimum
configurations of the adiabatic potential energy surface
\cite{dunn92,dunn12}. Naturally, due to the very large masses of the
ionic cores, the explicit symmetry-broken wave-packet state localized
within a single minimum can be observed in cases when tunneling is
suppressed \cite{bers16}. This is analogous to the observation of
pinned classical Wigner crystals of trapped ultracold ions
\cite{yann17,thom15}.

\subsection{Other emerging directions}

\subsubsection{Relation to entanglement and quantum information science.}
The emergence of modern quantum information theory is being built
around exotic and counterintuitive theoretical concepts, such as
entanglement \cite{aspe04,woot98} and quantumness
\cite{pian14,modi12}, which reflect the complexity of the structure
of the quantum wave functions (e.g., non-factorizability in the case
of two or more particles), or of quantum measurement. The
symmetry-breaking mean-field solutions are at a disadvantage in this
area because they do not conserve
symmetries of the many-body Hamiltonian. In this context, it was
shown \cite{zeng14} that the broken-symmetry BCS wave function
represents a class of wave functions where the required quantumness
was lost. It is noticeable that the lost quantumness [in the
form of proper description for the concurrence \cite{woot98}
and quantum discord \cite{zure01}]
is restored simultaneously with the restoration of the particle
number symmetry in the projected BCS wave function.

Earlier, the ability of the total-spin, symmetry-restored wave
function to describe properly the entanglement [in the form of concurrence
and von Neumann entropy \cite{woot98}] for two electrons in a double
quantum dot under the influence of an increasing magnetic field was
also investigated in detail \cite{yann06.1,yann06.4,yann07}.

\subsubsection{Time evolution in finite systems out of equilibrium.}
Apart from the small-amplitude harmonic vibrations, broken-symmetry
wave functions (single determinants or permanents) fail to describe
the proper time-evolution behavior when propagated in time with the
corresponding mean-field Hamiltonian \cite{grif76,yann17}; see also
Chapter 12.2.4 in \citeasnoun{ring2000}. This drawback of the
mean-field treatment of finite systems was earlier discussed in the
framework of heavy-ion collisions in nuclear physics \cite{grif76}.
It is easily overcome by expressing the broken-symmetry wave function
as a wave packet (superposition) of symmetry-restored wave functions
and evolving independently in time (by multiplying by a
time-dependent phase) each component of the wave packet. Using this
approach, other symmetry-broken wave packets (different from the UHF
solutions) can be envisaged that exhibit single-particle densities
with controlled periodicities in both space and time, as was recently
discussed in the framework of implementing a quantum space-time
crystal of ultracold atoms or ions in a ring-shaped trap
\cite{yann17}. If the initial wave packet reproduces the UHF or
Gross-Pitaevskii broken-symmetry solution, revival and recurrence
in-time behavior is generated \cite{kavo18}.

\section{Projected statistics}
\label{sec:statistics}

Symmetry restoration can also be introduced in the
context of quantal statistical ensembles.
The formal treatment of
statistical ensembles requires the replacement of mean values of operators  by traces over the
whole Fock space  "weighted" by a density matrix operator $\hat D$ responsible
for the probability distribution \cite{kadanoff1994quantum,huang1987statistical,schieve1987statistical,attard2015quantum}.
The form of this operator depends on the problem at hand but it is typically
defined as the exponential of the Hamiltonian plus
some additional terms. When
the problem is restricted to the mean-field level, the density matrix operator
is the exponential of the one-body  mean-field Hamiltonian including pairing fields. In the trace, all
possible multi-quasiparticle excitations of the mean-field ground state are considered.

In those mean-field applications in nuclear physics that require the inclusion of pairing
correlations, the density matrix operator to be used is the one of the
grand canonical ensemble. This is required to accommodate the possibility of
exchanging particles with the external "reservoir". When restricted to the
mean-field approximation at finite
temperature $T$, $\hat D$ is proportional to the exponential of the
one body HFB Hamiltonian $H_\mathrm{HFB}$
\begin{equation}
        \hat{D}_\mathrm{HFB} = Z^{-1} \exp \left[ - \beta (\hat H_\mathrm{HFB}-\lambda \hat N) \right]
\end{equation}
with $\beta=1/(k_{B}T)$ and $Z=\Tr [\hat{D}_\mathrm{HFB}]$ the partition function. As $\hat H_\mathrm{HFB}-\lambda \hat N$ is a quadratic form of creation
and annihilation operators
$$
\hat H_\mathrm{HFB}-\lambda \hat N= \frac{1}{2}
(a^{\mydagger}\, a) (\mathcal{H}
-\lambda \mathcal{N}) \left(\begin{array}{c}
a \\ a^{\mydagger}  \end{array}\right)
=\frac{1}{2}
(a^{\mydagger}\, a) \left(
\begin{array}
[c]{cc}%
h-\lambda & \Delta\\
-\Delta^{\ast} & -h^{\ast}+\lambda%
\end{array}
\right)
\left(\begin{array}{c}
a \\ a^{\mydagger}  \end{array}\right)
$$
its exponential $\hat{D}_\mathrm{HFB}$ is the operator of a canonical transformation acting
on the quasiparticle operators
satisfying
\begin{equation}
\left(\begin{array}{c} a \\ a^{\mydagger}  \end{array}\right) \hat D_\mathrm{HFB} =
\hat D_\mathrm{HFB}
 \exp \left[ -\beta \mathcal{H}'\right] \left(\begin{array}{c}
a \\ a^{\mydagger}  \end{array}\right)
\label{eq:stat1b}
\end{equation}
with $\mathcal{H}'=\mathcal{H}-\lambda \mathcal{N}$. The matrices $\mathcal{H}$ and
$\mathcal{N}$ have been introduced in {\Eq} \ref{eq:59}.
This identity allows the calculation of any statistical trace by using Gaudin's theorem
for Hartree-Fock states \cite{Gaudin60}, which is the extension to statistical ensembles of Wick's theorem.
Its generalization to HFB states is straightforward \cite{(Rin84a),ROSSIGNOLI1994350}.
Like Wick's theorem, it allows us to write the trace of any operator times
the HFB statistical density matrix as a contraction of the operator's matrix
elements with the density and pairing tensors for the statistical ensemble.
Introducing the set of operators
$a_{\mu}=(a_{1},\ldots,a_{N},a^{\mydagger}_{1},\ldots,a^{\mydagger}_{N})$
one gets for the contraction
\begin{equation} \label{eq:stat7}
        \mathrm{Tr}[a_{\mu} a_{\nu} \hat D_\mathrm{HFB}] =
        \sum_{\rho} \{a_\rho ,a_\nu \}
        \left( \frac{\mathbb I}{\mathbb{I}+\exp \left[ -\beta \mathcal{H}' \right]}
                \right)_{\mu \rho}
\end{equation}
The zero temperature limit of this result can be used to derive a sort of generalized Wick's theorem for
multiquasiparticle overlaps \cite{(Per07c)}. If we denote by
$$\mathcal{W} = \left( \begin{array}
[c]{cc}%
U & V^*\\
V & U^*%
\end{array}
\right)
$$
the matrix diagonalizing $\mathcal{H}'$ (see {\Eq} \ref{eq:59} ) then
\begin{equation}\label{eq:stat8a}
\frac{\mathbb I}{\mathbb{I}+\exp \left[ -\beta \mathcal{H}'\right] } = \mathcal{W} \left( \begin{array}
[c]{cc}%
1-f & 0\\
0 & f%
\end{array}
\right)
\mathcal{W}^+
\end{equation}
where $f_\mu=1/(1+\exp (\beta E_\mu))$ are the Fermi statistical occupation factors given in
terms of the quasiparticle energies $E_\mu$, eigenvalues of the HFB-equations, see {\Eq} \ref{eq:59}. Inserting
this in {\Eq} \ref{eq:stat7} we finally arrive to the standard definitions for the finite
temperature density $\rho=V^*(1-f)V^T+UfU^+$ and $\kappa =UfV^+V^*(1-f)U^T$.

\subsection{Symmetry restoration at finite temperature}
\label{sec:restoration-at-finite}

In the context of finite temperature or statistical ensembles, symmetry restoration means that
the density matrix operator $\hat D$ has to be able to select from all the states considered
in the statistical trace only those with the desired set of quantum
numbers. The easiest way to achieve this property is by sandwiching the
statistical operator with the projector onto the required quantum
numbers one wants to select. For instance, for particle-number restoration we have to replace
the density matrix operator $\hat D$ by
\begin{equation}
        \hat{D}_{N} = \hat{P}^{N \mydagger} \hat{D} \hat{P}^{N}.
\end{equation}
For non-abelian symmetry groups, like
the one of angular-momentum projection
the expression of the density matrix operator gets a bit more involved
and the reader is referred to  \citeasnoun{ROSSIGNOLI1994350}
for the technical details.
In general, the expression for $\hat D_N$ is rather involved, but it
simplifies enormously if $\hat D$ is restricted to be the exponential
of an one-body operator, as it is the case in the mean-field approximation to the exact $\hat D$.
In the following and just to illustrate the method, we will restrict the discussion to the abelian
case of particle-number projection. The projector $P^{N}$ is a linear
combination of exponentials of one body operators. Therefore, $ \hat{P}^{N \mydagger} \hat{D}_\mathrm{HFB} \hat{P}^{N}$
becomes a (very involved) linear combination of products of exponentials
of one body operators. As in {\Eq}~(\ref{eq:stat1b}) those products are
generators of canonical transformations
\begin{equation}\label{eq:stat8}
\hat T_{1}^\mydagger \hat T_{2}^\mydagger   \left(\begin{array}{c}
a \\ a^{\mydagger}  \end{array}\right) \hat T_{2} \hat T_{1} = \exp (\mathcal{T}_{1}) \exp (\mathcal{T}_{2}) \left(\begin{array}{c}
a \\ a^{\mydagger}  \end{array}\right)
\end{equation}
where $\mathcal{T}_{i}$ are the matrices defining the one body operators in the
exponents of $\hat T_{i}$,
$$
\hat T_i = \exp \left[ \frac{1}{2} \sum_{\mu \nu} a^+_\mu (\mathcal{T}_i)_{\mu \nu} a_\nu        \right]
$$
As a consequence of this property, Gaudin's theorem can still be used
just replacing the $\exp [-\beta \mathcal{H}']$ in {\Eq}~(\ref{eq:stat7}) by the appropriate product of
exponentials (see \citeasnoun{ROSSIGNOLI1994350} for details). The only difficulty
in carrying out this program is in the evaluation of the entropy, required to evaluate the
free energy $F=H-TS$. The standard definition of the entropy involves
the logarithm of the density matrix operator. In the standard
mean-field approximation, this logarithm can be evaluated analytically and
the final expression for the entropy in terms of quasi-particle energies is
straightforward. Unfortunately, the projected density matrix can not be
expressed in general as
the exponential of an  operator and therefore the evaluation of the entropy
becomes a very complicated task \cite{Esebbag1993}. In spite of these
difficulties, the use of projected statistics proved to be advantageous
over other techniques when applied in the spirit of projection after variation, that does not require
the evaluation of the entropy \cite{Fanto2017}. The intrinsic difficulty
associated with the sign ambiguity in the evaluation of the partition function
was addressed in \citeasnoun{Fanto2017} in a time reversal preserving
scenario and further generalized using the
Pfaffian method \cite{(Rob09)} to the more general case involving time reversal breaking
intrinsic states \cite{(Fan17)}.

The shell model Monte Carlo method \cite{(Lan93a),(Koo97d),(Koo97c)} is often used to
evaluate the partition function of nuclei with relatively large
configuration spaces. The shell model Monte Carlo requires the use of
particle-number projection to carry out calculations in the more
convenient canonical ensemble \cite{(Alh99b)} and therefore this
constitutes another field of application of the techniques discussed
here. Recently, the use of particle-number projection to carry out calculations
in the canonical ensemble was also explored in \citeasnoun{Magnus2017}.

\subsection{Thermo-field dynamics}
\label{sec:tfm}

To finish this section, let us briefly mention an alternative to the
traditional approach described above and known under the name of thermo-field dynamics.
It consists in computing the traces of statistical operators by using
mean values of pure states defined in an extended Fock space
including twice the original degrees of freedom. This approach was introduced in the context of quantum
field theory by Takahashi and Umezawa \cite{TFD}. In this approach, the statistical
average of an operator $\hat F$ with probabilities $p_{n}$
\begin{equation}
        \langle \hat F \rangle_{\text{stat}} = \sum_{n} \langle n | \hat F | n \rangle p_{n},
\end{equation}
is replaced by the mean value of $\hat F$ with the wave function
\begin{equation}
        |\Phi\rangle = \sum_{n} \sqrt{p_{n}} | n \tilde{n} \rangle =
        \sqrt{p_{1}} |1\tilde{1}\rangle + \sqrt{p_{2}} |2\tilde{2}\rangle + \cdots
\end{equation}
which is a linear combination of wave functions $|n\tilde{n}\rangle=|n\rangle \otimes |\tilde{n}\rangle$ defined
in an extended Fock space which is the tensor product of the original Fock space with itself.
The ket $|\tilde{n}\rangle$ represents a
new set of states with identical characteristics as $|n\rangle$.
In the extended Fock space all the operators are defined as $\hat F \otimes \mathbb{I}$ where
$\mathbb{I}$ is the identity in the space spanned by $|\tilde{n}\rangle$.
A new set of creation and annihilation single-particle operators $\tilde{a}^{\mydagger}_{k}$
and $\tilde{a}_{k}$ satisfying fermion canonical commutation relations and anti-commuting with
all the elements of the original ${a}^{\mydagger}_{k}$ and ${a}_{k}$ set is required too.
An advantage of the formalism is that $|\Phi\rangle $ can be written as
a HFB state, vacuum of a set of quasiparticles defined in terms of the $\tilde{a}^{\mydagger}_{k}$
,$\tilde{a}_{k}$, ${a}^{\mydagger}_{k}$ and ${a}_{k}$ by means of an appropriate BCS like
transformation. Therefore, we can use verbatim all the zero temperature formalism
developed before (including the generalized Wick's theorem) to restore symmetries but taking into account properly the
doubling of the single-particle Fock space. It is not clear, however, if
this method represents any advantage over the traditional one due to the
doubling of matrix sizes.
The procedure is analogous to the construction of statistical ensembles by
taking the trace over a subsystem of the whole Hilbert space of a pure
state density matrix operator. Applications to nuclear physics in the
context of symmetry restoration are given in \citeasnoun{Tanabe2005} but only
formal expressions are developed in the mentioned reference.

\section{Summary, conclusions, and perspectives}
\label{sec:summary}
In recent years, research in mesoscopic many-body systems
witnessed a discernable progress with the development of the
state-of-the-art models and methods. In particular, {\it ab initio} methods
and mean-field theory based on effective interaction are now widely
used to elucidate rich and fascinating properties of these
quantum many-body systems. On the one hand, applications of the {\it ab initio}
methods are still restricted to lighter systems only, whereas the
mean-field approaches can be applied to investigate mesoscopic systems of any size.
In particular, in nuclear physics, the density functional
theory was employed to investigate ground-state properties of all
nuclear species predicted to exist in the Segr\'e chart.

Spontaneous symmetry breaking mechanism, inherent to the mean-field
based approaches,  played an important role in our understanding of
many-body systems. For instance, in rotating nuclei, the breaking of rotational
symmetry  led to the fundamental concept of deformation in nuclei.
Nevertheless, the quantal fluctuations of the observables, absent in mean-field
approaches, become quite important for mesoscopic systems.

To build quantal fluctuations on top of the mean-field solutions, several
approaches were developed. A powerful method to include these
fluctuations is through the restoration of the broken symmetries.
In the prelude section of this review, the spontaneous
symmetry breaking mechanism was illustrated through three simple
examples. The main purpose of the present review was to provide an
overview of the recent developments and, more importantly, to bring to
focus the bottlenecks in the application of symmetry-restoration
methods.

The general formalism of symmetry restoration, whose origin can be traced in
group theory and generator coordinate method, was laid out in
{\Sec}~\ref{sec_srgf}. There we
distinguished between the symmetry restoration for abelian groups (particle number, linear momentum, parity),
which have mathematical properties of projection operators, versus the symmetry restoration
for non-abelian groups, as is the case of the rotational symmetry, relevant for
the spatial coordinates, spin, or isotopic spin. In the later case, the
symmetry-restoration operators do not obey properties of a
projection operator, but they nevertheless project out the relevant symmetry quantum numbers.
Further, we discussed the fact that the symmetry restoration can be performed either
before minimization or after minimization of the energy
functional. In the former approach, commonly referred to as variation
after projection, symmetry breaking states
(often called ``intrinsic states") are determined by application of the variational
principle on the projected energy (i.e., energy computed with
the projected wave functions). In this procedure, different intrinsic
states are obtained for different quantum numbers of the restored
symmetry. In the latter approach, referred to as projection after variation,
the intrinsic state is determined without consideration of the subsequent
projection.

In general, implementation of the symmetry restoration is numerically quite
challenging, especially in realistic applications where several
quantum numbers need to be restored simultaneously, and calculations need to be
performed in the spirit of variation-after-projection method. Due to
these numerical challenges, development of approximate projection methods
were actively pursued  by exploiting
the sharp character of the overlap kernel when the intrinsic state strongly
breaks the underlying symmetry. This resulted in the development of
popular methods of Lipkin-Nogami and Kamlah, and were
discussed in {\Sec}~\ref{sec:approximate}. We showed that these
methods lead to successful approximations, like the mean-field cranking model --
a useful concept to understand the physics of rotational
bands. Symmetry-restoration methods were also successfully applied to
simple nuclear models, where the Hamiltonian is separable or the configuration
spaces are limited to a few oscillator shells. These applications were
discussed in {\Sec}~\ref{sec:simple}.

Although the mechanism of symmetry restoration can be consistently formulated
for systems described in terms of a Hamiltonian operator, this is not the
case for the energy density functionals, which are commonly employed in
nuclear physics to provide a description of low energy observables all over the
Segr\'e chart. The complexity of nuclear interaction, and in-medium effects
that characterize many-body systems, required the introduction of phenomenological
density-dependent interaction terms, for which symmetry-restoration methods cannot be uniquely
defined. This is further aggravated when separate
interactions are being considered in particle-particle and
particle-hole channels.
Recently, there were several attempts to overcome these difficulties, but a
satisfactory solution, covering both sources of problems, is still not available. A possible
solution, which is being vigorously pursued, is to base the functionals on the
Hamiltonian picture with explicit three-body terms. However, this
approach is not yet sufficiently developed to give definite answers at this stage.
Nevertheless, many symmetry restored calculations performed with
present-day energy density functionals seem
to provide reasonable and a consistent picture of low-energy nuclear
phenomena, as was elucidated in {\Sec}~\ref{sec:functional}. The
results, however, should be taken with a pinch of salt as they might be
contaminated with spurious effects.

Mesoscopic condensed matter systems and the physics of atoms and molecules, as well as
assemblies of trapped ultracold ions and neutral atoms, are mostly free from the above
mentioned difficulties as the interaction is often just the Coulomb or a contact
interaction between the constituents of the system. Applications of symmetry
restoration to those areas share many technical details with the ones in
nuclear physics, but there are also clear differences like the fact that rotational
symmetry restoration can be carried out separately for the spatial coordinates
and the spin in condensed matter physics. Many applications were presented
where symmetry breaking and restoration represent an easy way to understand
the complexity of the problem, see {\Secs}~\ref{sec:functional} and \ref{sec:mesoscopic}.
Finally, in {\Sec}~\ref{sec:restoration-at-finite}
we demonstrated that the concept of
symmetry restoration can be extended to the realm of quantum statistical
mechanics where pure states are replaced by a set of quantum
states with a prescribed probability distribution.

Based on the results presented in this work, it can be concluded that the method of symmetry
restoration applied to mean-field wave functions provides a simple and fruitful
mechanism to incorporate important dynamic correlations, while
still using a simple framework of product wave
functions. Furthermore, in this approach one stays within a
fully quantum mechanical description from the beginning to end,
and the classical picture of collective motion does not have to be invoked.

The generator coordinate method can be
employed along with the symmetry restoration to provide a powerful framework
to describe quantal fluctuations of relevant degrees of freedom
around the mean-field values.
In future, we expect development of more advanced configuration
interaction approaches with symmetry
projected states as the basic building blocks. For instance, projected multi-quasiparticle
basis configurations can be constructed around the optimal mean-field, in the
spirit of traditional shell-model approach, to describe the physics
of excited configurations and also to incorporate many-body correlations
in the ground state. Finally, inspired by the successes of symmetry
breaking and restoration in mean-field approaches, in nuclear physics
and quantum chemistry analogous methods were also recently
implemented in the context of the so-called {\it ab initio}
calculations, see, e.g., \citeasnoun{(Dug14a),(Dug16),(Qiu19),(Yao19)}
and {\Sec}~\ref{sec:ab-initio}.

\section{Acknowledgments}
JD would like to thank Micha{\l} B\k{a}czyk for a collaboration on
the doubly-symmetric-potential-well model.
The work of JD was partly supported by the STFC Grants No.~ST/M006433/1
and No.~ST/P003885/1 and by the Polish National Science Centre under
Contract No.~2018/31/B/ST2/02220.
The work of LMR was partly supported by Spanish
MINECO Grant No. PGC2018-094583-B-I00.
PR acknowledges partial support from the Deutsche Forschungsgemeinschaft (DFG, German Research Foundation)
under Germany's Excellence Strategy -- EXC-2094 -- 390783311.
CY wishes to thank his co-authors at Georgia Tech, and
especially Uzi Landman, head of the Center for Computational Materials
Science. The work of CY was supported over the years by the
Office of Basic Energy Sciences of the US D.O.E. (Grant No.
FG05-86ER45234) and by the Air Force Office of Scientific
Research (USA) (Grant No. FA9550-15-1-0519).

\appendix

\section{Overlaps and matrix elements between HFB states}
\label{sec:AppA}

The restoration of the symmetries broken by intrinsic HFB states requires to consider matrix elements
of various operators between different HFB states. This is a rather general statement because the
action of any operator belonging to the symmetry group on a given HFB state is again another HFB
state. The origin of this property lies on the fact that the algebra of
one-body operators can be used to provide a representation of any matrix Lie
algebra \cite{gilmore2008}. Therefore, the symmetry operations (which are members of the
group spanned by the corresponding Lie algebra) are equal to exponentials of one
body operators and the Thouless theorem \cite{(Tho62a),Man75} applies.

The evaluation of the required matrix elements is best carried out with the help
of the generalized Wick's theorem. The theorem, that can be derived in many different
ways \cite{Lowdin1955,ONISHI1966367,(Bal69a),(Har95a),(Ber12b),ring2000},  states that the matrix elements
of an arbitrary operator $\hat O$ between arbitrary, non orthogonal, HFB
states $|\Phi_0\rangle$ and $|\Phi_1\rangle$, $\langle \Phi_0 | \hat O | \Phi_1\rangle / \langle \Phi_0 | \Phi_1 \rangle $,
can always be written in terms of the sum of all possible two-quasiparticle contractions,
\begin{equation}\label{eq:cont}
\frac{\langle \Phi_0 | \beta_\mu \beta_\nu | \Phi_1 \rangle}{\langle \Phi_0 | \Phi_1 \rangle} =  C_{\mu \nu},
\quad\quad
\frac{\langle \Phi_0 | \beta_\mu \beta_\nu^\mydagger | \Phi_1 \rangle}{\langle \Phi_0 | \Phi_1 \rangle} =  \delta_{\mu \nu},
\end{equation}
where the $\beta_\mu$ and $\beta^\mydagger_\mu$ are annihilation and creation operators
associated with $|\Phi_0\rangle$.
The only non-trivial contraction is given in terms of the skew-symmetry matrix $C_{\mu \nu}$ which is
the product of the inverse of $A=U^\mydagger_0 U_1 + V^\mydagger_0 V_1$ times
$B=U^\mydagger_0 V_1 + V^\mydagger_0 U_1$, i.e., $C = A^{-1}B$.
The $U_0$, $V_0$ and $U_1$, $V_1$ are the Bogoliubov transformation
amplitudes of the corresponding HFB states.
As an example, one of the terms entering  the Hamiltonian overlaps is given by
\begin{equation}
\frac{\langle \Phi_0 | \beta_\mu \beta_\nu \beta_\sigma \beta_\rho | \Phi_1 \rangle }
{\langle \Phi_0 | \Phi_1 \rangle } =
C_{\mu \nu} C_{\sigma \rho} - C_{\mu \sigma} C_{\nu \rho} + C_{\mu \rho} C_{\nu \sigma}
\end{equation}
In the general case, where the matrix element of a product of $n$ creation and annihilation
quasiparticle operators is required, the number of terms in the sum grows very quickly and
is given by $(n-1)!!$. This is the so called
combinatorial explosion problem (see \citeasnoun{Hu2014} for an example) that
hampers applications where multi-quasiparticle excitations have to be considered.  This
difficulty can be avoided using the formulation of \citeasnoun{(Ber12b)} in terms of
Pfaffians (see below).

The overlap $\langle \Phi_{0} | \Phi_{1} \rangle$ given by the Onishi formula
\cite{ONISHI1966367,(Bal69a),(Har95a)},
\begin{equation}
\langle \Phi_{0} | \Phi_{1} \rangle = \pm \sqrt{\det A},
\label{eq:onishi}
\end{equation} 
suffers from a sign indeterminacy that requires further consideration.
The sign of the overlap affects the quantities to be integrated in  the symmetry restoration or configuration-mixing methods. A
wrong assignment of the sign even in a small integration interval can substantially change the value of the integral.
The sign problem was addressed in the past using different strategies like
continuity arguments or determining pairwise degenerate eigenvalues of a
general matrix \cite{(Nee83)}. However, the use of techniques of fermion
coherent states allows us to avoid the sign problem by expressing the overlap in terms of the Pfaffian of
a skew-symmetric matrix \cite{(Rob09),Robledo2011,(Ber12b),(Ave12)},
\newcommand{\openone}{\mathbb{I}}
\begin{equation}
\langle\Phi_{0}|\Phi_{1}\rangle=s_{N}\textrm{pf}(\mathbb{M})=s_{N}\textrm{pf}\left(\begin{array}{cc}
M^{(1)} & -\openone\\
\openone & -M^{(0)\,*}\end{array}\right)\label{eq:Over_1},\end{equation}
for the skew-symmetric matrices $M^{(i)}=(V_{i}U_{i}^{-1})^{*}$ given in terms of the Bogoliubov amplitudes.
Formula (\ref{eq:Over_1}) assumes that both HFB states are normalized as $\langle0|\Phi_{i}\rangle=1$
and that their Bogoliubov amplitudes are expressed in a common single-particle basis
of dimension $N$, which defines the phase factor $s_{N}=(-1)^{N(N+1)/2}$; see \citeasnoun{(Ave12)}
for a generalization relaxing this assumption.
We also note that recently \citeasnoun{(Miz18),(Bal18a)} have provided
additional perspectives into the sign problem.
The Pfaffian of a skew-symmetric matrix is a quantity similar to the determinant and shares
with it many properties.
The numerical evaluation of the Pfaffian
can be carried out using the traditional algorithms of linear
algebra with a cost similar to the one of the determinant \cite{(Gon11a),(Wim12)}.

\newpage

\bibliographystyle{jphysicsB-withTitles}

\end{document}